\input harvmac
\newcount\figno
\figno=0
\def\fig#1#2#3{
\par\begingroup\parindent=0pt\leftskip=1cm\rightskip=1cm\parindent=0pt
\global\advance\figno by 1
\midinsert
\epsfxsize=#3
\centerline{\epsfbox{#2}}
\vskip 12pt
{\bf Fig. \the\figno:} #1\par
\endinsert\endgroup\par
}
\def\figlabel#1{\xdef#1{\the\figno}}
\def\encadremath#1{\vbox{\hrule\hbox{\vrule\kern8pt\vbox{\kern8pt
\hbox{$\displaystyle #1$}\kern8pt}
\kern8pt\vrule}\hrule}}

\overfullrule=0pt

%macros
%
\def\tilde{\widetilde}
\def\bar{\overline}
\def\Z{{\bf Z}}

\def\R{{\bf R}}

\font\zfont = cmss10 %scaled \magstep1
\font\litfont = cmr6

\def\bigone{\hbox{1\kern -.23em {\rm l}}}
\def\ZZ{\hbox{\zfont Z\kern-.4emZ}}
\def\half{{\litfont {1 \over 2}}}

\def\CM{{\cal M}}

\def\Im{{\rm Im ~}}

%%Greg's macros

% Something to deal with sub-sub-sections

\def\unlockat{\catcode`\@=11}
\def\lockat{\catcode`\@=12}

\unlockat
% Something to deal with sub-sub-sections

\def\newsec#1{\global\advance\secno by1\message{(\the\secno. #1)}
\global\subsecno=0\global\subsubsecno=0\eqnres@t\noindent
{\bf\the\secno. #1}
\writetoca{{\secsym} {#1}}\par\nobreak\medskip\nobreak}
\global\newcount\subsecno \global\subsecno=0
\def\subsec#1{\global\advance\subsecno
by1\message{(\secsym\the\subsecno. #1)}
\ifnum\lastpenalty>9000\else\bigbreak\fi\global\subsubsecno=0
\noindent{\it\secsym\the\subsecno. #1}
\writetoca{\string\quad {\secsym\the\subsecno.} {#1}}
\par\nobreak\medskip\nobreak}
\global\newcount\subsubsecno \global\subsubsecno=0
\def\subsubsec#1{\global\advance\subsubsecno by1
\message{(\secsym\the\subsecno.\the\subsubsecno. #1)}
\ifnum\lastpenalty>9000\else\bigbreak\fi
\noindent\quad{\secsym\the\subsecno.\the\subsubsecno.}{#1}
\writetoca{\string\qquad{\secsym\the\subsecno.\the\subsubsecno.}{#1}}
\par\nobreak\medskip\nobreak}

\def\subsubseclab#1{\DefWarn#1\xdef
#1{\noexpand\hyperref{}{subsubsection}%
{\secsym\the\subsecno.\the\subsubsecno}%
{\secsym\the\subsecno.\the\subsubsecno}}%
\writedef{#1\leftbracket#1}\wrlabeL{#1=#1}}% Macros for boxes
\lockat

\font\cmss=cmss10 \font\cmsss=cmss10 at 7pt
\def\IR{\relax{\rm I\kern-.18em R}}

\def\Tr{{\rm Tr}}
\def\vol{{\rm vol}}

\def\IL{\relax{\rm I\kern-.18em L}}
\def\IH{\relax{\rm I\kern-.18em H}}
\def\IR{\relax{\rm I\kern-.18em R}}
\def\IC{\relax\hbox{$\inbar\kern-.3em{\rm C}$}}
\def\IZ{\relax\ifmmode\mathchoice
{\hbox{\cmss Z\kern-.4em Z}}{\hbox{\cmss Z\kern-.4em Z}}
{\lower.9pt\hbox{\cmsss Z\kern-.4em Z}}
{\lower1.2pt\hbox{\cmsss Z\kern-.4em Z}}\else{\cmss Z\kern-.4em
Z}\fi}
\def\IB{\relax{\rm I\kern-.18em B}}
\def\IC{{\relax\hbox{$\inbar\kern-.3em{\rm C}$}}}
\def\ID{\relax{\rm I\kern-.18em D}}
\def\IE{\relax{\rm I\kern-.18em E}}
\def\IF{\relax{\rm I\kern-.18em F}}
\def\IG{\relax\hbox{$\inbar\kern-.3em{\rm G}$}}
\def\IGa{\relax\hbox{${\rm I}\kern-.18em\Gamma$}}
\def\IH{\relax{\rm I\kern-.18em H}}
\def\II{\relax{\rm I\kern-.18em I}}
\def\IK{\relax{\rm I\kern-.18em K}}
\def\IP{\relax{\rm I\kern-.18em P}}
\def\IQ{\relax\hbox{$\inbar\kern-.3em{\rm Q}$}}

\def\hat{\widehat}
\def\CM {{\cal M}}
\def\CN {{\cal N}}

\def\CD {{\cal D}}
\def\CF {{\cal F}}
\def\CJ {{\cal J}}

\def\CO {{\cal O}}

\def\CE {{\cal E}}
\def\CG {{\cal G}}
\def\CH {{\cal H}}

\def\CB {{\cal B}}
\def\CS {{\cal S}}

\def\CQ{{\cal Q}}

\def\mod{{\rm mod}}
\def\p{\partial}

\def\gof{ \Gamma^0(4)}

\def\duab{ {d \bar u \over  d \bar a} }
\def\dau{ {da \over  du} }
\def\daub{ {d \bar a \over  d \bar u} }
\def\vt#1#2#3{ {\vartheta[{#1 \atop  #2}](#3\vert \tau)} }

\def\inbar{\,\vrule height1.5ex width.4pt depth0pt}

%%% END GREG'S MACROS

%%% REFERENCES

\lref\AFT{I. Antoniadis, S. Ferrara, T.R. Taylor,
``N=2 Heterotic Superstring and its Dual Theory in Five Dimensions,''
hep-th/9511108; Nucl.Phys. B460 (1996) 489-505.}

\lref\bateman{A. Erdelyi et. al. , {\it Higher
Transcendental Functions, vol. I,
Bateman manuscript project} (1953)
McGraw-Hill}

\lref\borcherds{R. Borcherds,
``Automorphic forms with singularities on
Grassmannians,'' alg-geom/9609022}

\lref\cohen{H. Cohen, ``Sums Involving the
Values at Negative Integers of L-functions of
Quadratic Characters,''
Math. Ann. {\bf 217}(1975) 271.}

\lref\dkl{L. Dixon,  V. S. Kaplunovsky and J. Louis, ``Moduli-dependence of
string
loop corrections to gauge coupling constants, ''Nucl. Phys. {\bf B329} (1990)
27. }

\lref\doncobord{S. Donaldson,
``Irrationality and the $h$-cobordism conjecture,''
J. Differential Geometry {\bf 26}(1987)141.}

\lref\donaldson{S. Donaldson, ``Polynomial Invariants For Smooth
Four-Manifolds,''
Topology {\bf 29} (1990) 257.}

\lref\donint{S.K. Donaldson, ``Connections,
Cohomology and the intersection forms of
4-manifolds,'' J. Diff. Geom. {\bf 24 }
(1986)275.}

\lref\DoKro{S.K.~ Donaldson and P.B.~ Kronheimer,
{\it The Geometry of Four-Manifolds},
Clarendon Press, Oxford, 1990.}

\lref\donrev{S.K. Donaldson, ``The Seiberg-Witten
equations and 4-manifold topology,''
Bull. Amer. Math. Soc. {\bf 33} (1996) 45.}

\lref\eg{G. Ellingsrud and L. G\"ottsche,
``Wall-crossing formulas, Bott residue formula and
the Donaldson invariants of rational surfaces,''
alg-geom/9506019.}

\lref\ez{M. Eichler and D. Zagier, {\it The theory of
Jacobi forms}, Birkh\"auser, 1985.}

\lref\finstern{R. Fintushel and R.J. Stern,
``The blowup formula for Donaldson invariants,''
alg-geom/9405002;Annals of Math. {\bf 143} (1996) 529.}

\lref\FrMor{R. Friedman and J.W. Morgan,
{\it Smooth Four-Manifolds and Complex Surfaces},
Springer Verlag, 1991.}

\lref\fre{D. Anselmi and P. Fr\'e, ``Topological $\sigma$-Models In Four
Dimensions And Triholomorphic Maps,'' Nucl. Phys. {\bf B416} (1994) 255,
``Gauged Hyper-Instantons And Monopole Equations,'' Phys. Lett. {\bf B347}
(1995)
247.}

\lref\gottsche{L. G\"ottsche, ``Modular forms and Donaldson
invariants for 4-manifolds with $b_+=1$,'' alg-geom/9506018.}

\lref\gottzag{L. G\"ottsche and D. Zagier,
``Jacobi forms and the structure of Donaldson
invariants for 4-manifolds with $b_+=1$,''
alg-geom/9612020}

\lref\greenseib{Green and Seiberg}

\lref\hm{J. A. Harvey and G. Moore,
``Algebras, BPS states, and strings,''
hep-th/9510182; Nucl. Phys. {\bf B463}(1996)315.}

\lref\kontsevich{M. Kontsevich,
``Product formulas for modular forms on
$O(2,n)$,'' alg-geom/9709006, Sem. Bourbaki
1996, no. 821}

\lref\km{P. Kronheimer and T. Mrowka, ``Recurrence
Relations And Asymptotics For Four-Manifolds Invariants,'' Bull. Am. Math. Soc.
{\bf 30} (1994) 215, ``Embedded
Surfaces And The Structure Of Donaldson's Polynomial Invariants,'' J. Diff.
Geom. {\bf 33} (1995) 573.}

\lref\lilu{T.J. Li and A. Liu, ``General wall crossing
formula,'' Math. Res. Lett. {\bf 2} (1995) 797.}

\lref\labadescent{J.M.F. Labastida and P. M. Llatas,
``Topological Matter in Two Dimensions,'' hep-th/9112051 Nucl. Phys.  B379
(1992) 220;
M. Alvarez and J.M.F.L , `Topological Matter in Four Dimensions',
hep-th/9404115, Nucl. Phys. B437 (1995) 356-390;
J.M.F.Labastida and C. Lozano, `Lectures on Topological Quantum Field Theory',
hep-th/9709192}

\lref\marino{J. M. F. Labastida and M. Mari\~no,
``A Topological Lagrangian For Monopoles On Four-Manifolds,''
hep-th/9503105, Phys. Lett. {\bf B351} (1995) 146;
``Nonabelian Monopoles On Four-Manifolds,'' hep-th/9504010,
Nucl. Phys. {\bf B448} (1995) 373;
``Twisted N=2 Supersymmetry with Central Charge and Equivariant Cohomology,''
hep-th/9603169, Commun.Math.Phys. 185 (1997) 37-71.}

\lref\marinop{J. M. F. Labastida and M. Mari\~no,
``Polynomial invariants for $SU(2)$ monopoles,''
hep-th/9507140, Nucl. Phys. {\bf B456}(1995)633}

\lref\okonek{C. Okonek and A. Teleman,
``Seiberg-Witten Invariants for Manifolds with $b_+=1$, and the Universal Wall
Crossing Formula,''
alg-geom/9603003, to appear in Int. J. Math.}

\lref\swi{N. Seiberg and E. Witten,
``Monopole Condensation, And Confinement In $N=2$ Supersymmetric Yang-Mills
Theory,''
hep-th/9407087;Nucl. Phys. {\bf B426} (1994) 19.}

\lref\swii{N. Seiberg and E. Witten,
``Monopoles, Duality and Chiral Symmetry Breaking in N=2 Supersymmetric QCD,''
hep-th/9408099;Nucl. Phys. {\bf B431} (1994) 484.}

\lref\vw{C. Vafa and E. Witten,
``A strong coupling test of S-duality,''
hep-th/9408074; Nucl. Phys. {\bf B431}(1994)3-77}

\lref\verlinde{E. Verlinde, ``Global Aspects of Electric-Magnetic Duality,''
hep-th/9506011; Nucl.Phys. B455 (1995) 211.}

\lref\tqft{E. Witten,
``Topological Quantum Field Theory,''
Commun. Math. Phys. {\bf 117} (1988)
353.}

\lref\monopole{E. Witten, ``Monopoles and
four-manifolds,'' hep-th/9411102,
 Math. Research Letters {\bf 1} (1994) 769.}

\lref\abelS{E. Witten, ``On $S$-Duality In Abelian Gauge Theory,''
hep-th/9505186; Selecta Mathematica {\bf 1} (1995) 383. }

\lref\kahler{E. Witten, ``Supersymmetric Yang-Mills theory
on a Four-manifold,''  hep-th/9403193;
J. Math. Phys. {\bf 35}(10) 5101.}

\lref\zagi{D. Zagier, ``Nombres de classes et formes
modulaires de poids 3/2,'' C.R. Acad. Sc. Paris,
{\bf 281A} (1975)883.}

\lref\zagii{F. Hirzebruch and D. Zagier,
``Intersection numbers of curves on Hilbert modular
surfaces and modular forms of Nebentypus,''
Inv. Math. {\bf 36}(1976)57.}

\lref\joeloop{J. Polchinski, ``Evaluation of the one-loop
string path integral,'' Commun. Math. Phys.
{\bf 104} (1986) 37.}
\lref\mcclain{B. McClain and B. D. B. Roth,
``Modular Invariance For Interacting Bosonic Strings at Finite Temperature,''
Comm. Math. Phys. {\bf 111} (1987) 539.}

\lref\husemoller{See, e.g., D. Husem\"oller,
{\it Elliptic Curves}, Springer, 1987.}
\lref\ganor{O. Ganor, ``Toroidal Compactification of Heterotic 6D Non-Critical
Strings Down to Four Dimensions,'' hep-th/9608109; Nucl.Phys. B488 (1997)
223-235}
\lref\gms{O. Ganor,  David R. Morrison,  Nathan Seiberg,
``Branes, Calabi-Yau Spaces, and Toroidal Compactification of the N=1
Six-Dimensional $E_8$ Theory,'' hep-th/9610251; Nucl.Phys. B487 (1997) 93-127}
\lref\senpillow{A. Sen, ``F theory and orientifolds,'' hep-th/9605150 }
\lref\vafa{C. Vafa, ``Evidence for F-theory,'' hep-th/9602022}
\lref\vfmr{D. Morrison and C. Vafa, ``Compactifications of F-theory on
Calabi-Yau threefolds -I,'' hep-th/9602114;``Compactifications of F-theory on
Calabi-Yau threefolds -II,'' hep-th/9603161}
\lref\bds{T. Banks, M.R. Douglas, and N. Seiberg, ``Probing
F-theory with branes,'' hep-th/9605199}
\lref\matone{M. Matone, ``Instantons and recursion relations in N=2 Susy gauge
theory,'' hep-th/9506181, Phys. Rev. {\bf D53}(1996) 7354;
G. Bonelli and M. Matone, ``Nonperturbative Renormalization Group Equation and
Beta Function in N=2 SUSY Yang-Mills,''  hep-th/9602174, Phys.Rev.Lett. 76
(1996) 4107-4110;
 M. Matone, ``Modular Invariance and Structure of the Exact Wilsonian Action of
N=2 SYM,''
hep-th/9610204, Phys.Rev.Lett. 78 (1997) 1412-1415.}
\lref\parkpark{S. Hyun, J. Park, and J.S. Park,
hep-th/9503201,
Nucl. Phys. {\bf B453}(1995)199}

\Title{ \vbox{\baselineskip12pt\hbox{hep-th/9709193}
 \hbox{IASSNS-HEP-97/97}
\hbox{YCTP-P9-97}}}
{\vbox{\centerline{INTEGRATION OVER THE $u$-PLANE}
\bigskip
\centerline{ IN DONALDSON THEORY}}}
\smallskip
\centerline{Gregory Moore\foot{Research supported in part by DOE grant
DE-FG02-92ER40704.}}
\centerline{\it Department Of Physics, Yale University}
\centerline{\it New Haven, Connecticut 06520, USA}\bigskip
\centerline{Edward Witten\foot{Research supported in part
by NSF  Grant  PHY-9513835.}}
\smallskip
\centerline{\it School of Natural Sciences, Institute for Advanced Study}
\centerline{\it Olden Lane, Princeton, NJ 08540, USA}\bigskip

\medskip

\noindent
We analyze the $u$-plane contribution to Donaldson invariants of a
four-manifold $X$. For $b_2^+(X)>1$, this contribution vanishes, but for
$b_2^+=1$, the Donaldson invariants must be written as the sum of a $u$-plane
integral and an SW contribution.  The $u$-plane integrals are quite intricate,
but can be analyzed in great detail and even calculated.  By analyzing the
$u$-plane integrals,  the relation of Donaldson theory to
$\CN=2$ supersymmetric Yang-Mills theory can be described much more fully,
the relation of Donaldson invariants to SW theory can be generalized
to four-manifolds not of simple type, and interesting formulas can be
obtained for the class numbers of imaginary quadratic fields.
We also show how the results generalize to extensions
of Donaldson theory obtained by including hypermultiplet
matter fields.

\Date{September 19, 1997}
%\draft
%text of paper

\newsec{Introduction}

Donaldson theory can be formulated  \tqft\ as a twisted version of $\CN=2$
supersymmetric Yang-Mills theory.
Accordingly, new understanding
of  $\CN=2$ supersymmetric
quantum field theory \refs{\swi,\swii}
has led to new insights about Donaldson theory  \refs{\monopole,\donrev}.
In this paper we continue
this development, the main goal being to apply the understanding
of supersymmetric Yang-Mills theory to determine the Donaldson invariants
of four-manifolds with $b_2^+=1$.

Let $X$ be a smooth, compact, oriented
four-manifold with Riemannian metric $g$, and let $E\rightarrow X$
be an $SO(3)$ bundle over $X$ (that is, a rank three real vector
bundle with a metric).
As originally formulated, the Donaldson polynomials are polynomials
on the homology of $X$ with rational coefficients:
\eqn\donp{
\CD_E: H_0(X,\IQ)
\oplus H_2(X,\IQ) \rightarrow \IQ \ .
}
Assigning degree  $4$ to
$p\in H_0(X,\IQ)$ and $2$ to $S \in H_2(X,\IQ)$,
the degree $s$  polynomial  may be expanded as:
\eqn\degree{
\CD_E(p,S) = \sum_{2n + 4 t = s} S^n p^t d_{n,t}
}
where $s$ is the dimension of the moduli space ${\cal M}$
of instanton connections
on $E$.
The numbers $d_{n,t}$ were defined by
Donaldson in terms of intersection theory
on this moduli space
\refs{\donaldson, \FrMor,\DoKro}.
  It is useful
to  assemble the Donaldson polynomials into
a generating function.  To do so, one sums over all topological
types of bundle $E$ with fixed $\xi=w_2(E)$ but varying $p_1(E)$
(that is, varying instanton number), to define
\eqn\genfun{
\Phi_{\xi}^{X,g}(p,S) \equiv
\sum_{n \geq 0, t\geq 0} {S^n \over  n!}{ p^t \over  t!} d_{n,t}.
}
This quantity is often the most useful
way of organizing the $d_{n,t}$'s.  Here
$\Phi_{\xi}$ depends on the characteristic class $w_2(E)$
 but not on the instanton number $p_1(E)$  (as this
has been summed over).

If $b_2^+>1$, $\Phi$ is independent of the metric $g$
and thus defines ``topological invariants'' of $X$ (or more precisely
invariants of the smooth structure of $X$).
If $b_2^+=1$, $\Phi$ is only piecewise constant as a function of $g$
\donaldson; its detailed dependence on $g$
will be analyzed in section 4.

In \tqft, the Donaldson invariants were
identified physically as
the correlation functions of
certain operators in a topologically twisted
$\CN=2$ supersymmetric Yang-Mills (SYM) theory
with gauge group $SU(2)$ or $SO(3)$.
(The $SU(2)$ theory can be regarded as
the special case of the $SO(3)$ theory
in which one considers an $SO(3)$ bundle
$E$ with $w_2(E)=0.$)
One introduces the  fundamental observable
\eqn\zerobs{
\CO(P) = {1 \over  8 \pi^2} \Tr \phi^2(P)
}
where $P$ is a point in $X$, and $\phi$
is a complex scalar field, valued in the adjoint representation of $SU(2)$, and
related to the gauge field by supersymmetry.
\foot{For gauge group $SU(n)$, we mean by $\Tr$ simply the trace
in the $n$ dimensional representation.  Equivalently, for $SO(3)$
or $SU(2)$, $\Tr$ is 1/4 of the trace in the adjoint representation.
With this normalization, $\CO$ is related to the restriction
to $P\times {\cal M}$ of the second Chern class of
the universal instanton bundle over $X\times {\cal M}$
 -- in case there
is such a universal bundle.}  By a fairly standard
``descent'' procedure, one derives from \zerobs\ a family
of $k$-form valued observables for $k=1,\dots,4$.  For $X$ simply-connected,
the important case is the two-form valued observable
\eqn\twoobs{
I(S)= {1 \over  4 \pi^2} \int_S \Tr[  {1 \over  8} \psi\wedge \psi
-  {1 \over  \sqrt{2}} \phi F]
}
We will sometimes refer to $\CO$ and $I(S)$ as the zero-observable
and two-observable, respectively.

One of the main results of \tqft\ is that
\eqn\identfy{
\Phi_{\xi}^{X,g}(p, S) = \bigl \langle
e^{ p \CO + I(S)} \bigr \rangle_{\xi}
}
where the right hand side is the path integral  in a topologically
twisted version of the
supersymmetric Yang-Mills  theory (summed over all $SO(3)$ bundles $E$
with a fixed value of $\xi=w_2(E)$ and varying instanton number). This proves
to be an effective approach to evaluating Donaldson
invariants once one understands the vacuum structure
of the supersymmetric gauge theory.

The supersymmetric
field theory in question has a family of vacuum states parametrized
by a complex parameter $u$ which is defined\foot{The factor
of two in this formula
is meant to take care of a slight mismatch in conventions between
the mathematical and physical literature on this problem.  As written
by Kronheimer and Mrowka \km,
the ``simple type'' condition for $X$ reads
$[{\p^2 \over \p p^2}-4]\Phi=0$, where $p$ is the variable that appears in the
definition
of the generating functional $\Phi$.  This is
an insertion of $\CO^2-4$ in the correlator.
In the physics
literature, $u$ is defined so that the discriminant of the elliptic
curve that governs the $\CN=2$ supersymmetric gauge theory is $u^2-1$
(in other words, massless monopoles and dyons appear at the points
$u=\pm 1$ where the discriminant vanishes).  To reconcile a
vanishing discriminant
condition $u^2-1=0$  with a simple type
condition $\CO^2-4=0$, we
require a factor of 2 in the relation between $u$ and $\CO$.}
by
$2u=  \langle \CO\rangle $ where here $\langle \CO\rangle$ denotes
the expectation value computed in a normalized vacuum state on
flat ${\bf R}^4$.
As was shown in \refs{\swi,\swii}, the
complex $u$-plane can be identified as the modular
curve of the subgroup $\Gamma^0(4)$ of $SL(2,\IZ)$ consisting of integral
unimodular matrices whose upper right entry is divisible by four.  As such,
the complex variable $u$ parametrizes a family of elliptic curves that
can be described by a Weierstrass equation\foot{This equation describes
an elliptic curve with a distinguished subgroup of order four, generated
by the points $x=1/2$, $y=\pm \sqrt{(1-u)}/2$.  Note that $\Gamma^0(4)$ is
conjugate in $GL(2,\IQ)$ to $\Gamma(2)$, the subgroup of $SL(2,\IZ)$ consisting
of matrices congruent to the identity modulo two.  Hence the $u$-plane
could be identified (as in \swi) as the modular curve of $\Gamma(2)$,
but we use instead (as in \swii) the $\Gamma^0(4)$ description (which
differs by a two-isogeny), to make some
formulas slightly more natural and to facilitate comparison to recent
papers such as \gottzag.  }
\eqn\hubbo{y^2=x\left(x^2-ux+{1\over 4}\right).}
The cusps of $\Gamma^0(4)$ are the points at $u=\infty, 1$, and $-1$ where
the elliptic curve $C_u$ defined by \hubbo\ degenerates to a rational curve.
\foot{Some technical details about elliptic curves and
their associated modular functions are collected in
Appendix A.}

To compute Donaldson invariants of a four-manifold $X$ -- in other words, to
compute certain correlation functions of the twisted $\CN=2$ theory on $X$ --
one
can use any Riemannian metric on $X$.  It is
convenient to consider the one-parameter
family of metrics $g_t=t^2g_0$ with $t\in {\bf R}$ and some fixed $g_0$.
If $t$ is taken large,
on general grounds one can compute the correlation functions using a knowledge
of the infrared behavior in the various vacua of the theory.
If there are only finitely many vacua, one writes the correlation functions
as a sum of contributions of the different vacua.  In the present case, there
is a continuous family of vacua, and one should expect to represent the
correlation functions as some sort of ``integral'' on the $u$-plane.

We have put the word ``integral'' in quotes because this is not entirely
a continuous
integral; the measure on the $u$-plane has delta functions supported at $u=1$
and $u=-1$.  This occurs because \swi\ at $u=\pm 1$ there are massless
monopoles (or dyons) transforming as hypermultiplets of the supersymmetric
theory;
the twisted topological theory receives contributions from supersymmetric
configurations (obeying the equations $F_+=(\bar M M)_+,\,\Gamma\cdot
D M = 0$, introduced in  \monopole)
which are possible only at $u=\pm 1$.

Moreover, for many and in some sense most four-manifolds, the contributions
from $u=\pm 1$ are the only ones.  This, as will become clear,
is the physical interpretation of
the ``simple type'' condition
\km,
 which has
played an important role in the mathematical analysis of Donaldson theory.
In fact, let $b_i=b_i(X)$ be the Betti
numbers of $X$, and write $b_2=b_2^++b_2^-$, where $b_2^{\pm}$ are respectively
the dimensions of the spaces of self-dual and anti-self-dual harmonic
two-forms on $X$.
For $b_2^+>1$, the $u$-plane, away from $u=\pm 1$, does not contribute, as we
will show in section 2.3.
This is because there are ``too many fermion zero modes.''
Hence for this very large class of four-manifolds,
the Donaldson invariants can be written just in terms of monopole solutions,
via a formula that is presented in \monopole\ for four-manifolds of simple
type, and which we will generalize in section 7 for arbitrary four-manifolds.

Our main interest in the present paper is to explore what happens for
$b_2^+=1$, where the $u$-plane definitely does contribute.  (We actually
will mainly limit ourselves to the case  $b_1=0$, although the general case
is similar, as we will briefly discuss in section 10.   The $u$-plane
will also contribute for $b_2^+=0$ and $b_1$ odd, but this case has a very
different flavor and will not be considered here.)
The Donaldson invariants are therefore the sum of  a continuous integral over
the $u$-plane plus delta function contributions from $u=\pm 1$.  If
we write $Z_D$ for a Donaldson theory path integral or correlation
function,
$Z_{SW}$ for the analogous contribution from monopole solutions at $u=\pm 1$
(how to obtain the precise formula for $Z_{SW}$ in terms of the conventional
monopole or SW invariants will be explained in section 7),
and $Z_u$ for the continuous integral over the $u$-plane, then the
general structure is
\eqn\hobo{Z_D=Z_{SW}+Z_u.}

We will show that
for $b_2^+=1$, the contribution of the $u$-plane to Donaldson invariants
is given by quite complicated-looking integrals which nevertheless, because of
their
interpretation as integrals over a modular domain, can be analyzed very
effectively
and even calculated.
The integrals involved are similar to integrals that have been studied
in work of R. Borcherds in representation theory
\borcherds\kontsevich\  (and were
conjectured in \borcherds\
 to be related to Donaldson invariants of four-manifolds of
$b_2^+=1$)
and also in analyses of one-loop threshold corrections in string
theory (for example, in \hm).

Once the $u$-plane integrals have been constructed, our
 analysis of them will involve the  following main ingredients:

{\it (i)} Homotopy invariance.

{\it (ii)} Wall crossing formula.

{\it (iii)} Vanishing in certain chambers.

{\it (iv)} Behavior under blowups.

{\it (v)} Explicit evaluation and verification of invariance.

A fuller explanation of these points is as follows.

\bigskip\noindent{\it Homotopy Invariance}

One of the first important points is that the $u$-plane integral $Z_u$,
despite its considerable complexity and subtlety, depends on only elementary
topological information.   $Z_u$ is completely determined by
the cohomology ring of $X$ (in fact, by the intersection form on $H^2(X,{
\IZ})$ if $X$ is simply-connected).  This will be completely clear from the
structure of the integrand in the $u$-plane integral.

One therefore gets the same $Z_u$ if $X$ is replaced by any four-manifold
with the same cohomology ring.  For $X$ simply-connected, it follows,
given what is known about the intersection pairing on $H^2(X,{\IZ})$ for
smooth
four-manifolds $ X$,
that in the evaluation of $Z_u$, $X$ could be replaced by a rational algebraic
surface,
either $\IP^2$ blown up at $n$ points or $\IP^1\times \IP^1$.

\bigskip\noindent{\it Wall Crossing Formula}

For $b_2^+=1$, the Donaldson ``invariants'' are not quite invariants
\refs{\doncobord,\donaldson}; as
the metric of $X$ is varied, $Z_D$ generically is  constant but ``jumps''
when certain ``walls'' are crossed in the space of metrics.  Analogous
wall-crossing is known for the monopole or SW contributions $Z_{SW}$.
\hobo\ clearly implies that the wall-crossing of $Z_u$ is determined by
the Donaldson and SW wall-crossing.  If we denote the wall-crossing
of $Z_D$, $Z_{SW}$, and $Z_u$ as $\delta_D$, $\delta_{SW}$, and $\delta_u$,
then \hobo\ implies that
\eqn\nobo{\delta_D=\delta_{SW}+\delta_u.}

This can be better understood as follows.
{}From a physical point of view, it is clear that wall-crossing in $Z_D$ must
involve the behavior at $u=\infty$.  In fact, the proof of invariance under
change of metric in the twisted topological field theory involves a fermionic
symmetry whose validity depends on integration by parts in field space.
Invariance can fail only due to a lack of compactness of field space,
which, once one  reduces to integration over the space of vacua, means
lack of compactness of the $u$-plane.  But compactness of the $u$-plane
fails only at $u=\infty$.  Thus one should aim to understand $\delta_D$ in
terms of the behavior near $u=\infty$.

On the other hand, as $Z_{SW}$ is supported
at $u=\pm 1$, its wall-crossing $\delta_{SW}$  is likewise a contribution
from $u=\pm 1$.  The structure of $\delta_u$ implied by \nobo\ is therefore
clear; the wall-crossing behavior of the $u$-plane integral must be a sum
of a contribution from $u=\infty$ (which in \nobo\ will cancel $\delta_D$)
and a contribution from $u=\pm 1$ (which in \nobo\ will cancel $\delta_{SW}$).
This is just the structure we will find.

To be more precise, we will write $\delta_u$ as a sum
\eqn\junglo{\delta_u=\delta_{u,\infty}+\delta_{u,1}+\delta_{u,-1}}
where the three terms are the contributions to wall-crossing from $u=\infty,
1,$
and $-1$, respectively.
$\delta_{u,\infty}$ will be shown to coincide with the wall-crossing formula
for $\delta_D$ as determined in greatest generality
in  \refs{\gottsche,\gottzag}.
As for $\delta_{u,1}$ and $\delta_{u,-1}$, we will see, as expected, that these
contributions to wall-crossing are supported exactly where
 wall-crossing occurs in the monopole or SW invariants.

However, the details of the formulas for $\delta_{u,1}$ and $\delta_{u,-1}$
involve several universal functions of $u$ (universal in the sense that they
do not depend on the choice of four-manifold $X$) which have not been computed
previously.  As we will see in section 7, the same functions arise in
expressing
Donaldson invariants for hypothetical
four-manifolds of $b_2^+>1$  that are
not of simple type in terms of monopole or SW invariants.
A knowledge of the formulas for $\delta_{u,\pm 1}$ will
determine all the requisite universal functions and
enable us to get the general formula for $Z_{SW}$ in terms of monopole
or SW invariants, generalizing a formula presented in \monopole\ in the simple
type case.

The analysis of $\delta_u$ is comparatively easy in that one can calculate the
change in $Z_u$ upon crossing a wall much more easily than one can actually
evaluate $Z_u$; the change in $Z_u $ in crossing any given wall comes
entirely from one relatively simple term in a complicated sum.

\bigskip\noindent{\it Vanishing In Certain Chambers}

\nref\qin{Z. Qin, ``Complex Structures On Certain Differentiable
4-Manifolds,'' Topology {\bf 32} (1993) 551.}
\nref\hoppe{H. J. Hoppe and H. Spindler, ``Modulraume Stabiler 2-Bundel
Auf Regeflachen,'' Math. Ann. {\bf 249} (1980) 127.}
\nref\a{J. Morgan and Z. Szab\'o, ``Embedded Tori In Four-Manifolds,'' preprint
(1996).}

Donaldson invariants of four-manifolds with $b_2^+=1$ generally
do not exhibit simple type behavior.  But
it is known mathematically  that certain $SO(3)$ Donaldson
invariants for certain four-manifolds $X$ do exhibit such behavior
in certain chambers.  This occurs if $X$ maps to a two-dimensional base $B$
with generic fiber $F$ a two-manifold.  If $E$ is an $SO(3)$ bundle with
$(w_2(E),F)\not= 0$, then simple type behavior is found
in a chamber in which the fiber has a very small area relative to the base.
If moreover $F$ has genus zero, in which case $X$ is said to be
a rational ruled surface, then the Donaldson invariants
 actually vanish because
\hoppe\ there are no stable bundles.  In this situation, the SW invariants
also vanish (because there is a metric of positive scalar curvature),
so the $u$-plane integrals must vanish.
(Simple type behavior, but not vanishing of the Donaldson or SW invariants,
is also found for $F$ of genus one \a.)  Simple examples are
$X=\IP^1\times \IP^1$ or a Hirzebruch surface $\IF_1$.

Vanishing of the $u$-plane integrals for $X$ a rational
ruled surface of very small fiber area and with $(w_2(E),F)\not= 0$
 is again an easy consequence of the quantum field theory
formulation.  Simple type behavior (or at least generalized simple
type behavior, the
vanishing of $(\partial^2/\partial p^2-4)^r$
for some integer $r$) is
a consequence of vanishing of the $u$-plane integral.  (This will become
clear in section 7.)
Vanishing of the $u$-plane integrals for $X$ in the indicated
situation follows from pointwise vanishing of the integrand,
plus some analysis of behavior near $u=1,-1$, and $\infty$.
The pointwise vanishing holds because on the $u$-plane, $SO(3)$
is broken to $U(1)$, and the $SO(3)$ bundle $E$ with $(w_2(E),F)\not= 0$
becomes a line bundle $T$ with $(c_1(T),F)\not= 0$.  For
any connection on such a line bundle, the ``magnetic energy'' diverges as the
area of $F$ goes to zero, causing the $u$-plane integrand to vanish.

\bigskip\noindent{\it Blow-up Formula}

By ``blowing up'' a point in a four-manifold $X$ (or in topological
language, taking the connected sum with a copy of $\overline{{\IP}}^2$), one
gets a new four-manifold $\widehat X$.  The  Donaldson invariants of $\widehat
X$ (in a chamber
in which the exceptional divisor $b$ produced by the blowup has a very small
area)  are related
to those of $X$ by a blow-up formula that has been much studied mathematically
\refs{\finstern,\gottzag}.
There are two cases of the blowup formula, involving
$SO(3)$ bundles $E$ with $(w_2(E),b)=0$ and with $ (w_2(E),b)\not=0$.

It is natural to expect that the $u$-plane integral
will obey a similar blowup formula.  In fact, if there is a universal blowup
formula for $Z_u$, it must precisely coincide with the blowup formula for
$Z_D$,
since it can be determined by considering the special case that $X$ is
$\IP^2$ with a small number of points blown up.  Such an $X$ admits
a metric of positive scalar curvature, so $Z_{SW}$ vanishes in some chamber;
and if $b_-(X)<9$, there is no SW wall-crossing on $X$ so $Z_{SW}$ actually
vanishes everywhere.  Hence for such manifolds $Z_D=Z_u$, so that, if $Z_u$
has a blowup formula of the same general structure as $Z_D$, it must be
precisely
the same formula with the same universal functions.

In fact, we will see that blowing
up a point (and assigning a very small area to the exceptional divisor) has
a very simple effect on the $ u$-plane integrand; by analyzing this effect,
we get a blowup formula for $Z_u$ that is in perfect agreement with the
blowup formula for $Z_D$ as presented in \gottzag.
This result is easy in the sense that it can be seen before evaluating
the $u$-plane integrals; it comes from a relation between the integrands.

\bigskip\noindent{\it Computations}

The basic properties that have been explained up to this point completely
determine $Z_u$ for all four-manifolds of $b_2^+=1$
in all chambers, at least for the case
that $X$ is simply-connected.  Indeed, if $\pi_1(X)=0$, one can use the
homotopy
invariance to reduce to the case that $X$ is a rational surface.  Any two
rational surfaces, with any two given metrics, can be related to each other
by a succession of blowups, blowdowns, and wall-crossings.  (There is no
obstruction to this involving $w_2(E)$ since one case of the wall-crossing
formula involves a change in  $w_2$.)  So one can reduce to the special case
just described of $X=\IF_1$ (or $\IP^1\times \IP^1$)
in a chamber with $Z_u=Z_{SW}=Z_D=0$.

The most extensive mathematical computations of Donaldson invariants
for four-manifolds of $b_2^+=1$, such as those in \gottzag,
are based on the blowup and wall-crossing formulas
and the vanishing in certain chambers.  We will establish all of these
properties
for the $u$-plane integrals, so we can assert without any detailed calculation
that the $u$-plane integrals plus monopole contributions
agree with Donaldson theory for rational surfaces.

However, computations based only on blowup and wall-crossing formulas
and reduction to $\IF_1$ tend to be ineffective in the following sense.
To determine any given Donaldson invariant of $X$ by using
wall-crossing and blowup formulas to reduce to a vanishing invariant on $\IF_1$
involves only finitely many steps.  But as one considers Donaldson invariants
of $X$ associated with $SO(3)$ or $SU(2)$ bundles of greater and greater
instanton
number, the number of walls that must be crossed diverges, and it can be
hard to get a general and illuminating formula.

One possibility  to get effective formulas for $Z_u$ is simply to
evaluate the integrals.  It turns out that,
despite their complexity, the integrals defining $Z_u$ have special modular
properties that make this possible, though the calculations are certainly
much harder than the ones alluded to so far.

In this paper we will perform in detail
two direct computations of $Z_u$.
The first is a general computation of $Z_u$ for
any four-manifold whose intersection form contains as a summand the lattice
\eqn\plipo{H= \left(\matrix{ 0 & 1 \cr 1 & 0 \cr}\right).}
This includes any rational surface except $\IP^2$ or $\IF_1$.
In this computation we consider bundles with $w_2(E) =0$, and certain chambers.
This computation is performed using methods of \refs{\borcherds,\hm} together
with the Rankin-Selberg method
(familiar in string theory
\refs{\joeloop,\mcclain,\dkl}) of ``unwrapping'' a modular integral,
and certain additional tricks.  In this computation, we will explicitly
show that, for the chambers we consider, $Z_u$ is metric-independent within
a chamber.  Also, for the case that $ X =\IP^1\times \IP^1$, we will
recover formulas of G\"ottsche and Zagier \gottzag.

The other computation we perform is for $X=\IP^2$.  This computation
depends on techniques of a quite different sort.  The main technical tool
is a non-holomorphic modular form of weight $3/2$, related to Eisenstein
series of half-integral weight, that was introduced by Zagier
\refs{\zagi,\zagii}.   We will obtain a closed formula for Donaldson invariants
of $SU(2)$ bundles on $\IP^2$, in terms of Hurwitz numbers (essentially
class numbers of imaginary quadratic fields).  The formula
agrees with the special cases that have been computed previously \eg\ and, when
compared with the general
expression obtained (via wall crossing and vanishing theorems)
by G\"ottsche \gottsche, yields
interesting and perhaps even new formulas for class numbers.

\bigskip\noindent{\it Other $u$-plane integrals}

The basic twisting procedure that relates
Donaldson theory to the $SO(3)$ super
Yang-Mills theory
can be applied to other theories
with $d=4,\CN=2$ supersymmetry. In
the case of the  $SU(2)$ theory
with  ``quark'' hypermultiplets,
the resulting topological field theory involves an
integral on the $u$-plane for a family of curves
described in \swii. The integral is similar to
the case without matter, and can be
studied using the techniques discussed above.
The results are qualitatively similar for the
asymptotically free theories with $N_f<4$ flavors.
There are some surprises for the asymptotically
conformal theories, e.g., for   $N_f=4$.
In this case, there is no wall-crossing for $b_2^+=1$; a finite
set of correlation functions in the theory
turn out to be topological, and others vary continuously with the metric.

\bigskip\noindent{\it Organization Of The Paper}

This paper is organized as follows.  In section two, we present essential
physics background.  In section three, we work out the detailed form of the
$u$-plane integral. In sections  four, five, and six we derive the
wall-crossing, vanishing, and blowup properties of this integral. In section
seven we use these results
to derive the universal form of the SW contributions
to the Donaldson invariants. In section eight we
  perform the detailed computation of $Z_u$ for
four-manifolds
whose intersection form contains a summand $H$.  In section nine, we compute
the $SU(2)$ Donaldson invariants of $\IP^2$. In
section ten we indicate briefly how the results generalize
to nonsimply connected manifolds. In section eleven
we describe the generalization of these results to topological theories arising
from twisting $SU(2)$ SYM
with matter.  Some technical
details and definitions are collected in appendices A,B,C.

%%%
Extension of the $u$-plane integrals considered in the present paper to
the case of higher-rank gauge groups
(with $I(S)$ still derived from the quadratic
Casimir) is under investigation  by M. Mari\~no and G. Moore.
Relations between integrable systems
and contact terms such as $T(u)$ and its
generalizations will be addressed in
\ref\lns{A. Losev, N. Nekrasov, and S. Shatashvili,
to appear.}.

%%%
%%%

\newsec{Physics Background}

\subsec{Generalities}
\def\1{{\bf 1}}
\def\2{{\bf 2}}
\def\R{{\bf R}}
We begin with some generalities about $\CN=2$ supersymmetric theories in four
dimensions.  We start out on flat ${\bf R}^4$, where the double cover
$Spin(4)$ of
the rotation group is isomorphic to $SU(2)_-\times SU(2)_+$; the two factors
of $SU(2)$ act respectively on the $-$ and $+$ spin representations of
$Spin(4)$,
which we call $S_-$ and $S_+$.  The $\CN=2$ theories of interest here also
possess
an additional $SU(2)$ group of $R$ symmetries, which we call $SU(2)_R$.  Under
$SU(2)_-\times SU(2)_+\times SU(2)_R$, the supersymmetries transforms
as $(\2,\1,\2)\oplus (\1,\2,\2)$, where $\1$ and $\2$ represent respectively
the trivial representation and the two-dimensional representation of $SU(2)$.
We introduce $SU(2)_-$ indices $A,B,C=1,2$, $SU(2)_+$ indices
$\dot A,\dot B,\dot C=1,2$, and $SU(2)_R$ indices $I,J,K=1,2$,
and write the supersymmetries as $Q_A^I$ and $\bar Q_{\dot A J}$.  The
coordinates
of $\R^4$ transform as $(\2,\2,\1)$ and will be called $x^{A\dot A}$.

The non-zero anticommutators of the $Q$'s
(modulo central terms that will not be important here) are
\eqn\hogo{\{Q^I_A,\bar Q_{\dot A J}\}=4i \delta^I{}_JP_{A\dot A},}
where $P_{A\dot A}=\partial/\partial x^{A\dot A}$ is the translation generator.

To construct a twisted topological field theory, one introduces a diagonal
subgroup $SU(2)'$ of $SU(2)_+\times SU(2)_R$, and one introduces a new action
of
the Poincar\'e group of $\R^4$ in which rotations act via not
$Spin(4)=SU(2)_-\times
SU(2)_+$ but $Spin(4)'=  SU(2)_-\times SU(2)'$.  Among the supersymmetries
there is  the $Spin(4)'$-invariant object $\bar \CQ=\epsilon^{\dot A \dot B}
\bar Q_{\dot A \dot B} $
and
the $Spin(4)'$ vector $K_{A\dot A}=-{i \over  4}
\delta_{ A}^IQ_{\dot AI}$.
They obey
\eqn\hobo{\bar{\CQ}^2=0}
and
\eqn\nobo{P_{A\dot A}=\{\bar{\CQ},K_{A\dot A}\}.}
\nobo\ is an integrated version of a formula that asserts locally that
the stress tensor $T$ is of the form
\eqn\polo{T=\{\bar{\CQ},L\}}
for some $L$.  Note that $K$ obeys
\eqn\kobo{\{K_{A\dot A},K_{B\dot B}\}=0}
on gauge invariant quantities, as a consequence of \hogo.

If the underlying $\CN=2$ theory has a $U(2)_R$ group of $R$ symmetries (and
not just $SU(2)_R$), then the center $U(1)_R$ is a symmetry  of the
topologically
twisted theory.  Under this symmetry, $\CQ$ has charge 1, and $K$ has charge
$-1$.  In Donaldson theory, the $U(1)_R$ is a symmetry classically, but quantum
mechanically has an anomaly proportional to the dimension of instanton moduli
space
and is conserved only modulo 8.  The $R$ (or $U(1)_R$) charge is often called
``ghost number'' in the context of topological field theory.  In the mapping
from physical operators to differential forms on instanton moduli space, an
operator of ghost number or $R$ charge $q$ is mapped to a $q$-form.

Given a four-dimensional supersymmetric theory with the
properties described above, one can aim to formulate the
same theory on a general Riemannian
four-manifold $X$ in such a fashion
that $\bar{\CQ}$ is still conserved and \hobo\ and \polo\ still
hold.  This was done for the pure $\CN=2$ gauge theory (without
hypermultiplets)
%%%
in \tqft, and generalized to include hypermultiplets in \refs{\fre,
\marino,\parkpark}.
The fact that $\bar{\CQ}$ is conserved means that one can consistently restrict
to
$\bar{\CQ}$-invariant observables, and the fact that $\bar{\CQ}^2=0$ means that
if one makes
this restriction, only the $\bar{\CQ}$ cohomology class of a given observable
is
relevant.
The fact that the stress tensor (which is the change in the integrand of the
Feynman path integral  under a change in metric)
is of the form $\{\bar{\CQ},L\}$ means that
correlation functions of $\bar{\CQ}$-invariant observables are invariant under
a change
in metric.  The theory is therefore a topological field theory.

In constructing the $\bar{\CQ}$-invariant observables, an important step is the
``descent'' procedure, in which one starts with a $\bar{\CQ}$-invariant
zero-form
operator ${\cal O}^{(0)}$. By
inductively solving the equations
\eqn\joggo{d{\cal O}^{(j)} =\{\bar{\CQ},{\cal O}^{(j+1)}\},\,\, {\rm
for}\,\,j=0,\dots, 3}
one then finds $k$-form valued observables ${\cal O}^{(k)}$ for $k=1,\dots, 4$
which are $\bar{\CQ}$-invariant modulo exact forms.
This property ensures that for $\Sigma^{(k)}$ a $k$-cycle in $X$, the
integral
\eqn\opoggo{I(\Sigma^{(k)}) =\int_{\Sigma^{(k)}}{\cal O}^{(k)} }
is $\bar{\CQ}$-invariant and
depends only on the homology class of $\Sigma^{(k)}$.

\bigskip\noindent
{\it A Canonical Representative}

So far we have merely summarized standard facts about the construction
of a certain class of topological field theories.  Now we come to a point
that is less well-known and is helpful in analyzing the $u$-plane
integrals in Donaldson theory.
\foot{
%%%
The use of the canonical solution was
suggested in this context by N. Seiberg.
The existence of a canonical solution to the
descent equations was also  investigated in \labadescent.
%%%
}
 This is that there is actually a canonical
solution of \joggo.  That is because of \nobo, which in the twisted topological
field theory becomes the statement that there is a one-form valued operator
$K$ such that
\eqn\uggu{d=\{\bar{\CQ},K\}.}
This means that we can solve \joggo\ via
\eqn\ikoggo{{\cal O}^{(j)}=K^j{\cal O}^{(0)}.}
In interpreting the right hand side of \ikoggo, one understands
that $K$ acts on an operator ${\cal O}$ by conjugation, that
is $K{\cal O}$ is short-hand for $[K,{\cal O}\}=K{\cal O}-(-1)^{{\cal O}}{\cal
O}K$.
For ${\cal O}$ a zero-form
valued operator,  the $j$-fold iterated action of $K$ on ${\cal O}$ gives
an operator, called $K^j{\cal O}$ in \ikoggo, that transforms as a $j$-form
on $X$; the terms that are not completely antisymmetric (and so do not
transform as a $j$-form) vanish according to \kobo.

There are at least two reasons that in the present paper it is useful to have
a canonical solution of the descent equations:

(1) The choice of a concrete low energy Lagrangian to describe physics on the
$u$-plane is not unique, but is subject
to duality transformations that enter the theory in an important way.
It is essential to have duality-invariant solutions of the descent equations.
The canonical solution, since it can be described without committing oneself
to any particular Lagrangian description of the low energy theory, is
duality-invariant.

(2) Having this canonical procedure simplifies the task of matching
$\bar{\CQ}$-invariant operators defined in a microscopic description with
$\bar{\CQ}$-invariant
operators in a macroscopic description.  For instance, in the case that
the $\CN=2$ theory we start with is an $SU(2)$ or $SO(3)$ gauge theory,
the basic zero-form observable is $\CO ={1 \over  8 \pi^2} \Tr \phi^2$, where
$\phi$ is a
complex field in the adjoint representation that is part of the $\CN=2$ vector
multiplet.  (For gauge theory with a gauge group of rank higher than one,
one must also include higher order Casimir invariants of $\phi$.)  The
expectation value $\langle \CO \rangle = 2u$ is the
basic order parameter in the low energy theory,\foot{The
factor of $2$ was explained in a footnote in the introduction.}
so the operator in the low
energy theory corresponding to the microscopic operator $u$ is ``known.''
Since one also knows what the microscopic
supersymmetry generators, and in particular $K$,
correspond to in the low energy theory, there is no problem in identifying
the descendants $K^nu$ as computed in the microscopic theory with
corresponding operators in the low energy or macroscopic theory.

\bigskip\noindent
{\it Auxiliary Fields}

The last preliminary that we wish to discuss concerns the utility
of describing the low energy theory on
the $u$-plane in a formalism in which, by including auxiliary fields,
the supersymmetry algebra is closed off shell.

In this paper, we will mainly consider simply-connected four-manifolds,
so we specialize to the case that the cycles $\Sigma^{(k)}$ of the previous
discussion are two-dimensional Riemann surfaces embedded in $X$.  We will
denote such a Riemann surface as $S$.
The existence of a canonical
solution of the descent equations enables one to associate with an operator
$I(S)=\int_SK^2u$
of the microscopic theory a corresponding operator $\tilde I(S)=\int_SK^2u$
in the effective theory on the $u$-plane.

Now we would like to make a similar correspondence for products
$I(S_1)I(S_2)\cdots I(S_n)$ with distinct (though perhaps
homologous) surfaces $S_i$.  It is not the case that if a microscopic
operator $I(S_i)$ maps to an operator $\tilde I(S_i)$ in the
low energy description, then the product $I(S_1)I(S_2)\cdots I(S_n)$
maps to $\tilde I(S_1)\tilde I(S_2)\cdots \tilde I(S_n)$.
Rather, at intersections of the $S_i$, ``contact terms'' will appear.
\foot{Such  contact terms appeared in \kahler,
for much the same reason,
in using $N=1$ super Yang-Mills theory to compute Donaldson invariants of
Kahler surfaces.}
One important simplification is that, as we can pick the $S_i$ to have
only pairwise intersections, only pairwise contact terms will appear.
Moreover, we can assume that the intersections of the $S_i$ are generic
or ``transverse.''

The basic structure therefore appears
in a product of only two operators:
\eqn\hubbu{I(S_1)I(S_2) \to \tilde I(S_1)\tilde I(S_2)
+\sum_{P\in S_1\cap S_2}\epsilon_P T(P) .}
Here $T$ is some operator,  the sum over $P$ runs over points in the
intersection
of $S_1$ and $S_2$, and $\epsilon_P$ is $\pm 1$ depending on
whether $S_1$ and $S_2$ meet with positive or negative orientation
at $P$.
The operator $T(P)$ must be such that the right hand side of \hubbu\
is $\bar{\CQ}$-invariant and duality-invariant and obeys some more detailed
conditions
that
will be stated later.

If auxiliary fields are included so as to close the supersymmetry algebra
off-shell, then the condition for an operator, such as $\tilde I(S_1)$, to be
$\bar{\CQ}$-invariant is independent of the choice of a specific Lagrangian and
in particular is invariant under adding a multiple of $\tilde I(S_2)$ to the
action.
In that case, if $\tilde I(S_1)$ and $\tilde I(S_2)$ are separately
$\bar{\CQ}$-invariant,
so is their product.  If the supersymmetry algebra is {\it not} closed
off-shell, then the condition for $\tilde I(S_1)$ to be $\bar{\CQ}$-invariant
can change if $\tilde I(S_2)$ is added to the action -- or, what is closely
related,
if one takes an operator product with $\tilde I(S_2)$.

Thus, off-shell closure of the supersymmetry algebra makes $\tilde I(S_1)
\tilde I(S_2)$ automatically $\bar{\CQ}$-invariant, so that the operator
$T$ is separately $\bar{\CQ}$-invariant.  This is a major simplification, and
for
that reason we will use a formalism in
which  the supersymmetry algebra
is closed off-shell.  Of course, in a different formalism, one would
obtain equivalent results after lengthier analysis!

In the case of the $u$-plane theory of Donaldson theory,
$\bar{\CQ}$-invariance of $T$ means (after dropping an irrelevant term of the
form
$\{\bar{\CQ},\cdot\}$) that $T$ is a holomorphic ``function''
of $u$.  We have put the words ``function'' in quotes for the following reason.
We recall from \swi\ that one of the main points in the understanding
of ${\cal N}=2$ super Yang-Mills theory is that the low energy theory
has many possible Lagrangian descriptions that differ from each other
by duality transformations.  No one such description is valid globally
throughout the $u$-plane.
As we will
see, the product $\tilde I(S_1)\tilde I(S_2)$, in a formalism
with the supersymmetry algebra closed off shell, though $\bar{\CQ}$-invariant,
is not duality-invariant.  As a result, though in any Lagrangian description
of the
low energy theory, $T$ corresponds to a holomorphic function $T(u)$, in order
to achieve duality-invariance of the right hand side of \hubbu, one
must require $T$ to  transform non-trivially
under duality transformations.  After determining the requisite transformation
law, we will see that $T$ can be readily and uniquely determined.

Another benefit of holomorphy of $T$ is that it means that in the topological
field theory, the point $P$ at which one inserts the operator $T(P)$
(or more precisely $T(u(P))$) is irrelevant.  As a result, once one
has determined the object $T$, one can write the formulas in a much more
convenient fashion.  A useful way to proceed is as follows.
Let $S_i$, $i=1,\dots,b_2(X)$ be cycles representing a basis of
$H_2(X)$; let $\lambda_i$  be complex numbers; and let $S$
be a formal sum $S=\sum_i\lambda_iS_i$.  Thus $S$
represents an arbitrary element of $H_2(X,{\bf C})$.  We let
$S^2=\sum_{i,j}\lambda_i\lambda_jS_i\cdot S_j$
(where $S_i\cdot S_j$ is the intersection number of
$S_i$ and $S_j$; thus $S^2$ is simply the square of $S$
using the intersection pairing on $H_2(X,{\bf C})$), and we set
$I(S)=\sum_i\lambda_iI(S_i)$, $\tilde I(S) =
\sum_i\lambda_i\tilde I(S_i)$.
Then the formula \hubbu, together with the absence of higher order contact
terms and the separate $\bar{\CQ}$-invariance of each term in the formula we
are
about to write, enables us to put the transformation from microscopic
to macroscopic two-observables in its most convenient form:
\eqn\omigosh{\exp(I(S))\to \exp(\tilde I(S)+S^2T(u)).}
Here the point at which $T$ is inserted is irrelevant, given the
$\bar{\CQ}$-invariance, so we have written $T(u)$ instead
of $T(u(P))$.

\subsec{The Effective Theory On The $u$-Plane}

We have gotten about as far as we can with generalities.  At this point it
is time to describe in detail the theory of a single $\CN=2$ vector
multiplet in four dimensions, of which a special case is the theory of the
$u$-plane.  (The physical, untwisted model with a general prepotential
is described in \ref\gates{S. J. Gates, Jr., ``Superspace Formulation Of
New Nonlinear Sigma Models,'' Nucl. Phys. {\bf B238} (1984) 349.}.
The following formulas can be obtained by performing the $\theta$
integrals to reduce to an ordinary Lagrangian and ``twisting.'')

The bosons in the $\CN=2$ vector multiplet are a $U(1)$ gauge field $A$ and
a complex scalar $a$ (with its complex conjugate $\bar a$).  The fermions
are, in the topologically twisted version of the theory, a zero-form $\eta$,
a one-form $\psi,$ and a self-dual two-form $\chi$.  Under the $U(1)_R$
symmetry
(``ghost number''), $A$ has charge 0, $a$ and $\bar a$ have charges 2 and $-2$,
$\eta$ and $\chi$ have charge $-1$, and $\psi$ has charge 1.  In order to
close the supersymmetry algebra off-shell, one also introduces an auxiliary
field
$D$; in the topologically twisted theory, $D$ is a self-dual two-form,
of $U(1)_R$ charge zero.  In what follows, given a two-form such as the $U(1)$
field strength $F=dA$, we write $F=F_++F_-$, with $F_+$ and $F_-$ the self-dual
and anti-self-dual projections.  Note that as $D$ is self-dual, $D=D_+$ and
$D_-=0$.

In the topologically twisted model, the $\bar\CQ$ or BRST transformations
are
\eqn\toptmns{
\eqalign{
[\bar{\CQ}, A ] =  \psi
\quad & \quad
[\bar{\CQ}, \psi ]   =  4 \sqrt{2} da \cr
[\bar{\CQ}, a]   = 0
\quad & \quad
[\bar{\CQ}, \bar a] =   \sqrt{2}i \eta\cr
[\bar{\CQ}, \eta] = 0
\quad & \quad
[\bar{\CQ}, \chi]   = i( F_+  - D_+)  \cr
[\bar{\CQ}, D] & =   (d_A \psi)_+    \cr }
}
and the action of $K$ is
\eqn\actkay{
\eqalign{
[K , a]   = {1 \over 4 \sqrt{2}} \psi
\quad & \quad
[K, \bar a]   = 0 \cr
[K , \psi ] = - 2(F_-  + D )
 \quad & \quad
[K,  A ]   = - 2i \chi\cr
[K, \eta]   = -{i \sqrt{2} \over 2} d \bar a
\quad & \quad
[K, \chi]   = -{3i \sqrt{2} \over 4} * d \bar a\cr
[K, D]  = -{3i \over 4} * d \eta
&
+ {3 i \over 2} d \chi\cr
}
}

The Euclidean Lagrange density  is the 4-form:
\eqn\manfsttop{
\eqalign{
{i \over 6 \pi} K^4 \CF(a) + { 1 \over 16 \pi} \{ \bar \CQ, \bar \CF''
&
\chi (D+ F_+) \}  - { i \sqrt{2} \over 32 \pi}
\{ \bar \CQ, \bar \CF'   d * \psi  \}\cr
& - {\sqrt{2} i \over  3\cdot 2^5 \pi   }   \{ \bar{\CQ} ,
\bar\CF''' \chi_{\mu\nu} \chi^{\nu\lambda} \chi_\lambda^{~\mu} \} \sqrt{g}d^4
x\cr}
}
where $\CF(a)$ is a holomorphic function called the prepotential.
The free theory (quadratic action) corresponds to the case $\CF=\half\tau_0a^2$
for some constant $\tau_0$.

Using \toptmns\ and \actkay\ we may expand
out \manfsttop\ to get  the Lagrange density:
\eqn\lowenlag{
\eqalign{{\cal L}
 =  { i \over  16 \pi}   \bigl( \bar \tau F_+ \wedge F_+
+   \tau F_- \wedge F_-\bigr)
&
 +  {1 \over  8 \pi}   {\rm Im} \tau da \wedge * d\bar a
     - {1 \over  8 \pi}
({\rm Im} \tau) D \wedge *D \cr - {1 \over  16 \pi}
\tau  \psi \wedge * d \eta
+ {1 \over  16 \pi} \bar \tau \eta \wedge d * \psi
&
+ {1 \over  8 \pi}  \tau \psi
\wedge d \chi  - {1 \over  8 \pi} \bar \tau \chi \wedge d \psi +
\cr
+ {i \sqrt{2}   \over  16 \pi } {d \bar \tau \over  d \bar a} \eta \chi\wedge
(D_+ + F_+ )
&
 - {i \sqrt{2}   \over  2^7 \pi } {d \tau \over  da}
(\psi\wedge \psi) \wedge  ( F_-  +  D_+)
\cr
 + {i \over  3 \cdot 2^{11} \pi  }      {d^2 \tau \over  da^2} \psi\wedge \psi
\wedge \psi\wedge \psi
& - {\sqrt{2} i \over  3\cdot 2^5 \pi   }   \{ \bar{\CQ} , {d \bar \tau \over
d \bar a}\chi_{\mu\nu} \chi^{\nu\lambda} \chi_\lambda^{~\mu} \} \sqrt{g}d^4
x\cr}
}
where $\tau(a) = \CF''(a)$.
In addition there are extra terms
\eqn\grvcpl{
e(u) \Tr R\wedge R^*
 + p(u) \Tr R \wedge R  + {i \over  4} F \wedge w_2(X)
}
which must be taken into account when
coupling to gravity \abelS. For the case of the $u$-plane in Donaldson theory,
explicit expressions were
found for $e(u),p(u)$ in \abelS; these expressions
are further discussed in section 3.1.

\bigskip\noindent{\it Observables And Contact Term}

We now want to work out the description in the low energy theory of the
observable associated with a Riemann surface $S$ and the associated contact
term.
The mapping of observables from the high
energy theory to the low energy theory is
\eqn\highlowi{
\eqalign{
\CO & \rightarrow 2 u \cr
I(S) & \rightarrow \tilde I(S)= {i \over  \pi\sqrt{2}}
 \int_S K^2 u =
{i \over  \pi\sqrt{2}} \int_S
\Biggl\{
{ 1 \over 32} {d^2 u \over  d a^2   } \psi  \wedge \psi  -
{ \sqrt{2}\over  4} {d u \over  d a  } (F_-  +   D_+)
\Biggr\}
 \cr}
}
The two-observable is obtained simply
 by computing $K^2u$, with the above description of $K$ in the twisted
theory.  The normalization constants in these formulas have been
fixed by matching to known results on Donaldson invariants
(for instance, the factor of 2 in the first equation is discussed in a footnote
in the introduction).  In principle, by a more careful understanding of the
relation
between the $\CN=2$ theory as normalized physically and Donaldson theory
as defined mathematically, one should be able to make an
{\it a priori} computation of these normalization factors.

To determine the function called $T(u)$ in \omigosh,
consider integrating out the auxiliary field $D$ to
describe $e^{I(S)}$ in terms of physical fields only.  Since the
$D$ propagator is $\langle D(x)D(y)\rangle \sim \delta(x,y)/{\rm Im}\,\tau$,
this generates a term $\sim (du/da)^2/{\rm Im}\,\tau$.  So after integrating
out
$D$, $e^{\tilde I(S)}$ becomes
\eqn\jormo{ \exp\left(-{i \over  4 \pi} \int_S
\left({du\over da}F_-\right) +(S_+)^2{(du/da)^2\over
8\pi
{\rm Im}\,\tau}+{\rm fermion\,terms}\right),}
where $S_+$ is the  self-dual part of the cohomology class $S$, and
for the moment we need not concern ourselves with the fermion terms.

Equation \jormo\ is guaranteed {\it a priori} to be $\bar{\CQ}$-invariant, but
is not
modular-invariant.  For the terms involving $S_+$, the lack of modular
invariance is clear in \jormo: it comes because the function
$(du/da)^2/{\rm Im}\,\tau$ is not modular-invariant.
However, the full integral involves the contact
term discussed previously, and this effectively replaces
 $e^{\tilde I(S)}$ by
\eqn\gugu{\exp({\tilde I(S)+T(u)\cdot S^2}),}
where $T(u)$ is a function
that will be determined.

The part of the exponent in \jormo\ that involves $S_-$ (the anti-self-dual
part of $S$) is $(du/da)(S_-,F)$.  The lack of modular invariance here
is less obvious.  To see it
 involves analyzing a certain theta function that will enter when we study the
$u$-plane
 integral in detail.
We will in due course study this   theta function;
for now it suffices to note that according to general considerations leading
to \hubbu\ and \omigosh,
the contact term must, as written in \gugu,
be proportional to the intersection number
$S^2=S_+^2+S_-^2$, so that we can determine the contact term by computing
the $S_+^2$ term and then replacing $S_+^2$ by $S^2$.

The contact term will be a holomorphic ``function'' $T(u)$, appearing in the
low
energy theory as in \gugu, with the following properties:

(1) In any description by a special coordinate $a$ and photon $A$, $T(u)$
is a holomorphic function.  Under a duality transformation to a different
description, $T(u)$ changes in such a way that
\eqn\impobo{{(du/da)^2\over 8\pi {\rm Im}\,\tau}+T(u), }
which is the total coefficient of $S_+^2$ in the exponent in \gugu,
is invariant.

(2) $T$ has no singularity away from cusps in the $u$-plane. The behavior at
cusps is as follows.  If one
works in the appropriate local coordinate near $u=\pm 1$ (for instance,
$a_D$ at the monopole point, and $a+a_D$ at the dyon point) then $T$ has
no singularity at $u=\pm 1$.
For $u\to\infty$, if one computes using the special coordinate $a$ which is
valid near infinity, then  $T/u$ vanishes for $u\to \infty$.

(3) $T$ is odd under $u\to -u$.

The statements in (2) about the behavior of $T$ at cusps are justified as
follows.
Integrating out massless monopoles or dyons does not produce a singularity
in $T$ (we will see in section seven that $T$ coincides with an analogous
function
$T^*$ defined without integrating out the monopoles),
so $T$ has no singularity at $u=\pm 1$.  The behavior at infinity
follows by dimensional analysis and asymptotic freedom.  Dimensionally,
$T$ takes the form $T=u f(\Lambda^2/u)$, where $\Lambda$ (generally set to 1
in this
paper) is the scale parameter of the $\CN=2$ theory.  $T$ vanishes in the tree
approximation, so $f(0)=0$ and hence $T/u$ vanishes for $u\to\infty$.

Oddness of $T(u)$ under $u\to -u$ holds because the microscopic $SU(2)$ theory
has a classical $U(1)_R$ symmetry which is broken to ${\bf Z}_8$ by a quantum
anomaly.  Let  $w$ be a generator of
the ${\bf Z}_8$ which multiplies  a  quantum operator
of degree (or ghost number, or $R$ charge) $d$
by $e^{2\pi i d/8}$. Thus, $w$
   maps $u\to -u$ (because $u=\Tr \,\phi^2$ has $R$
charge four) and $\tilde I(S) \rightarrow i \tilde I(S)$.
Thus $T$ must be    odd under $w$, that is under
$u\to -u$.

Given these properties, $T$ can be determined as follows.
The main point is to determine how the function $G(u)=(du/da)^2/(8\pi {\rm
Im}\,\tau)$
transforms under $SL(2,\Z)$.  Under $\tau\to \tau+1$, that  is
$a\to a$ and  $a_D\to a_D+a$, clearly $G(u)$ is  invariant.  Under $\tau\to
 -1/\tau$,
we have
\eqn\pokko{{\rm Im}\,\tau\to {{\rm Im}\,\tau\over \bar\tau \tau}}
and
\eqn\nokko{{du\over da}\to {du\over da_D}={du/da\over {da_D/da}}={1\over \tau}
{du\over da}.}
Combining these results, we find that under $\tau\to -1/\tau$ we have
\eqn\okko{G(u)\to G(u) -{i\over 4\pi \tau}\left({du\over da}\right)^2.}
Note that the inhomogeneous term in this equation
(unlike $G$ itself) is holomorphic in $u$,
a crucial property that enables a contact term with the desired properties to
exist.

Modular invariance of $G+T $ now amounts
to the statement that $T$ is invariant under $\tau\to \tau+1$
and transforms under $\tau\to -1/\tau$ as
\eqn\juffy{T\to T+{i\over 4\pi \tau}\left({du\over da}\right)^2.}
A comparison to the standard transformation law for the Eisenstein series
$E_2(\tau)$ shows that these conditions are equivalent to the statement that
\eqn\jokko{T=-{1 \over  24} E_2(\tau)\left({du\over da}\right)^2 +H(u)}
where $H$ is modular invariant and so is an ordinary  holomorphic function
of $u$.  Conditions (2) and (3) assert that $H$ has no singularities on the
finite part of the $u$-plane, grows precisely as $u/3$ for $u\to \infty$,
and is odd under $u\to -u$.  Hence $H(u)=u/3$ and
\eqn\pokko{T=-{1 \over  24}\biggl(
E_2(\tau)\left({du\over da}\right)^2-8u \biggr) .}

The relation between a microscopic operator $\exp(I(S))$ and macroscopic
observables on the $u$-plane is hence
\eqn\bombo{\exp(I(S))\rightarrow \exp(\tilde I(S)+S^2T(u))}
with $T(u)$ given in \pokko.

\subsec{Vanishing Of The $u$-Plane Contribution For $b_2^+>1$}

We will now establish a fundamental result: the vanishing of the
the $u$-plane contribution for four-manifolds with $b_2^+>1$.  In the process
we will also learn what contributions do survive for $b_2^+\leq 1$.

The contribution of the $u$-plane cannot be evaluated by the usual topological
field theory technique of reducing to supersymmetric field configurations
and then evaluating their contributions.
The reduction is usually made by adding to the Lagrangian a term
$\lambda\{\bar{\CQ},V\}$,
where $\lambda$ is real and
$V$ is chosen so that $\{\bar{\CQ},V\}$ vanishes only for supersymmetric
configurations, and then taking $\lambda\to\infty$.  In the case of the
$u$-plane theory, such a $V$ cannot be chosen in a duality-invariant fashion.
In fact, $V$ would have to be chosen to achieve $\tau\to i\infty$, and this
notion is certainly not duality-invariant.  (Any choice of $V$ depends
on a choice of a particular ``photon'' multiplet.)
Covering the $u$-plane with open sets and using different $V$'s in different
patches would be unhelpful, because the proof of invariance of the correlation
functions under addition of $\{\bar{\CQ},V\}$ to the Lagrangian involves an
integration
by parts in field space which in particular involves integration by parts
on the $u$-plane; so one would run into serious complications in the
intersections
of  different patches.

The alternative is to exploit the fact that the theory is expected to
be metric-independent (within a chamber, for reasons that will be clear) and
to take advantage of
this by looking at the behavior in a one-parameter family
of metrics $g_t=t^2g_0$, for fixed $g_0$, with  $t\to \infty$.  What
contributions
survive as $t\to\infty$?  The one-loop determinants of the various fields
cancel, by supersymmetry.  Almost all Feynman diagram contributions vanish
because -- the theory on the $u$-plane being unrenormalizable and without
marginal or relevant
couplings in the renormalization group sense -- the vertices scale as
negative powers of $t$.  We will analyze presently which contributions do
survive.
Finally, in the path integral over abelian connections, one must sum over
the various line bundles and the classical solutions (connections with harmonic
curvature) on each line bundle.  Since free $U(1)$ gauge theory is conformally
invariant, the generalized theta function coming from the sum over line bundles
survives as $t\to\infty$; it will be analyzed in some detail later.

To illustrate how the quantum theory works without any technicality, first
consider
the case that $b_1=0$ and $b_2^+=1$.  There is always a single $\eta $ zero
mode, with wave-function $1$.  For $b_1=0$ and $b_2^+=1$, there
are no $\psi$ zero modes and one $\chi$ zero mode.  The $\eta$ and $\chi$
zero modes, being zero-forms and two-forms, respectively, are naturally
of dimension 0 and 2.  The bosonic fields $F$ and $D$ are of dimension 2.  The
$\eta$ zero mode and a single
$\chi$ zero mode can be soaked up using the terms $\eta \chi D$ or
$\eta\chi F_+$ in the low energy effective action.
As these terms are of dimension $0+2+2=4$ and we are in four dimensions,
this gives a way to soak up all
fermion zero modes with the overall power of $t$ being $t^0$.  (In doing this,
one sets
$d\bar \tau/d\bar a$ to its expectation value at the given point on the
$u$-plane.)
More explicitly, in performing the path integral, one must sum over line
bundles.  If $\eta $ and $\chi$ are set equal to harmonic forms, then
$\int \eta\wedge \chi\wedge F_+$ is equal to the integral of the wedge product
of the three cohomology classes in question, and is certainly invariant
under rescaling of the metric by $g_0\to t^2g_0$.  The $\eta\wedge
\chi\wedge  D$ term
is similar (after integrating out $D$ it will be replaced by $\int_S\eta
\wedge\chi$,
which again depends only on the cohomology classes and not the metric).
So these terms
give contributions that survive as $t\to\infty$.  As we will see, these
are the only contributions that survive.

Suppose that, still with $b_2^+=1$, we take $b_1>0$.  We should limit
ourselves to the case $b_1$ even, since everything vanishes in Donaldson
theory unless $1-b_1+b_2^+$ is even.  There are $b_1$ $\psi$ zero modes,
and as these are one-forms, they are naturally considered to be of dimension 1.
We can absorb $\psi$ zero modes in groups of four using the
interaction vertex $(d^2\tau/d a^2) \psi\wedge \psi\wedge \psi
\wedge \psi$, and we can absorb $\psi $ zero modes in groups of two
using the interaction vertex
$(d\tau/d a) \psi\wedge \psi\wedge (F_- + D) $.  Either
type of vertex gives a factor independent of $t$. Meanwhile
the (unique) $\eta$ and $\chi$ zero modes are absorbed by the $\eta\wedge \chi
\wedge F$ or $\eta\wedge \chi\wedge D$ terms.  So for $b_2^+=1$ and any
even $b_1$, there are contributions to the $u$-plane integrand that survive
for $t\to\infty$.

What about $b_2^+>1?$  For example, for $b_1=0$ and $b_2^+=3$, one could
try to soak up all fermion zero modes using the $\eta\chi^3$ term or
by using the $\chi^2(F_+-D)$ term (along with $\eta\chi(F_+-D)$) in the
Lagrangian.  These additional terms, however, are not topological and in fact
scale as $t^{-2}$.
Other contributions with the given Betti numbers behave similarly
or worse, as we will prove below, so the $u$-plane contribution vanishes
for these values of the Betti numbers.  The behavior is similar whenever
$b_2^+>1$.

For $b_2^+=0$ (and $b_1$ odd), there are surviving contributions as $t\to
\infty$
which actually come from one-loop diagrams.  Rather than explaining this
in an {\it ad hoc} fashion, we will now adopt a more systematic approach.

\bigskip\noindent{\it Scaling}

The non-zero modes in the path integral come in bose-fermi pairs related by
$\bar{\CQ}$ and so carry a natural measure.  Metric independence of the measure
thus means that
the zero modes (or classical solutions) of
 the fields $a,A,\eta,\psi,\chi$ should be  normalized in a fashion
invariant under rescaling of the metric of $X$. (The auxiliary field $D$
has no zero modes.)
For example, the expectation value of
$a$ determines a point on the $u$-plane; the labeling of such points is
completely independent of any metric on $X$.
The classical solutions for $A$ (the $U(1)$ connection of the low energy
theory)
are connections with harmonic curvature on various line bundles over $X$;
these are naturally labeled by topological data.
The zero modes of $\eta,\psi,$ and
$\chi$ are harmonic $q$-forms (for $q=0,1$, and $2$ respectively) which we take
to represent fixed cohomology classes.   (We are limited here to speaking
of invariance under conformal rescalings of the metric, not under arbitrary
changes in metric, since $\chi$ is a self-dual harmonic two-form, whose
cohomology class takes values in the
self-dual part of $H^2(X,\IR)$, which is
invariant under conformal changes of metric on $X$
but not under arbitrary changes.)

Now we expand the various fields as a sum of zero modes plus quantum
fluctuations.
For instance,
\eqn\milmo{a=a_0+a',}
where $a_0$ is a constant and $\int_X d^4x\sqrt g a'=0$,
so that $a'$ is orthogonal to the constants or zero modes.  Likewise, we
set
\eqn\ilmo{\eqalign{\eta&=\eta_0+\eta'\cr
                   \psi&=\psi_0+\psi'\cr
                   \chi&=\chi_0+\chi'\cr
                     A&=A_0+A',\cr}}
where $\eta_0,\psi_0$, and $\chi_0$ are harmonic forms, $A_0$ is a connection
with harmonic curvature, and $\eta'$, $\psi'$, $\chi'$, and $A'$ are orthogonal
to the space of zero modes.

To analyze the large $t$ behavior, it is convenient to assign dimensions
to all fields in such a way that the kinetic energy of all fields has dimension
four.  There is not a unique way to do this.  It is convenient to assign
the natural dimensions $1,1,2$ to the bosons $a',A',D$ while assigning
dimension 1 to $\psi$ and 2 to $\eta,\chi$.  This assignment of dimensions to
the fermions differs from the usual choice (dimension $3/2$ for all fermions),
but still has the property that every term in the fermion kinetic energy is
of dimension four.

It is now easy to see that every interaction vertex has dimension at least
four,
and that every such vertex that contains  $\eta$ or $\chi$ fields or contains
no
fermions at all has dimension greater than four.  Thus, every dimension four
vertex has $\psi$ fields and no other fermions.
Since the $\langle\psi\psi\rangle$
propagator vanishes (nonzero fermion propagators are $\langle \eta\psi\rangle$
and $\langle \chi\psi\rangle$), all tree and loop diagrams constructed using
the quantum fluctuations only vanish for $t\to\infty$.

What happens if we include insertions of fermion zero modes? (Insertions of
bose zero modes just give derivatives with respect to the coupling constants
at the
various vertices and
do not affect the assertions of the last paragraph, which did not depend on
details
of the couplings.)  The zero modes represent fixed cohomology
classes and so have geometrical dimensions -- dimension $q$ for a $q$-form.
Replacing a quantum
fluctuation by a fermion
zero mode can only help if the zero mode has a smaller dimension
than the corresponding quantum fluctuation.
The only field for which this is so is $\eta$, which has precisely one
zero mode (with constant wave-function).  So precisely one vertex will
be ``improved'' in its large $t$ behavior by setting $\eta$ equal to a
constant.
There are three choices for which sort of vertex this might be:

(1) If we set $\eta$ to a constant in the $\eta\chi\chi\chi$ interaction, we
find
that this interaction still scales as a negative power of $t$.  Since all other
vertices scale as nonpositive powers of $t$, all contributions of this sort
vanish.

(2) We could set $\eta=\eta_0$ in the $\eta\wedge \chi\wedge F$ or $\eta\wedge
\chi
\wedge D$ interaction.  These terms then scale as $t^0$ regardless of whether
for $\chi$ we take a zero mode or a quantum fluctuation.  There are now
basically three cases:

(a) If $b_2^+=1$, we must take the $\chi$ field in
the $\eta\wedge \chi\wedge F$ or $\eta\wedge\chi\wedge D$ vertex
to be a zero mode, since there are no other
vertices of dimension four or less that could soak up the $\chi$ zero mode.
In the case of the $\eta\wedge \chi\wedge F$ vertex,
we must set $F$ equal to a harmonic
form; for if $F=dA'$ with $A'$ a quantum fluctuation while $\eta$ is a
constant and
$\chi$ is harmonic, then $\int \eta\wedge\chi\wedge F
=0$.  Otherwise we must use only
vertices of dimension four (the alternative being
vertices of  higher dimension that give vanishing
contributions).  The only such vertices are $\psi^4$, $\psi^2 F$, and $\psi^2
D$;
as there is no $\langle\psi\psi\rangle$ propagator, all factors of $\psi$ must
be set equal to zero modes (and $F$ to a harmonic form, since if $F=dA'$ and
the $\psi$'s are harmonic then $\int \psi\wedge \psi \wedge F=0$).
\foot{It can be shown  that for $b_2^+ <3$ and all  $\psi$'s
equal to zero modes, $\int_X\psi^4=0$, so this interaction can be dropped.}
This gives the non-vanishing contributions described earlier
for four-manifolds with $b_2^+=1$ and any even $b_1$.

(b) If $b_2^+>1$, there is no way to absorb the $\chi$ zero modes
without negative powers of $t$ from vertices of dimension bigger than four.

(c)   If $b_2^+=0$, then in the $\eta\wedge\chi\wedge F$
or
$\eta\wedge \chi\wedge D$ vertex, as there are no $\chi$ zero modes, $\chi$
equals a quantum fluctuation $\chi'$.  (Hence $F$ must likewise be a quantum
fluctuation $F=dA'$.)
There must therefore be additional vertices, and these must be dimension
four vertices $\psi^4$, $\psi^2F$, or $\psi^2D$.  As the
$\langle\psi\psi\rangle$
propagator vanishes, all $\psi$ fields except one must be zero modes.
(The $\psi^4$ term can hence be dropped for the reason given in the footnote.)
By following these rules, one can find several one-loop diagrams
that contribute for four-manifolds with $b_2^+=0$ and $b_1$ odd.
These diagrams come from $\eta\chi F\cdot \psi\psi F$ or
$\eta\chi D\cdot \psi\psi D$ with a $\langle \chi \eta\rangle$ propagator
and either an $\langle F F\rangle $ propagator or a (delta function)
$\langle D D\rangle$ propagator.

\def\bar{\overline}
(3)  The remaining possibility is to set $\eta$ to a constant in the $\bar\tau
\,\eta
d^*\psi$ term in the Lagrangian.  Since $\int_Xd^4x \sqrt g\, d^*\psi=0$,
this gives a vanishing contribution if we set $a$ (and hence
$\bar\tau$) to a constant.
So the lowest dimension term that arises from this source
 is a dimension three operator
$\bar {a'} d^*\psi$.  Moreover, to get a non-vanishing contribution one must
use
precisely this vertex, since the dimension four operator $(\overline {a'})^2
d^*\psi$ would
give a negative power of $t$ by the time one absorbs the $\bar {a'}$
fields, and higher
powers of $\bar {a'}$ are of course only worse.  So relevant terms of this
kind come
precisely from the $\bar{a'} d^*\psi$ operator.  In this operator, the
$\bar{a'}$
field
is a quantum fluctuation, of course (rather than a zero mode), and the same
is true of $\psi$ as $d^*\psi=0$ for $\psi$ a harmonic one-form.  So we need
to absorb an $\bar {a'}$
field and a $\psi'$ field using at most one vertex of dimension
five (more vertices of dimension five or any of dimension greater than five
would give a vanishing contribution for $t\to\infty$) together with dimension
four vertices.  The only way to do this is to use a dimension five vertex
$  a' d\eta \psi$ (coming from expansion of the $\eta\psi$ kinetic energy
in powers of $ a'$) together with any number of dimension four operators
$\psi^4$, $\psi^2 F$, and $\psi^2D$.
All $\psi $ fields in the dimension five
and four vertices just described must be zero modes since there is no
$\langle\psi\psi\rangle$ propagator.  None of these vertices contain $\chi$
fields,
so these terms only contribute for four-manifolds with $b_2^+=0$ (and any
odd $b_1$). These contributions actually involve the one-loop diagram
$\langle  a' d\eta(x) \bar {a'} d^*\psi(y)\rangle$.  (The $\psi^4$ term
can again be dropped.)

In short, certain one-loop diagrams contribute for $b_2^+=0$, some simple tree
diagrams
contribute for $b_2^+=1$, and
there are no surviving contributions at all for $b_2^+>1$.   This hierarchy
is reminiscent of the progressive simplification found in a certain class
of three-dimensional topological field theories as $b_1$ is increased
\ref\fw{L. Rozansky and E. Witten, ``Hyper-Kahler Geometry And Invariants
Of Three-Manifolds,'' hep-th/9612216.}.  The derivation is also
more or less similar.

\newsec{Explicit expression for the $u$-plane integral}

Our goal in the rest of this paper is to understand the $u$-plane and SW
contributions to
Donaldson invariants, focussing on the case $b_1=0$.  (After working out
some formal properties of the $u$-plane contributions in sections 4-6, we will
then in section 7 analyze the SW contributions to Donaldson invariants.)
We would like to calculate the value, in the twisted $\CN=2$ theory with
gauge group $SO(3)$, of the path integral with an insertion of the operator
\eqn\ububu{\exp\left(p\CO +I(S)\right),}
where $p$ is a complex number, $\CO ={1 \over  8 \pi^2} \Tr\phi^2$, and as
before $I(S)$ is
an arbitrary two-observable.  We consider the partition function with this
operator
inserted summed over
$SO(3)$ bundles $E$ with a fixed value of $\xi=w_2(E)$.   We call this object
$\langle \exp\left(p\CO   +I(S)\right)\rangle_\xi$.
It is the generating functional
of Donaldson invariants for bundles of the given value of $w_2(E)$.

As explained in the introduction, the answer will be the sum of a contribution
from the $u$-plane and a contribution from monopole or SW solutions at
$u=\pm 1$.
In this section, we
work out the contribution of the $u$-plane, the analysis of
which is the main focus of the present paper.  As we have seen, this
contribution
vanishes for $b_2^+>1$.

\subsec{Form Of The Integral For $b_2^+=1$, $b_1=0$ }

In analyzing the $u$-plane integrals,
the first task is simply to write down the $u$-plane integrand for $b_2^+=1$,
$b_1=0$.
A number of factors need to be considered, including:

{\it (i)} Some interactions that vanish on flat ${\bf R}^4$ but are present
in the twisted theory on a curved four-manifold; and a factor involving
the center of the gauge group.

{\it (ii)} The integration measure for the zero modes.

{\it (iii)} The transformation of \ububu\ to the macroscopic theory.

{\it (iv)} The path integral of the photons.

{\it (v)} The absorption of fermion zero modes and elimination of auxiliary
fields.

We consider these factors in turn and then put the pieces together.

\bigskip\noindent{\it Effect Of Curved Space  }

On the $u$-plane, there are
interactions of topological importance that do not
appear on flat ${\bf R}^4$ but
do appear if one works on a curved four-manifold
$X$.
For the case of the topologically twisted theory,
these interactions were analyzed in section 3.3 of \abelS\ and multiply
the
measure by a factor
\eqn\dono{
A^\chi B^\sigma = \alpha^\chi \beta^\sigma
\left((u^2-1){d\tau\over du}\right)^{\chi/4}\left(u^2-1\right)^{\sigma/8}}
where $\chi $ and $\sigma$ are the Euler characteristic and the signature of
$X$, and $\alpha$ and $\beta$ are universal constants (independent of $X$)
that were not determined in \abelS.
(These constants
  could in principle be computed
by careful computations in the semiclassical region of large $u$, where
asymptotic
freedom prevails, but this has not been done.)

Also, in going from quantum field theory to $SU(2)$ Donaldson theory, there is
an extra factor
of 2, because the center of $SU(2)$, which is of order 2, acts trivially on
instanton
moduli space; in quantum field theory one divides by this factor of 2 (as part
of
the Fadde'ev-Popov gauge fixing), but in Donaldson theory
it is not customary to do so.\foot{Since we are working on $SO(3)$ bundles
of non-zero $w_2(E)$, the reader should ask why the center of $SU(2)$ is
relevant.  The answer is that in standard physical
$SO(3)$ gauge theory, one would sum over all values of
$w_2(E)$; we instead are calculating the value of the path integral for a fixed
value of $w_2(E)$.  This gives a sort of shifted version of $SU(2)$ gauge
theory (the standard version of which has $w_2(E)$ fixed to be zero).
In such a shifted version of
the $SU(2)$ theory, the path integral computes the natural
topological intersection theory on moduli space, divided by the order of the
center of $SU(2)$, just as in the ordinary $SU(2)$ theory.}

For $b_1=0$, $b_2^+=1$, we have $\chi+\sigma=4$,
so the additional factors in the path integral, including the factor of 2 just
mentioned,
become
\eqn\joggo{2 \alpha^\chi \beta^\sigma (u^2-1){d \tau \over  du}
\Biggl( { (   {du \over  d \tau} )^2 \over  u^2-1 } \Biggr)^{\sigma/8}}

\bigskip\noindent{\it Zero Mode Integration Measure  }

With $b_1=0$, the bosonic zero modes are purely the choice of a point on the
$u$-plane.  The metric  on the $u$-plane can be read off from
the Lagrangian, and is up to a constant multiple ${\rm Im}\,\tau |da|^2$.
So the zero mode measure for $a$ is a constant multiple of
\eqn\gogo{{\rm Im}\,\tau \,\,da\,d\bar a.}
We need not be precise in determining a universal multiplicative factor here
(and similar factors below); this would be part of the determination of the
factors
$\alpha,\beta$ in \joggo.

There is a single $\eta$ zero mode, with constant wave-function.
We write $\eta=\eta_0+\eta'$, where $\eta_0$ is a constant anticommuting
$c$-number and
$\eta'$ is orthogonal to the constants.  For $b_2^+=1$, there is a single
$\chi$ zero mode, the wave function being a harmonic self-dual two-form
$\omega$ which we normalize so that $\int_X\omega\wedge \omega=1$.
Note that this condition leaves the sign of $\omega$ undetermined.
We write $\chi=\chi_0\omega+\chi'$, where $\chi_0$ is an anticommuting constant
and $\chi'$ is orthogonal to $\omega$ (we are making a slight change in
notation
as $\chi_0\omega$ was earlier called simply $\chi_0$).
The integration measure for the fermion zero modes is just
\eqn\ollo{ {d\eta_0\, d\chi_0\over {\rm Im}\,\tau}.}
The reason for the factor of ${\rm Im}\,\tau$ in the denominator is that the
kinetic
energy of every field has a factor of ${\rm Im}\,\tau$, which means that  the
measure
for every bose zero mode has a factor of $({\rm Im}\,\tau)^{1/2}$ and that
of every fermion zero mode has a factor of $({\rm Im}\,\tau)^{-1/2}$.

Notice that this measure is odd under a reversal of sign of $\omega$, which
changes
the sign of $\chi_0$.
This corresponds to the standard fact
\nref\dona{S. Donaldson, ``The Orientation Of Yang-Mills Moduli Space
And Four-Manifold Topology,'' J. Diff. Geom. {\bf 26} (1987) 397.}
\refs{\dona,\DoKro}
that defining the sign of the Donaldson
invariants requires a choice of a ``Donaldson orientation,'' which is an
orientation
of $H^{2,+}(X)\otimes H^1(X)$.  For $b_1=0$ and $b_2^+=1$, a Donaldson
orientation
is a choice of $\omega$.  Our formulas will thus depend on a choice of
$\omega$,
and will be odd under reversal of sign of $\omega$.

Combining the above, the zero mode measure is simply
\eqn\juggo{ \,da\,d\bar a \,d\eta_0\,d\chi_0}
with no factors of ${\rm Im}\,\tau$.

\bigskip\noindent{\it Observables of the Low Energy Theory  }

How to represent the microscopic observable $\exp(p\CO+I(S))$ in the low energy
theory has already been determined above.  The most subtle step was the
determination
of the contact term $T(u)$ in \bombo.
For $b_1=0$ and $b_2^+=1$, the $\psi\psi$ terms
in $I(S)$ can be dropped, by arguments
similar to those that we gave in explaining the vanishing of the $u$-plane
contribution for $b_2^+>1$.
The net result is   that
\eqn\ojurry{\exp(p\CO +I(S)) \rightarrow
\exp\biggl[
  2 p u - {i \over  4 \pi}  \int_S
 {d u \over  d a  } (F_-  +   D_+)
 + S^2 T(u)
\biggr]
}
where the right hand side is to be evaluated in the
low energy theory on the $u$-plane.

\bigskip\noindent{\it Photon Path Integral  }

An important factor in the $u$-plane integral is the partition function of the
free photons.  Actually, to be more precise, one will require the photon
path integral with an insertion of  a  certain operator.  However, many of the
subtleties
occur already if one writes simply the photon partition function (which would
enter
in some physical observables on the $u$-plane, though not in the topological
observables
considered in the present paper), and we will do this first.

As explained in \refs{\abelS,\verlinde}, the photon partition function on
a four-manifold with $b_1=0$ is of the form
\eqn\gurgle{Z(\tau)={ \theta_0(\tau,\bar \tau)\over ({\rm
Im}\,\tau)^{1/2}},}
where $ \theta_0$ is a sort of  Siegel-Narain theta function
of the lattice $\Gamma= H^2(X,{\bf Z})$.  If one simply took
$U(1)$ gauge theory with
the action appearing in \lowenlag\ we would
substitute
\foot{Recall that with our conventions, $\lambda\in \half w_2(E)+\Gamma$
is a half-integral class.  That is why $F=4\pi\lambda$ and not $2\pi\lambda$.}
$F \rightarrow 4 \pi \lambda$ and this theta function would be
\eqn\yoyo{\theta_0=\sum_{\lambda\in\Gamma}
\exp\left(-i\pi\bar\tau(\lambda_+)^2 - i\pi \tau
(\lambda_-)^2\right).}
Here $\lambda_+$ and $\lambda_-$ are the self-dual and anti-self-dual
projections
of $\lambda$; hence $\lambda_+^2>0$ and $\lambda_-^2<0$.  Note that
the self-dual projection is explicitly $\lambda_+=\omega(\omega,\lambda)$
with $\omega$ the normalized self-dual harmonic two-form introduced
above.

To obtain the desired lattice theta function for the $U(1)$ gauge field that
appears
on the $u$-plane, \yoyo\ must be modified in two ways.  First of all, suppose
that we are doing $SU(2)$ gauge theory, spontaneously broken at a generic
value of $u$ to $U(1)$.  Such breaking means that an underlying   $SU(2)$
bundle
$W$ reduces to $T\oplus T^{-1}$ where $T$ is a line bundle.
The exponent in \yoyo\ is normalized
to be the correct effective action for an $SU(2)$ bundle of this form in
Donaldson
theory if $\lambda$ is identified as $c_1(T)$.

Suppose, however, that one wishes to do $SO(3)$ gauge theory with a bundle $E$
of $w_2(E)\not= 0$.  In this case, the bundle $W$ does not exist and should be
replaced
by $E={\rm Sym}^2(W)$.  However, as far as the low energy theory on the
$u$-plane
is concerned, the effect of having $w_2(E)\not= 0$
is simply that $\lambda$ is no longer an integral cohomology
class but is shifted from being an element of $\Gamma$ by $\half w_2(E)$.
Thus $\lambda$
is now an element of $\half \Gamma$ that is congruent to $\half w_2(E)$ modulo
$\Gamma$.
This shift by itself would turn the theta function into
\eqn\noyoyo{\theta_1=\sum_{\lambda\in\Gamma+\half w_2(E) }
\exp\left(-i\pi\bar\tau(\lambda_+)^2 - i\pi \tau
(\lambda_-)^2\right).}

In addition, there is an important
phase factor in the lattice sum whose origin
was explained
in section 4.4 of \abelS.  This  factor may be described
as follows.  Pick an arbitrary and fixed $\lambda_0\in \half w_2(E)+\Gamma$.
  The  factor in question is
\eqn\joxoz{(-1)^{(\lambda-\lambda_0)\cdot w_2(X) }e^{2\pi i \lambda_0^2}.}
Of course, there is no canonical choice of $\lambda_0$ (unless $w_2(E)=0$, in
which case we take $\lambda_0=0$).  If $\lambda_0$
is replaced by $\tilde\lambda_0$, then \joxoz\ is multiplied
by
\eqn\pokko{(-1)^{\beta\cdot w_2(X)}}
where $\beta$ is the integral class $\beta=\lambda_0-\tilde\lambda_0$.
Thus, with the factor \joxoz\ included, the overall sign of the Donaldson
invariants depends on a choice of $\lambda_0$.
This fact actually mimics
standard facts in Donaldson theory.  In Donaldson theory, for $w_2(E)=0$, one
conventionally orients the instanton moduli spaces
using an orientation
of $H^1(X)\otimes H^{2,+}(X)$; such an orientation entered our discussion
above as a choice of sign for the integration measure of the fermion zero modes
on the $u$-plane.  When $w_2(E)\not= 0$, the conventional way of orienting
the instanton moduli spaces depends in addition on a choice of integral lift
of $w_2(E)$, which in our above notation should be identified
with $2\lambda_0$.  Moreover, when the integral lift of $w_2(E)$ is
changed, the usual orientation of the moduli space is multiplied
by the factor \pokko.  (See \DoKro, pp. 281-3, for a summary of these matters.)
Thus, when the factor
\joxoz\ is included, the $u$-plane integral depends on the same choices,
and transforms in the same way when the choices are changed, as the orientation
of instanton moduli space.

Equally important is the fact that
when $\lambda$ is changed by an element of $\Gamma$, the phase factor \joxoz\
changes by a factor of  $\pm 1$.
Thus, in its dependence on $\lambda$, this phase factor behaves as a sign
factor.
A field theory explanation of the $\lambda$ dependence of this
 factor was given in
\abelS, where it was also shown to be crucial
in the appearance of ${\rm Spin}^c$ structures near $u=\pm 1$ after duality.
In writing the phase factor precisely as in \joxoz, we are fixing an overall,
$\lambda$-independent factor that was not analyzed in \abelS, in such a way
as to agree with standard mathematical conventions in Donaldson theory.
\foot{In particular, that is the reason for the
$e^{2\pi i\lambda_0^2}$ factor in \joxoz.  That factor
is the same as $e^{i\pi w_2(E)^2/2}$, and is completely independent
of the choice of $\lambda_0$, as $w_2(E)^2$ has a well-defined
value modulo 4.  This factor is included to agree with standard
mathematical conventions in Donaldson theory.  It would have been equally
natural to include instead a factor $e^{-2\pi i\lambda_0^2}$.  The
change would multiply the Donaldson invariants by a sign factor
$(-1)^{w_2(E)^2}$, corresponding to  a reversal of the orientation on
instanton moduli space.  It is necessary to include one or the other
factor $e^{\pm  i\pi w_2(E)^2/2}$ in order for the Donaldson invariants
to come out to be real after performing the $u$-plane integral.}
Multiplication by a different $\lambda$-independent factor would simply
multiply the generating function of the Donaldson invariants by that factor.
\foot{The convention used in \monopole\ actually differed from the present
choice (which as we have stated is chosen to agree with standard mathematical
conventions) by a  sign factor.  This sign factor depends on an integral
lift of $w_2(X)$ and is $(-1)^{(2\lambda_0^2+\lambda_0\cdot w_2(X))}$.}

Putting these factors together,
the photon partition function on the $u$-plane is
\eqn\lateth{
Z={e^{2 \pi i \lambda_0^2} \over \sqrt {{\rm
Im}\,\tau}}\sum_{\lambda\in\Gamma+\half w_2(E) }
(-1)^{( \lambda-\lambda_0)\cdot w_2(X) }
\exp\left(- i\pi \bar \tau(\lambda_+)^2-i\pi \tau
(\lambda_-)^2\right).}
For application to Donaldson theory,
we will actually require not the partition function but the path integral
with a certain operator insertion,
so it is a related but different function that will
appear in the $u$-plane integrals.

\bigskip\noindent{\it Absorption Of Fermion Zero Modes And Elimination Of
Auxiliary
Field }

For $b_1=0$ and $b_2^+=1$, there are precisely two fermion zero modes.
We have already determined in section 2.3 which interactions should be used
to absorb these zero modes.  The relevant part of the path
integral contains precisely
one insertion of the $\eta\chi (F+D)$ interaction vertex.  This vertex
corresponds
to a factor in the path integrand that reads
\eqn\momo{
\exp\biggl[ - {i \sqrt{2} \over 16\pi}\int_X
{d\bar \tau\over d\bar a}\eta\wedge\chi\wedge(F_++D_+)
\biggr] .}

Now is a good time to integrate out $D_+$ and eliminate it from further
discussion.
The only $D$-dependent factor in the path integral, other than \momo, appears
in
\ojurry.  Combining this with \momo, the $D$-dependence of the path integral
is in
a factor
\eqn\hogo{ - \exp\left(-{i\over  4 \pi} {du\over
da}\int_SD_+\right)\cdot\left({i \sqrt{2} \over 16\pi} \int_X
{d\bar \tau\over d\bar a}\eta
\wedge\chi\wedge (F_++D_+)\right).}
One can integrate $D$ out of this expression, using the fact that $D$ is a
Gaussian
field with propagator
$\langle D(x)D(y)\rangle\sim \delta(x,y)/{\rm Im}\,\tau$.
Upon integrating out $D$,   \hogo\ becomes
\eqn\nogo{- \exp\left(S_+^2\left({ (du/da)^2\over 8 \pi  {\rm
Im}\,\tau}\right)\right)
\cdot \left({  \sqrt{2} \over 16\pi} \int_X {d\bar\tau\over d\bar a}
\eta\wedge \chi\wedge (F_+ + i { (du/ da)\over {\rm Im}\tau}
                        S_+)\right).}

To   reduce this further, note that $F_+ $ coincides with
$4\pi \lambda_+$.
Also, upon integrating over the fermion zero modes, we can replace $\eta$ by 1
and
$\chi$ by $\omega$.  The resulting factor is hence
\eqn\onogo{-{\sqrt{2} \over  4}
{d\bar \tau\over  d\bar a}
 \cdot
\exp\left(S_+^2\left({  (du/da)^2\over 8 \pi y} \right)\right)\cdot
\left(  (\omega,\lambda) + {i \over  4 \pi y}  {du\over  da} (\omega,S)
 \right).}
where $\tau = x + i y$.

The first factor in \onogo\ depends on the lattice vector $\lambda$ and should
be
included in defining the  lattice sum $  \Psi$
that appears in  the $u$-plane integrands
of Donaldson theory.
It is also  convenient to include in the definition of $ \Psi$
an additional $\lambda$-independent factor of
$\exp\left(-S_-^2(d  u/ d  a)^2/ 8\pi y \right)$; this factor is
chosen
to simplify the modular behavior of $  \Psi$.
The lattice sum in the $u$-plane integrand is thus
\eqn\newtheta{
\eqalign{
\Psi = \exp\bigl[  - { 1 \over  8 \pi y}({d   u \over  d  a})^2 S_-^2 \bigr]
&
e^{2\pi i \lambda_0^2}
\sum_{\lambda\in H^2+ \half w_2(E) }(-1)^{(\lambda-\lambda_0)\cdot w_2(X)}\cr
\biggl[ (\lambda ,\omega) +{i \over  4 \pi y} {d u\over  d a} (S,\omega )
   \biggr]  \cdot  &
\exp\biggl[ - i \pi \bar\tau (\lambda_+)^2-i \pi   \tau(\lambda_-  )^2-i
{ d    u \over  d   a} (S,\lambda_-) \biggr]  \cr }
}

The net effect is that the photon
path integral relevant to the $u$-plane integral,
combining what appears in \gurgle, \onogo, and \newtheta, is
\eqn\momobo{Z=-{\sqrt{2} \over  4}
{d\bar \tau\over  d\bar a}
y^{-1/2}\exp\left(S^2{(d u/ d a)^2\over 8 \pi y
}\right)\cdot \Psi.}

\bigskip\noindent{\it Putting The Pieces Together   }

Combining what we have obtained in \juggo, \ojurry, \onogo, and \momobo, the
integral over the $u$-plane is simply
\eqn\hoggo{
Z_u = \int_{\CM}
{dx dy \over  y^{1/2}} \mu(\tau) e^{2 p u + S^2 \hat T(u)} \Psi
}
where $\tau=x+i y$,
$\hat T = T +{1 \over  8 \pi y} \bigl({du \over  da}\bigr)^2$, and the measure
factor is:
\eqn\measfact{
 \mu(\tau)   = - {\sqrt{2} \over  2} {da \over  d \tau} A^\chi B^\sigma =-4
 \sqrt{2}i (u^2-1){da \over  du}
\Biggl( { ( {2i \over  \pi} {du \over  d \tau} )^2 \over  u^2-1 }
\Biggr)^{\sigma/8}
}
We have here fixed the normalization factors $\alpha$ and $\beta$ so as to
agree
(in the computations that follow) with
known results on Donaldson invariants.  We have also set $\chi+\sigma=4$,
since this is so for four-manifolds of $b_1=0,\,b_2^+=1$.

Notice that various factors have combined neatly so that the bosonic
integration
measure
$da\,d\bar a$ in \juggo\ is transformed to $d\tau\,d\bar \tau$.  This is very
convenient since in terms of $a$ there is no reasonable description of the
integration
region.  In terms of $\tau$, however, there is a natural answer: the
integration
is to be taken over the modular region ${\cal M}$ of the group $\Gamma^0(4)$,
that is, over the quotient $\Gamma^0(4)\backslash {\cal H}$, where
${\cal H}$ is the upper half plane on which the subgroup $\Gamma^0(4)$
of $SL(2,\Z)$ acts in the usual fashion.  This is simply the assertion that
the $u$-plane is the modular curve of $\Gamma^0(4)$.

The formal proof that this integral (regularized as in the next subsection)
is metric-independent follows from the fact that the stress tensor
of the twisted theory is of the form $\{\bar\CQ,\dots\}$, as a result of which
the derivative of the integral with respect to $\omega$ is a total
derivative on the $u$-plane.
We postpone this argument to section 11.3, where we make this
argument in a wider context and show directly that the integral is a locally
constant function of $\omega$ with wall-crossing.

\subsec{ Definition Of The Integral}

 At this point we must
face the fact that the integral \hoggo\  does not converge, because of
bad behavior near $u=\infty$ and in some cases also near $u=\pm 1$.  It is
quite
clear that if one expands \hoggo\ in powers of $p$, so as to compute Donaldson
invariants
of increasing order, then as $u$ diverges at infinity one will eventually run
into a divergent integral.  There is a similar problem near $u=\pm 1$ if
$\sigma$
is sufficiently
negative.

To complete the definition of the integrals that will be studied in the rest
of the paper,
we must therefore explain how the divergences will be cut off.  We do this in
a standard
and natural way, as follows.
First we expand \hoggo\ out to a given order in $p$ and $S$, to obtain
an integral that should give a Donaldson invariant of some given order.  To
define
that particular integral,
after writing $\tau=x+iy$, we perform the integral for $y<y_0$,
for some cutoff $y_0$, and then take the limit as $y_0\to\infty$ only at the
end.
A similar procedure
is followed near the cusps at $u=\pm 1$, introducing the dual $\tau$-parameters
at the other
cusps and integrating first over ${\rm Im}\,\tau_D<y_0$, before taking the
limit
as $y_0\to\infty$.
This eliminates the infinities, for the following reason.  Set $q=
\exp(2\pi i \tau)$.   Then the term in \hoggo\ that is of any given order in
$p$ and
$S$  is a sum of terms, each of which
is a power of $y$ times a sum of the form
\eqn\jucci{\sum_{\nu,\mu}q^\nu \bar q^\mu.}
$\nu$ and $\mu$ are not integers (or even rational numbers, in general) but
obey
$\nu-\mu \in {1 \over  4} \IZ $.

The important point is that, though $\nu$ has no lower bound, $\mu$ is bounded
below by zero.  The reason for this is that negative exponents in \jucci\ come
only
from
factors in \hoggo, such as $u$ and
$(d\tau/du)^{-\sigma/4}$, which are singular at the
cusps; these factors are holomorphic and so contribute to $\nu$ but not $\mu$.

Under these conditions, consider an integral of the following form:
\eqn\hippo{\lim_{y_0\to\infty}\int_{y_1}^{y_0}{dy\over y^c} \int_0^kdx
\sum_{\nu,\mu}q^\nu \bar q^\mu.}
Here $y_1$ is an irrelevant lower cutoff, say $y_1=3$, that is included so as
to study one
cusp while keeping away from others.  The interest is in whether the integral
converges
for $y_0\to \infty$.
The $x$ integral runs from 0 to $k$  where (for $\Gamma^0(4)$) $k=4$ for the
cusp at
infinity, and $k=1$ for the other cusps.
A detailed examination of \hoggo\ and of the definition of the function
$ \Psi$
shows that in all cases either $c>1$ or there are, for a generic metric on $X$
(the metric
enters in the definition of $ \Psi$), no terms with
$\nu=\mu=0$.  Now integrating first over $x$ projects the sum in \hippo\ onto
terms with
$\nu=\mu$, and hence (as $\mu$ is non-negative) onto terms that vanish
exponentially
or, if $\nu=\mu=0$, are constant at infinity.  For a generic metric on $X$,
the $y$ integral converges as  $y_0\to\infty$, since all terms that have
survived
the $x$ integral have $c>1$ or $\nu,\mu>0$.
Via this procedure, the integral becomes for generic metric a well-defined
formal power series in $p,S$.

For special metrics, on the other hand, there are terms with $c<1$, in fact
$c=1/2$,
and $\nu=\mu=0$.  That is where wall-crossing will occur, as we discuss in
section 4.

The cutoff we have given is certainly  quite natural and will lead to elegant
formulas
that agree with Donaldson theory as it
has been formulated mathematically.  However, one is reluctant to think of any
cutoff as
fundamental.
Since the behavior near $u=\infty$ is linked to the ``bubbling'' phenomena in
Donaldson theory,
one might guess that a different but still
``reasonable'' cutoff might correspond to a different recipe from the one
usually
used in Donaldson theory for handling the singularities of instanton moduli
space.
The usual experience in quantum field theory is that upon making a change in
the cutoff
recipe (within a class of ``reasonable'' cutoffs) one gets the same theory
with a different
parametrization.  In the present case, for instance, such a reparametrization
might
mean replacing the function $e^{pu}$ by a function
$e^{pu+\alpha(u)\chi+\beta(u)\sigma}$,
where $\chi$ and $\sigma$ are the Euler characteristic and signature of $X$
and $\alpha$
and $\beta$ are some universal functions.  We will
not, however, investigate the extent to which either the cutoff-dependence of
the
$u$-plane integrals, or the dependence of Donaldson theory on how
the singularities are treated, can be so written.

Curiously, the regularization relevant to Donaldson
theory also coincides with that needed to define
one-loop amplitudes in string theory.
In particular \hoggo\ bears a striking resemblance to
threshold corrections in compactifications of heterotic
string on ${\rm K3} \times S^1$.
It would be
interesting to understand this
more deeply.

\subsec{Verification of modular invariance}

Having defined the integration at the cusps it
is still
worth checking that the integral actually
 makes sense, namely, that
the integrand is single-valued. This is equivalent to
checking modular invariance of the integrand under
$\gof$.  Verifying this is a test of our calculation since
the underlying $SU(2)$ gauge theory is intrinsically defined,
but to compute the explicit $u$-plane integrand we have had to use
an effective low energy $U(1)$ description that is only uniquely
determined up to a modular transformation.

Modular invariance is most readily checked by
relating $\Psi$ in \hoggo\ to the standard
Siegel-Narain theta functions which transform
simply under modular transformations. Our
notation is explained in Appendix B.
We introduce the theta function
\eqn\thta{
\Theta  =\kappa^{-(w_2(X), w_2(E)) } \Theta_{H^2}(\tau, \half w_2(X),\half
w_2(E) ;P_\omega,\xi)
}
with $\kappa=e^{2\pi i/8}$ and
\eqn\defofx{
\xi = \rho y \daub \omega + {1 \over  2 \pi}  \duab S_-
}

Defining
\eqn\effhat{
\eqalign{
\hat f (p, S, \tau,y) & \equiv \CN
\bigl((u^2-1){d \tau \over  d u} \bigr)^{\chi/4} (u^2-1)^{\sigma/8} {du \over
d \tau} \exp\bigl[ 2p u+S^2 \hat T(u)
\bigr]\cr}
 }
for an appropriate normalization constant $\CN$
we now introduce the auxiliary integral $\CG(\rho)$
\eqn\brght{
\CG(\rho)  \equiv
\int_{\gof \backslash\CH} {dx dy \over  y^{3/2}}
\hat f(p, S, \tau,y) \bar \Theta
}
 related to the
Coulomb partition function by:
\eqn\rowiii{
 Z_u = (S,\omega)  \CG(\rho) \biggr\vert_{\rho=0}
 + 2
 {d \CG  \over  d   \rho}   \biggr\vert_{\rho=0}
}

Denote the   integrand of \brght\
by ${dx dy \over  y^2} \CJ$, where
$\CJ = \hat f \cdot y^{1/2} \bar\Theta$.
We obtain a fundamental domain for $\gof$
from  a
fundamental domain $\CF$
for $ PSL(2,\IZ)$ by
\eqn\fnddom{
\gof \backslash\CH \cong
\biggl[ \CF \cup T\cdot \CF \cup T^2\cdot\CF
\cup T^3 \cdot \CF\biggr] \cup S \cdot \CF \cup T^2 S \cdot \CF
}
The first four domains give the region of the cusp at
$\tau \rightarrow i \infty$ and correspond to the semiclassical
region. The region $S \cdot \CF$ surrounds the
cusp near $\tau=0$ and will be referred to as the
monopole cusp. The region
$T^2 S \cdot \CF$ surrounds the cusp near $\tau=2$
and corresponds to the massless dyon.

Mapping the integrand in these 6 regions to the
domain $\CF$ we get six functions:
\eqn\sxfn{
\eqalign{
\CJ_{(\infty,0)}(\tau )    &\equiv \CJ(\tau)  \cr
\CJ_{(\infty,1)}(\tau )  &\equiv \CJ(\tau+1)\cr
\CJ_{(\infty,2)}(\tau )   &\equiv \CJ(\tau+2) \cr
\CJ_{(\infty,3)}(\tau )   &\equiv \CJ(\tau+3) \cr
\CJ_{M}(\tau ) &   \equiv \CJ(-1/\tau) \cr
\CJ_D (\tau ) &  \equiv \CJ(2 -1/\tau) \cr}
}

In general we  will  denote $\gof$-modular
forms $F$ transformed as in \sxfn\ by
$F_I$ where
\eqn\cosetnot{
I=(\infty,0), (\infty,1),(\infty,2),(\infty,3), M, D
}
These will form representations of the
permutation group $S_3 \cong \bar \Gamma/\gof$.
\foot{Actually, one encounters modular forms of
half-integer weight and hence occasionally one
must work with the metaplectic double-cover. }

It is now straightforward to bring the integral
to the  form
\eqn\brght{
\CG(\rho)  =
\int_{\CF} {dx dy \over  y^{3/2}}
\sum_I
\hat f_{I}(p, S, \tau) \bar \Theta_I }
where
\eqn\thetcoset{
\Theta_I = e^{i \phi_I} \Theta_{H^2}(\tau, \alpha_I , \beta_I ; \xi_I)
}
are the transforms of the Siegel-Narain theta function
implied by \sxfn. It is easy
   to check that $\hat f_I$ and $\Theta_I$ transform
in the same unitary representation of the modular
group.  Hence $\CG$ is modular invariant, and therefore,
so is $Z_u$.

\subsec{The four basic properties}

Here and in sections 4-6 we will  examine, in light of what we have
learned,  the basic formal
properties mentioned in the introduction.

The homotopy invariance of the $u$-plane integral
is  manifest from the form of the integrand in \hoggo.
The
$u$-plane integral for a simply-connected four-manifold is completely
determined
by the lattice $\Gamma=H^2(X,\Z)$ with its intersection pairing.
Thus, these integrals, while extremely subtle, capture only elementary
topological information.  Only because there are additional contributions
from $u=\pm 1$ is this
compatible with the fact that Donaldson invariants
of four-manifolds contain further information beyond the
intersection form.
Those contributions involve the SW invariants
and will be the subject of section 7.

As explained in the introduction, beyond homotopy invariance, the $u$-plane
integrals possess three additional formal
properties that determine them completely.
These are  the chamber dependence,
the vanishing in certain chambers, and the
blowup formulas.    The next three sections are devoted to these properties
in turn.

\newsec{Chamber-dependence of $Z_u$ and Wall-Crossing Formulae}

We are now in a position to give a comparatively simple explanation
of the wall-crossing formula for the $u$-plane integrals.
It is useful to think in terms of the analogy of \hoggo\
to one-loop integrals in string theory. In this analogy
the wall-crossing discontinuities arise when --
in the language of string amplitudes - a massive
particle   becomes massless on a subvariety
of Narain moduli space, leading to an
infrared divergence in the integral.

In the present integral, there are three cusps
$\tau = i \infty, \tau = 0 , \tau = 2$. The first
cusp leads to the Donaldson wall-crossing
formulae. The other two lead to the SW
wall-crossing formulae.

As explained in section 3.2, any discontinuity
in $Z_u$  arises from a
finite number of terms in $\Psi$ and from
negative powers of $q$ in the
Fourier expansion of the nearly holomorphic
modular forms.
The relevant terms are of the form
\eqn\wcri{
\CI(\omega) \equiv
\int_{\CF} {dx dy \over  y^{1/2}} c(d) e^{2 \pi i x d - 2 \pi yd}
e^{-i \pi x (\lambda_+^2 + \lambda_-^2)}
e^{- \pi y (\lambda_+^2 - \lambda_-^2)}
 (\omega,\lambda) }
 for some integer $d$ and some $\lambda$.
(The  $u$-plane integrand also contains additional terms proportional
to $y^{-3/2}$ instead of $y^{-1/2}$ and lacking the factor
$(\omega,\lambda)$.  It will, however, become clear that such
terms produce no singularity.)
In \wcri, $c(d)$ is the coefficient of
some modular object. It is also  a function of
$p$ and $(S_-, \lambda_-), (S_+, \lambda_+)$, but this has not been
indicated explicitly.

We want to study the integral in \wcri\ for fixed $\lambda$
as the decomposition $\lambda=\lambda_++\lambda_-$ varies.  We recall
that $\lambda_+=\omega (\omega,\lambda)$, so this decomposition is
determined by $\omega$.
The issue is to find a discontinuity in $\CI(\omega)$ when a ``wall'' is
crossed.  Such a discontinuity occurs only if $\lambda^2<0$, since
otherwise the integral in \wcri\ (with the regularization described in
section   3.2) is too well-behaved to produce a discontinuity.
When, however, $\lambda^2<0$, there is a discontinuity at $\lambda_+=0$.
This can be computed as follows.  Upon doing the $x$ integral, one projects
onto $d$ such that $2d=\lambda^2$.  For this value of $d$, the $y$ integral
looks like
\eqn\jurdo{\int_{y_1}^\infty {dy\over y^{1/2}} c(\lambda^2/2) e^{-2\pi y
\lambda_+^2}\lambda_+.}
This is an elementary integral (if one replaces $y_1$ by 0) and converges
for all non-zero $\lambda_+$, but is discontinuous at $\lambda_+=0$.
The discontinuity comes from the large $y$ part of the integral and
so is independent of $y_1$.
The discontinuity in $\CI(\omega)$
as $\omega$ crosses  from
$(\omega, \lambda) = 0^-$ to
$(\omega,\lambda) = 0^+$
is easily computed to be\foot{
This is the contribution of a single copy of the $SL(2,\IZ)$
fundamental
domain near the cusp.  For the cusp at infinity, there is an extra
factor of four
%%%
from the summation over four copies of the
fundamental domain, or equivalently from the fact that $x$ runs from
$0$ to 4.}
\eqn\wcriv{
\CI(\omega_+) - \CI(\omega_-) =
\sqrt{2} c(d)  = \sqrt{2} \bigl[ q^{-\lambda^2/2} c(q) \bigr]_{q^0}
}
The notation
$[ \cdot ]_{q^0}$ indicates the constant term in
a Laurent expansion in powers of $q$. It may also be
expressed as a residue. Since $\lambda_+=0$
we may put $(S_+, \lambda_+)=0$
and $(S_-, \lambda_-)= (S, \lambda)$
in the function $c(q)$.

The conditions $\lambda^2<0$, $\lambda_+=0$ for a discontinuity
are very familiar in Donaldson theory.  They are the conditions
that the line bundle with Chern class $\lambda$ admits an instanton
connection, which (upon embedding of $U(1)$ in $SU(2)$ or $SO(3)$)
appears as a singular point in instanton moduli space.  The discontinuity
of
$Z_D$ is usually computed by studying the behavior near
this singularity.  The conditions $\lambda^2<0$, $\lambda_+=0$
for a wall are also very natural in string theory.  They are the
conditions that in toroidal compactification of the heterotic string,
a massive particle with Narain vector $\lambda$ becomes
massless and gives an infrared singularity.
The above computation exhibiting the discontinuity has a direct
analog in heterotic string threshold computations. See,
for examples,  \refs{\hm,\AFT}.

The equation \wcriv\ is of central importance.
It shows (modulo an analysis we give presently
showing that the contributions of the different cusps do not cancel)
that the partition function
$Z_u$ is {\it not} topologically
invariant. Indeed, the conditions
 $\lambda^2<0$, $\lambda_+=0$  define   chambers in the
forward light cone
$V_+= \{\omega\in H^2(X;\IR): \omega^2 > 0 \}$.
Any $\lambda$ with $\lambda^2<0$ defines a
wall in $V_+$   by
\eqn\walldef{
W_\lambda \equiv \{ \omega: \lambda\cdot \omega =0\}.
}
The chambers are the complements of the walls.

When $\omega$ crosses such  a wall there is a
discontinuity in $Z_u$
given by \wcriv\ for an appropriate $c(q)$. Any given correlation
function  $\sim \langle \CO^\ell  \CI(S)^r \rangle$
will involve an integral with a holomorphic form
with pole growing linearly with $r, \ell$.
Thus, any such
correlation function will only be piecewise
constant as a function of
$\omega$. The number of chambers for such
a correlator grows with $r, \ell$.
Let us now examine in more detail the chamber-dependence
coming from the singularities at the three cusps of
$\gof\backslash \CH$.

\bigskip\noindent{\it Comparison To Donaldson Theory Wall-Crossing
Formulas}

The four cosets forming the cusp at
$\tau\rightarrow i \infty$ contribute the semiclassical
wall-crossing formula.  As explained in the introduction,
this contribution should coincide with  the wall-crossing
formula for the Donaldson invariants.

For these cosets, the shifts $\beta_I$ defined in
\thetcoset\ are all given by
$\beta_I=\half w_2(E)$ and the formula
\wcriv\ for the quantity
$\delta_{u,\infty}$ of equation \junglo\ becomes:
\eqn\semiwc{
 \eqalign{
Z_{u,+}  - Z_{u,-}
& = -   32i     (-1)^{ (\lambda-\lambda_0,   w_2(X))   }
e^{2\pi i\lambda_0^2}\cr
\cdot \Biggl[q^{- \lambda^2/2}  (u^2-1) h(\tau) &
\Biggl( { ( {2i \over  \pi} {du \over  d \tau} )^2 \over  u^2-1 }
\Biggr)^{\sigma/8}
 \exp\biggl\{ 2p u  +  S^2 T(u)
  -  i  (\lambda , S)/h
\biggr\}
\Biggr]_{q^0} \cr}
}
Here $h \equiv {da \over  du}$. In
appendix A we give expressions for the
 various modular
forms in \semiwc\   in terms of Jacobi
$\vartheta$-functions.  Using these expressions
\semiwc\ simplifies to:
\foot{The right hand side of \semiwc\ is {\it odd} under
$\lambda \rightarrow -\lambda$. Therefore
at any wall the contributions of $\pm \lambda$
to the discontinuity in $Z_u$ {\it add}, rather than
cancel because this is a formula for
$Z_+ - Z_-$.}
\eqn\semiwcp{
\eqalign{
 Z_{u,+}  - Z_{u,-}   & = -   {i \over  2}   (-1)^{(\lambda-\lambda_0,w_2(X))}
 e^{2\pi
i\lambda_0^2}
\cr
\Biggl[q^{- \lambda^2/2}    {\vartheta_4^{8 + \sigma} \over
h(\tau)^3}
&
\exp\biggl\{ 2p u  +  S^2 T(u)
  -  i  (\lambda , S)/h
\biggr\}
\Biggr]_{q^0}  \cr}
}
with $h= \half \vartheta_2 \vartheta_3$ and
$$
u   = \half { \vartheta_2^4 + \vartheta_3^4 \over  (\vartheta_2 \vartheta_3)^2
}
$$

In fact,
the semiclassical wall-crossing formula
\semiwcp\ is identical to the wall-crossing formula
given in \refs{\gottsche,\gottzag}
for Donaldson invariants of four-manifolds with $b_1=0, b_2^+=1$.
In making this comparison
we must note the following.  Modular forms not of $\Gamma^0(4)$ but
of a group conjugate to it by $\tau\to\tau+1$ are used in \refs{\gottsche,
\gottzag}, and one must make this transformation in \semiwcp\ before
matching the modular forms. (The shift $\tau\to\tau+1$ does not affect
the residue or $q^0$ coefficient in \semiwcp.)
Also,  the formula in \gottzag\ should be
corrected by a factor of $1/2$.

\bigskip\noindent{\it SW Wall Crossing}

We will now analyze the wall-crossing behavior associated with
the monopole and dyon
cusps at $\tau=0,2$.  As explained in the introduction, this wall-crossing
contribution
is related to the wall-crossing behavior of SW invariants -- a connection
that we will explore further in section 7.

The ``monopoles'' that appear near $u=1$
are not spinors (sections of a spin bundle $S_+$) but  are
sections of a ${\rm Spin}^c$ bundle which we write somewhat
symbolically as $S_+\otimes   L$, where
$  L$ does not exist as a line bundle but
$  L^{\otimes 2}$ does.  We define $\lambda=\half c_1(L^2)$; thus,
$\lambda\in\half w_2(X)+\Gamma$.
Hence, in the previous notation,
  $\beta=-\half w_2(X)$ for these cusps.  The walls
are still defined by
\eqn\monwall{
\eqalign{
\lambda^2 & < 0 \cr
(\lambda   , \omega) & = 0. \cr}
}

At the monopole cusp, the formula \wcriv\  for the
discontinuity $\delta_{u,1}$ of equation \junglo\ becomes:
\eqn\mwcrssing{
 Z_{u,+}  - Z_{u,-}   = -   {i \over  8}
e^{ 2i\pi (\lambda_0
\cdot\lambda+\lambda_0^2) }  \Biggl[q^{-
\lambda^2/2}
{\vartheta_2^{8+\sigma} \over  (h_M )^3}
\exp\biggl\{ 2p u_M  +  S^2 T_M
  -  i  (\lambda , S)/h_M
\biggr\}
\Biggr]_{q^0}  }
where
\eqn\yooemm{
\eqalign{
h_M & = {1 \over  2i} \vartheta_3 \vartheta_4\cr
u_M   & = \half { \vartheta_3^4 + \vartheta_4^4 \over
(\vartheta_3 \vartheta_4)^2 }\cr
T_M & = - {1 \over  24}\bigl[   {  E_2\over  h_M(\tau)^2}  - 8 u_M\bigr] \cr}
}
There is a similar expression for the dyon cusp.
Recall that $\lambda_0$ is a fixed element of $\half w_2(E)+\Gamma$
(or in other words that $2\lambda_0$ is a fixed integral lift of
$w_2(E)$) which entered in defining the $u$-plane integrals in section 3
(and which in the usual mathematical theory enters in orienting the
instanton moduli spaces).

We would like to stress that the functions
$u,h,T$ in
\mwcrssing\ are the same functions occuring in
the integrand of $Z_u$; however, they are most
usefully expressed in terms of the   expansion
relevant to the cusp at $\tau=0$, namely the
expansion in powers of $q_D \equiv \exp(2 \pi i \tau_D)$
where  $\tau_D \equiv -1/\tau$. To avoid
cluttering the notation, we have simply written
$\tau, q$ in \mwcrssing.

It is interesting to derive a condition for
\mwcrssing\ to be nonvanishing. Defining
\eqn\swdimi{
d_\lambda  = {1 \over  4} \biggl[ (2\lambda)^2 - (9-b_-)\biggr]
}
we see that the leading power of $q$ in
\mwcrssing\ is $q^{-\half d_\lambda}$. Thus,
\eqn\mwcond{
d_\lambda  \geq 0
}
is a necessary condition for SW wall-crossing.
In particular, note that we must have
$8+\sigma = 9-b_- < 0$.
As we explain in section 7 below, \mwcond\ is in
perfect correspondence with SW theory.
Indeed, $d_\lambda$ is the virtual dimension of
SW moduli space. We will show that
  a knowledge of the wall crossing formula for the cusps
at $u=\pm 1$
even allows us to learn about Donaldson invariants
for (hypothetical) four-manifolds of any $b_2^+$
that are not of simple type.

\newsec{Vanishing In Certain Chambers}

Another important
 property of the $u$-plane integrals is that
they vanish for certain
four-manifolds $X$ in certain chambers, for $SO(3)$
bundles $E$ with certain values of $w_2(E)$.
In conjunction with the  wall-crossing
formulae explored in section 4,
this determines the values of
$Z_u$ for such $X,E$.

We recall from the introduction that the appropriate
  four-folds $X$ are ``rationally ruled surfaces,''
which  map to a two-dimensional base $B$, with  generic fiber $F$ of genus
zero.
The required bundles are
bundles such that $(w_2(E),F)\not= 0$. The
vanishing chamber is defined by the
requirement that   the area of the
fibers goes to zero.  As we will see, the vanishing occurs because
the lattice theta function vanishes in this limit, and hence the $u$-plane
integrand vanishes pointwise.

To fix
ideas, we will focus on the
Hirzebruch surface $\IF_1$.
The general case is similar.
Let us set up some notation.
We regard $X=\IF_1$ as a blowup
of projective space
$\IF_1 = Bl_P(\IP^2)$.  The blowup produces an exceptional divisor
$B$.  $X$ fibers over $\IP^1$ with $\IP^1$ fibers which we call $F$,
and $B$ is a section
of this fibration.
There are two natural bases for $H^2(X;\IZ)$.  One basis consists of the pair
$\langle F,B \rangle $.  In this basis, the intersection form is:
\eqn\intfrmi{
\pmatrix{0 & 1 \cr 1 & -1 \cr}.
}
Alternatively, we can introduce $H=B+F$, the pullback of a hyperplane
class on $\IP^2$.  In the basis
$\langle H, B\rangle$, the intersection
form is
\eqn\intfrmi{
\pmatrix{1& 0 \cr   0 & -1 \cr}.
}
We choose an integral lift of $w_2(X)$ by setting $w_2(X) = F$.
The Kahler cone   is $ \{ x B + y H$: $x\leq 0 , x+y\geq 0\} $.
Any Kahler metric of unit volume has a Kahler class of the form
\eqn\perpt{
\omega = \cosh \theta H - \sinh \theta B \qquad 0 \leq \theta < \infty
}
$\omega$ is a self-dual harmonic two-form with $(\omega,\omega)=1$.
Define $ \epsilon \equiv e^{-\theta} $. We are interested
in the limit
\eqn\volfb{
\eqalign{
\omega \cdot F & = \epsilon \rightarrow 0 \cr
\omega \cdot B & = \sinh \theta \rightarrow \infty\cr}
}
in which the area of the fibers becomes small.  This is the limit
in which vanishing will occur.

The basic reason for the vanishing is the following.  Suppose
that instead of $\IF_1$ we took $X=\IP^1\times \IP^1=B\times F$.
(This choice of $X$ -- being fibered over $B=\IP^1$ by the projection
of $B\times F $ to the first factor -- is in any case a perfectly
acceptable example of the class of manifolds for which the vanishing
result holds.)
To analyze the lattice theta function in this case, note that if $B\times F$
is given a product metric, then a harmonic
two-form on $B\times F$ is the sum of a pullback from $B$ and a pullback
from $F$.
The Maxwell action is easily seen to be:
\eqn\prdct{
\int_{B \times F}  F \wedge * F =
{\vol(B) \over  \vol(F)} \biggl( \int_F F\biggr )^2 +
{\vol(F) \over  \vol(B)} \biggl( \int_B F\biggr )^2
}
Therefore, if the flux $\int_F F$ is forced to be
nonvanishing, which will be the case if $(w_2(E),F)\not= 0$,
and ${\vol(B) \over  \vol(F)} \rightarrow \infty$,
then the action goes to infinity and the path-integral is
suppressed. Confining a nonzero magnetic flux in
a small fiber costs a lot of action.

For $\IF_1$ or a more general rationally or elliptically fibered manifold,
the metric is not such a simple product.
Nevertheless, the same basic idea holds.  For instance, in the case
of $\IF_1$, if we take $w_2(E)=B$ (a special case of $(w_2(E),F)\not=0$),
then upon reduction to the abelian theory on the $u$-plane one gets
line bundles of first Chern class
\eqn\lambchrn{
\lambda = n H + (m + \half ) B\qquad {\rm with}~n, m\in \IZ.
}
The gauge theory action for such a line bundle
is:
\eqn\gaugactn{
\exp\biggl[- i \pi \bar \tau
\lambda_+^2 - i \pi \tau \lambda_-^2\biggr] =
\exp\biggl[  - \pi y \bigl[ n^2  + (m + \half)^2\bigr]
\cosh 2 \theta - i \pi x \bigl[ n^2 - (m + \half)^2 \bigr] \biggr]
}
Since $n,m$ are integral, \gaugactn\ always
leads to an exponential suppression in the
limit \volfb.
The only other metric dependence in the
integrand comes from the terms:
\eqn\metdep{
\exp[{1 \over  8 \pi y}S_+^2/h^2] \exp[- i (S_-,\lambda)/h]
\biggl[ (S,\omega) + 4 \pi i y (\lambda,\omega)/h \biggr]
}
Now,
\eqn\someg{
\eqalign{
(S, \omega) & = \cosh \theta S\cdot F + e^{-\theta} S\cdot B \cr
(S, \omega^\perp) & =-\sinh \theta S\cdot F + e^{-\theta} S\cdot B \cr}
}
To any given order in $S$,
this extra metric dependence contributes at most
a power $1/\epsilon^N$. This is killed by the exponential
suppression of the terms \gaugactn\
in the $\Psi$-function.
Therefore, the contribution of the integration over
any compact region vanishes for $\epsilon \rightarrow 0$.

To show vanishing of the $u$-plane integral in the given limit,
pointwise vanishing of the integrand is not quite enough.  This
is because the integration region is noncompact and the
convergence is not uniform throughout the $u$-plane.
Therefore, to complete the argument, we must
make some more careful estimates
at the three cusps.

The cusp at $\infty$ is easily handled. To study the behavior near
the cusp, we can replace the integral over
$\CF$ by the integral over the strip
$-\half\leq x \leq +\half, y\geq 1$.
We focus on
a term giving a fixed power $p^\ell S^r$.
The expression \gaugactn\ multiplies a modular
form times a polynomial in $1/y$.
After the  projection $\int dx$ the integral
has an absolute value bounded above by:
\eqn\finti{
\eqalign{
\sum_{n,m\in \IZ} \vert c(d(n,m)) \vert
\sum_{N,M}
&
{ a_{N,M} \over
 \epsilon^{M}}
  \int^\infty_1 {d y \over  y^{1/2}}
  y^{-N}
\cr
\exp\bigl[  -2 \pi y n^2
& (\cosh \theta)^2 - 2 \pi y (m+\half)^2(\sinh
\theta)^2\bigr]
\cr}
}
where the modular form is $\sum c(d) q^d$,
$2d(n,m) =n^2-(m+\half)^2$,   $N,M$ are nonnegative
integers, and the number of terms in the sum $\sum_{N,M}$
is bounded by $r$ for the contribution to $p^\ell S^r$.
Now we simply use the estimate:
\eqn\estmate{
\int^\infty_1 {d y \over  y^{1/2}} y^{-N} \epsilon^{-M}
\exp\bigl[  -y {K \over \epsilon^2}   \bigr]
\sim { 1 \over  \epsilon^{M-2}} \exp\bigl[  - {K \over \epsilon^2}   \bigr]
\biggl( 1 + \CO(\epsilon^2) \biggr)
}

The contribution of the monopole and dyon cusps
requires a little more care. The modular transformation
exchanges $w_2(X) \leftrightarrow - w_2(E)$.
It is more useful to work in the basis $\langle F, B \rangle$
for $H^2(X)$, so we now have a sum over Chern classes:
\eqn\monlamb{
\lambda= n B + (m- \half ) F \qquad n,m\in \IZ
}
The gauge action now becomes:
\eqn\gaugactnp{
\eqalign{
\exp\biggl[
- i \pi \bar \tau \lambda_+^2 - i \pi \tau \lambda_-^2\biggr]
&
=\qquad\qquad\qquad\cr
\exp\biggl[  - \pi y \bigl[
n^2  \cosh 2\theta
&
+ 2(m-\half)^2 e^{-2 \theta}
\bigr] - i \pi x \bigl[ n^2 - 2n (m - \half) \bigr] \biggr]\cr}
}
The metric dependence is as in \metdep\ with
$h \rightarrow h_M$ etc.
For $n\not= 0 $, only a finite number of terms
contribute to the integral and the argument
is identical to that used for the cusp at $\infty$.
However, for $n=0$ the entire
sum on $m$ survives the projection by $\int dx$.  Notice that in this
dual way of writing the theta function, the pointwise vanishing as $\epsilon
\to 0$ comes not because each term in the sum vanishes, but because
once we set $n$ to zero, the sum over $m$ is strongly oscillatory.  In
fact, the
sum on $m$ is a derivative of $\vartheta_1$ and we can
use the esimate:
\eqn\estthetone{
\sum_{m\in\IZ} e^{-2 \pi y \epsilon^2 (m-\half)^2}
e^{ i \pi (m-\half)}  (m-\half)^k
\sim  const.  \bigl( {1 \over  \epsilon^2 y} \bigr)^{k+1/2} e^{-\pi/(8
\epsilon^2 y) }}
(The constant vanishes for $k$ even.)

Working at fixed order in $S$, the integral after
the projection $\int dx$ becomes a finite sum of
terms of the form:
\eqn\moncsp{
\eqalign{
\int_1^\infty {dy \over  y^{3/2} } \bigl[ \tilde \mu(\tau) \bigr]_{q^0}
\sum_{m\in\IZ} e^{-2 \pi y \epsilon^2 (m-\half)^2}
e^{ i \pi (m-\half)}
&
\biggl( { (S \cdot F)^2 \over  y \epsilon^2}\biggr)^{t_1}
 \cr
\biggl( (S \cdot F) (m-\half) \biggr)^{t_2} \cdot \biggl[ K_1
( S\cdot F)/\epsilon
&
+ K_2 y (m-\half) \epsilon\biggr] \cr}
}
Here $t_1, t_2$ are nonnegative integers,
$\tilde \mu(\tau)$ is a certain modular form, and $K_1,K_2$ are constants.
But
$\tilde \mu(\tau) \sim \vartheta_2^8/(\vartheta_3 \vartheta_4)^4 e^{2pu_M
+S^2 T_M} \sim q+ \cdots $ actually vanishes at the cusp,
so in fact $[\tilde \mu(\tau)]_{{q^0}}$ vanishes.
This completes the proof of the vanishing theorem.  With the vanishing
factor removed, the rest of \moncsp\
behaves like the integral:
\eqn\mncspi{
{1 \over  \epsilon} \int_1^\infty {dy \over  y^{3/2} }
({1 \over  \epsilon^2 y})^{t_1} ({1 \over  \epsilon^2 y})^{t_2+\half}
e^{-\pi/(8 \epsilon^2 y)}
}
Making the change of variables $z=\epsilon^2 y$ shows that
this particular integral is non-zero and finite as $\epsilon \rightarrow 0$.

A similar reasoning
applies for other rational ruled surfaces.  A dangerous looking
factor $\vartheta_2^{\sigma}\sim q_D^{\sigma/8}$ coming from the
$A^\chi B^\sigma$ measure factor, which can be
singular if $\sigma$ is negative enough (and is responsible for
SW wall-crossing) is canceled by a vanishing of the theta function
near $q_D=0$.

In the somewhat analogous case that
$X$ is an elliptic surface of $b_2^+=1$, which maps to a two-dimensional
base (necessarily of genus zero)
with generic fiber $F$ of genus one, a slightly different situation holds.
In a chamber with nearly zero area for $F$ and with a bundle
such that $(w_2(E),F)\not=0$, the Donaldson invariants do not vanish
but obey a simple type condition \a\ (the SW invariants are likewise not
zero).  The $u$-plane integrals  hence
must obey a simple type condition.
In fact, an analysis as above shows a pointwise vanishing of the
$u$-plane integrand as the area of $F$ goes to zero,
but study of the behavior near $u=\pm 1$ shows a surviving contribution from
that region, as a result of which the $u$-plane integral does not
vanish.
Indeed, for $b_{2}^-=9$, the factor
$[\tilde\mu]_{q^0} =1+\CO(q) $  in \moncsp\
and since \mncspi\ is nonzero, there  can indeed be nonzero contributions.
However, acting on $Z_u$ by the operator
$[ {\p^2 \over  \p p^2 } -4 ]$, relevant to the
simple type condition,  is equivalent to an
insertion of
\eqn\spltypeii{
4(u^2-1) = {\vartheta_4^8 \over  (\vartheta_2 \vartheta_3)^2 }
}
in the integral. This factor increases the
order of the zero in $\tilde \mu$ by one and
hence, the integral for $b_{2}^-=9$  obeys a simple type condition, in
accordance with \a.

\newsec{The  Blowup Formula}

The blowup formula compares the Donaldson invariants of a four-manifold $X$
to those of a four-manifold $\widehat X$ that is obtained by blowing up a point
in $X$. Let $\pi:\widehat X\to X$ be the blowdown map.
Let $b$ be the exceptional curve contracted by $\pi$, $I(b)$ the
corresponding two-observable, and $t$ a complex number.
In the blowup formula, one seeks to compute
\eqn\gupo{\langle \exp(2pu+I(S)+tI(b) )\rangle_{\widehat\xi,\widehat X},}
in a limit in which the area of $b$ is small, in terms of
\eqn\nuppo{\langle\exp(2pu+I(S))\rangle_{\xi,X}.}
In trying to do so, one assumes
that  $\widehat\xi$ is a class that coincides with $\pi^*(\xi)$ away from
$b$.
The last condition means that $\widehat\xi=\pi^*(\xi)+jb$ for $j=0$ or 1.
Also, we are identifying a surface $S$ in $X$ (which we can assume does
not pass through the point that is to be blown up) with its pullback to
$\widehat
X$.

Let us first discuss on very general grounds
why a blowup formula exists and what its general form would be.
We scale up the metric of $X$ by $g\to t^2g$, with very large $t$.  Then
we blow up a point in $X$, producing an ``impurity'' that is supposed to be
very small, since in the blowup formula the area of $b$ is supposed to be
small.
To a distant observer, it must be possible to simulate the effect of the
impurity
by some local, $\bar{\CQ}$-invariant observable.  But in the twisted $\CN=2$
theory,
any local $\bar{\CQ}$-invariant observable (supported at a point as opposed to
the
$k$-form descendants) is a holomorphic function of $u$.  There must thus
be holomorphic functions $F_j(u,t)$, for $j=0,1$, such that
\eqn\ottop{\langle\exp
\left(2pu+I(S) +tI(b)\right)\rangle_{\widehat\xi,\widehat
X}
=\langle\exp\left(2pu+I(S)+F_j(u,t)\right)\rangle_{\xi,X}.}
Thus, the blowup formula is very similar to the
replacement in conformal field
theory of  a disk or handle by
a sum of local operators.  The blowup formula
replaces  a small region in $X$ which
has been modified by the blowup to produce $\widehat X$
by a local operator on $X$.

By applying the same sort of reasoning to the $u$-plane integrals, which simply
measure the contributions of certain vacua to the correlation functions, we see
that the $u$-plane integrals should obey a formula of the same structure.
Moreover, for reasons given in the introduction, the functions appearing in the
blowup formula on the $u$-plane are precisely the functions appearing in the
blowup
formula for the Donaldson invariants.

Actually computing the functions $F_j$ amounts to
comparing the $u$-plane integrands for $X$ with those for $\widehat X$. Let $B$
denote the cohomology class dual to $b$. We
work in a chamber $B_+=0$ (more precisely, for any
given correlator, in a chamber $B_+ < \epsilon$ for
sufficiently small $\epsilon$). Take $\tilde S = S + t b$
and substitute into the theta function \newtheta\
and use the condition $B_+=0$ to obtain:
\eqn\blwiip{
 \Psi_{\widehat X}  = \Psi_{X}
\exp[ {\pi   t^2 \over  8 \pi y   h^2}]
\sum_{n\in \IZ + \half w_2(\tilde E)\cdot B }
\exp\bigl[  i \pi   \tau n^2 +   i n t  /  h ]
e^{- i \pi n}
}
Similarly, the measure factor for the
blown-up manifold is related to that of the original
manifold by:
\eqn\blwiiip{
 \hat f_{\widehat X} = \hat f_X  \vartheta_4^{-1}
\exp \bigl[  -t^2 \hat T(u)  \bigr]
}
The $\vartheta_4^{-1}$ factor comes because the blowup changes
$\chi$ and $\sigma$, and the other factor comes because $\tilde S^2=S^2-t^2$.
Now, taking $\tau \rightarrow \tau +1$ on this expression
(so as to facilitate comparison to \gottzag, where a subgroup of
$SL(2,\IZ)$ that differs from $\Gamma^0(4)$ by conjugation by
$\tau\rightarrow \tau+1$ is used), we see that the blowup has the
effect of modifying the integrand
in a way which is equivalent to the substitution
\eqn\blwii{
e^{2p u_{(\infty,1)} } \rightarrow
e^{2p u_{(\infty,1)} }  \exp\bigl[ {t^2 \over  24}\biggl({E_2 \over
(h_{(\infty,1)})^2} - 8 u_{(\infty,1)}\biggr) \bigr]
{\vartheta_3\bigl({t\over  2 \pi h_{(\infty,1)} } \vert \tau\bigr)
\over  \vartheta_3(0 \vert \tau) }
}
in the case where $w_2(\tilde E)\cdot B =0~\mod ~2$.  Likewise, it is
equivalent to the substitution
\eqn\blwiii{
e^{2p u_{(\infty,1)} } \rightarrow - \kappa
e^{2p u_{(\infty,1)} }
\exp\bigl[ {t^2 \over  24}\biggl({E_2 \over
(h_{(\infty,1)})^2} - 8 u_{(\infty,1)}\biggr) \bigr]
{\vartheta_1\bigl({t\over  2 \pi h_{(\infty,1)} } \vert \tau\bigr)
\over  \vartheta_3(0 \vert \tau) }
}
when $w_2(\tilde E)\cdot B =1~\mod~ 2 $. The equations
\blwii, \blwiii\ are equivalent to
the expressions in \gottzag, eqs. 4.5.1, 4.5.2.
(Note there is a misprint in these formulae; they
should have $G(\tau)/f^2(\tau)$. Also their overall
phase is $\kappa^{-1}$ and differs from ours.)

In order to interpret \blwii\ and \blwiii, it is helpful to expand
the expressions in powers of $t$ and then {\it re-express }
the modular functions of $\tau$ as power series in the function
$u_{(\infty,1)}$. The resulting expression is a power
series in
$t$ whose coefficients are polynomials in $u_{(\infty,1)}$.
For example, when $w_2(\tilde E)\cdot B =1~\mod ~2$ we have
\eqn\blwiv{
(- \kappa )  \exp\bigl[ {t^2 \over  24}\bigl({E_2 \over (h_{(\infty,1)})^2} - 8
u_{(\infty,1)} \bigr]
{\vartheta_1\bigl({t\over  2 \pi h_{(\infty,1)} } \vert \tau\bigr)
\over  \vartheta_3(0 \vert \tau) }
= \sum_{k\geq 1}  t^k \CB_k(u_{(\infty,1)})
}
where (as we will see shortly) $ \CB_k$ is a polynomial.
Since $u_{(\infty,1)}$ multiplies $p$ in the integral
representation we arrive at the relation between
invariants:
\eqn\blwvi{
\eqalign{
\biggl\langle \exp\bigl[ I(S) + t I(B) + p \CO \bigr] \biggr\rangle_{\hat{X},
\widehat{w_2(E)} = w_2(E) +B  }
& =  \qquad\qquad\qquad \cr
\sum_{k\geq 1}  t^k   \CB_k({\p \over  \p p} ) \Phi_{X,w_2(E)}(S,p)
 = \sum_{k\geq 1}  t^k
\biggl\langle
\exp\bigl[ I(S) +
&
 p \CO \bigr]   \CB_k(\CO) \biggr\rangle_{
X,  w_2(E) } \cr}
}

We will now determine  the polynomials $\CB_k$.
We use the formula:
\eqn\sigfnc{
{\vartheta_1(z\vert\tau) \over  2
\pi \eta^3(\tau) } = - z \exp\biggl[-\sum_{k=1}^\infty
{G_{2k}(\tau) \over  2k} z^{2k}\biggr]
}
where $G_{2k} = 2 \zeta(2k) E_{2k}$ are Eisenstein
functions, and $E_{2k}$ are normalized Eisenstein
series of weight $2k$.  Using
$\vartheta_1'(0\vert\tau)=-2\pi  \eta^3 = - \pi \vartheta_2
\vartheta_3\vartheta_4$ we rewrite \blwiv\ as:
\eqn\fnbli{
\sum_{k\geq 1}  t^k   \CB_k(u_{(\infty,1)}) =
t \exp\biggl[ - {t^2 \over  3} u_{(\infty,1)} -
\sum_{k=2}^\infty {t^{2k} \over  2k} {G_{2k}(\tau) \over
(2\pi h_{(\infty,1)})^{2k} } \biggr]
}
Now we note that the Eisenstein functions $G_{2k}$
can be expressed as:
\eqn\eiesent{
G_{2k} = \sum_{4 s + 6 t = 2k} c_{k,s,t} (G_4)^s (G_6)^t
}
where $c_{k,s,t}$ are rational numbers. Using the
expression for Eisenstein series in terms of theta functions,
we now show:
\eqn\eisyoo{
\eqalign{
{G_4 \over  (2\pi h_{(\infty,1)})^4 }
& = {1 \over  45} (4  u_{(\infty,1)}^2 -3 ) \cr
{G_6 \over  (2\pi h_{(\infty,1)})^6 }
& = {2 \over  945} (8  u_{(\infty,1)}^3 -9 u_{(\infty,1)}  ) \cr}
}
and hence the $\CB_k$ are polynomials in $u_{(\infty,1)}$
with rational coefficients. The ratio of $\vartheta$ functions
can be expressed in terms of Weierstrass $\sigma$
functions and this is the form in which the blowup
formula was originally stated, e.g. in
\finstern.

As a simple example of \blwvi, we may consider the
first term in the expansion:
\eqn\blwvii{
\CD^{\hat X}_{\widehat{w_2(E)}
= w_2(E) +B} (S^n , p^m, B) =   \CD^{X}_{w_2(E)} (S^n, p^m)
}
For the other case one must expand \blwii\ to
order 4. One finds:
\eqn\blwvii{
\CD^{\hat X}_{\widehat{w_2(E)}
= w_2(E)  } (S^n , p^m, B^4) = -2  \CD^{X}_{w_2(E)} (S^n, p^m)
}
The factor of $2$ in this formula is the reason that in Donaldson theory,
invariants outside the so-called ``stable range'' are not integral
but have factors of $\half$.

\newsec{Universal Form Of SW Contributions}

We now analyze the SW contributions to $Z_D$, that
is, the contributions of the special vacua at $u=\pm 1$.
First, we outline the basic mechanism by which the contributions of the
$u=\pm 1$ vacua are computed in terms of SW invariants.
Then we obtain precise formulas, without assuming a simple type condition,
using the $u$-plane wall crossing formulas of section 4
to determine some universal functions that will be required.

Near $u=1$, there is a distinguished special coordinate on the $u$-plane,
namely $a_D$.  It is part of a vector multiplet that also contains a
distinguished
photon $A_D$, and fermions $\eta_D,\psi_D,\chi_D$ (which in the twisted
theory are a zero-form, a one-form, and a self-dual two-form).
The theory near $u=1$ has a $U(1)_R$ symmetry
(violated quantum mechanically by an anomaly involving the dimension of
SW moduli space) under which $A_D$ is invariant and $a_D$ has $R$ charge 2.

 Because there is a distinguished special coordinate near $u=1$ and no
issue of duality symmetry,
the theory near $u=1$
can be analyzed as a topological quantum field theory of a standard sort.
By picking a suitable functional $V$ and adding to the Lagrangian a term
$\lambda\{\bar{\CQ},V\}$ with
$\lambda\to\infty$, one localizes on supersymmetric configurations
(solutions of the SW equations) which must be counted, in a suitable way,
to give the correlation functions.  We have explained in section 2.3 that
duality
presents an obstruction to such a treatment of the $u$-plane contribution.

The theory at $u=1$ has,
in addition to the vector multiplet, a massless hypermultiplet
whose bosonic part will be called $M$.
In the topologically twisted theory on a four-manifold
$X$,  for reasons explained in section 4.2 of \abelS, the dual $U(1)$ gauge
theory
with connection $A_D$ involves not quite a connection on a line bundle but
a ${\rm Spin}^c$ connection.  For our purposes, we symbolically associate
a ${\rm Spin}^c$ structure with a bundle
$S_+\otimes L$, where $S_+$ is the positive
spin bundle and $L^2$
is a line bundle  such
that
$c_1(L^2)$ is congruent to $w_2(X)$ modulo 2.  (The factors $S_+$ and $L$
in $S_+\otimes L$ do not really exist separately.)  We will use the symbol
$\lambda$
to denote $\half c_1(L^2)$, regarded as an element of $\half \Gamma$
(as before $\Gamma=H^2(X,\Z)$) that is congruent to
$\half w_2(X)$ modulo 1.  We call $\lambda$ the first Chern class of the
${\rm Spin}^c$ structure, and  we identify a ${\rm Spin}^c$ structure
with its first Chern class.
To avoid cluttering the formulas, we will write in this section
simply $F$, instead of $F_D$, for the curvature of $A_D$, and $\lambda$,
instead of
$\lambda_D$, for the first Chern class of the ${\rm Spin}^c$ structure;
likewise
we will omit the subscript $D$ for the fermions $\eta,\psi,\chi$.
In the twisted theory, the bosonic part of the charged hypermultiplet is a
section
$M$ of the ${\rm Spin}^c$ bundle  $S_+\otimes L$.

The supersymmetric
configurations on which the contribution at $u=1$ is localized can be
described as follows.  The supermultiplet that contains $a_D$ and $A_D$ also
contains, along with the
fermions $\eta,\psi,\chi$, the familiar auxiliary field
$D$.  In the
presence of the hypermultiplet, the equation of motion for $D$ (keeping
only the bosonic terms) says
that $D$ is equal to the hyper-Kahler moment map of the
hypermultiplets; we write
this as $D=(\bar M M)_+$.
The supersymmetry transformation law of $\chi$ is
$\delta\chi=i( F_+-D)$, or, with
 $D$
integrated out, $\delta\chi=i(F_+-(\bar MM)_+)$.  So  for a bosonic
configuration
to be supersymmetric, it must satisfy
\eqn\ikko{F_+=(\bar MM)_+.}
This is one of the SW equations.  The other SW equation, which is the Dirac
equation
\eqn\nikko{\Dsl M=0,}
arises because the hypermultiplet contains a fermi field $\zeta$ whose
$\bar{\CQ}$-variation
is $\delta\zeta =\Dsl M$.
So supersymmetric configurations are simply solutions of the SW equations.
Let ${\cal M}_\lambda$ be the moduli
space of SW solutions of given $\lambda$.

The dimension of ${\cal M}_\lambda$ is (according to an index theorem)
\eqn\kippo{d_\lambda=-{2\chi+3\sigma\over 4}+\lambda^2.}
In the special case that $b_1=0$, $b_2^+=1$, this amounts to
\eqn\okippo{d_\lambda=-2-{\sigma\over 4} +\lambda^2.}

The basic topological observable in the $u=1$ theory is simply the zero-form
operator ${\cal O}^{(0)}=a_D$, which has $R$ charge two  and so
is associated topologically with a two-dimensional class on
${\cal M}_\lambda$.
It has a one-form descendant, which is simply $\psi$,
and a two-form descendant, which is simply $F$.  The  correlation functions of
the
two-form descendant hence measure simply the first Chern class $\lambda$, and
the one-form descendant of course does not enter if $b_1(X)=0$.  In that case,
therefore, the only  relevant quantum topological observable is $a_D$ itself.
If $d_\lambda=2n$, the SW invariant is defined as
\foot{Mathematically, there is a tautological $U(1)$
bundle over $\CM_\lambda$ defined by dividing
the space of solutions to the SW equations using
only  based gauge transformations. $a_D$ is
then, up to a normalization, the first Chern
class of this $U(1)$ bundle.}
\eqn\ollo{SW(\lambda)=\langle a_D^n\rangle_\lambda=\int_{{\cal
M}_\lambda}a_D^n.}
(In all known cases with $b_2^+>1$,
$SW(\lambda)$ is non-zero only for  $\lambda$ such
that $n=0$.  But all values of $n$
contribute for $b_2^+=1$.  This is a
consequence
of the fact that, for sufficiently negative $\sigma$, wall crossing can occur
for $\lambda$ with arbitrarily large $n$.)

When we speak of the SW contribution to Donaldson invariants from $u=1$, we
mean by definition
the contribution of SW solutions with $M\not=0$.  Solutions with $M=0$ can be
``confused''
with the continuous contribution of the $u$-plane.  The restriction to
solutions with
$M\not= 0$ is, except for  $b_2^+\leq 1$, a mild one in the following sense.
A solution with $M=0$ is an abelian instanton, and is possible only if
$\lambda_+=0$.
In this case $2\lambda$ is an integral and anti-self-dual cohomology class.
For $b_2^+>0$,
there is no such class for a generic metric,\foot{Except $\lambda=0$.
On the $u$-plane, the computation in section 4 shows that,
at least in the simply-connected case,
there is never wall-crossing associated with $\lambda=0$;
the wall-crossing came from a term $(\lambda,\omega)/y^{1/2}$ that
vanishes for $\lambda=0$.  In SW theory, $\lambda=0$ never contributes
for a simply-connected four-manifold $X$
with $b_2^+=1$.  In fact, $\lambda=0$
is only possible if $X$ is spin, in which case for $b_2^+=1$,
a theorem of Donaldson shows that $b_2^-=1$ and the intersection form
is that of $\IP^1\times \IP^1$.  But for $b_1=0$, $b_2^+=b_2^-=1$,
and $\lambda=0$, the virtual dimension of the SW moduli space is negative,
so $SW(\lambda=0)=0$.}
 so the SW contributions with $M\not=0$ and
the continuous contribution of the $u$-plane are well-separated.  However, for
$b_2^+=1$, the
condition that a given $\lambda$ obeys $\lambda_+=0$ puts only one condition
on the metric.
Hence in a generic one-parameter family of metrics, one may ``cross'' a metric
for which
$\lambda_+=0$.
It is known that $SW(\lambda)$ jumps by $\pm 1$ in crossing such a wall.
Such wall-crossing does not occur for $b_2^+>1$, because in that
case $\lambda_+=0$ puts
more than
one condition on the metric, so that in a generic one-parameter family of
metrics, the SW
and $u$-plane contributions never meet.

The intuitive picture of SW wall crossing that we want to justify is that in
crossing
a wall, some SW solutions move to (or from) $M=0$, and their contributions
to the Donaldson invariants disappear from the $u=1$ vacuum
and move onto the $u$ plane.
Thus, the jumping of the SW invariant in crossing a wall, if suitably
measured, will
just cancel the $u=1$ part of the wall-crossing formula of the $u$-plane
integral.
A full justification of this will occupy the rest of this section, but let us
first
make the following simple remarks.
For $d_\lambda<0$, ${\cal M}_\lambda$ is generically empty and the SW
invariant vanishes.
For $M$ to go to zero in an SW solution requires, as we saw above,
$\lambda_+=0$.  So $4\lambda^2$ is a negative integer with
\eqn\otippo{4\lambda^2 - (8 + \sigma) \geq 0.}
These then are the standard conditions for SW wall-crossing.  They also
are exactly the conditions \mwcond\
we found in section 4 for a contribution
to $u$-plane wall-crossing at $u=\pm 1$.  This is a first indication
that SW wall-crossing and the $u=\pm 1$ wall-crossing of the $u$-plane
integrals can be matched up.

\bigskip\noindent{\it Effective Couplings Of The $u=1$ Theory}

The Lagrangian of the twisted theory near $u=1$
has the form
\eqn\unnu{L=\{\bar{\CQ},W\} +\int_X\left(c(u)F\wedge F +p(u)\Tr R\wedge R
+\ell(u)\Tr
R\wedge \tilde R + \cdots \right).}
In fact, the  most general $\bar{\CQ}$-invariant Lagrangian that can be
constructed
with the multiplets in question takes this form,
with some $W$ and some holomorphic functions $c,p,$ and $\ell$.
The terms involving $c,p,$ and $\ell$ are close cousins of terms
studied in section 2 on the $u$-plane.  The $c(u)F\wedge F$ term is the
fourth descendant of a zero-form observable
${\cal O}^{(0)}=\tilde \CF_M(u)$ which is quite analogous to the prepotential
${\cal F}$ of equation \manfsttop.
(To be more precise, this descendant contains additional terms involving $\psi$
which do not
contribute for $b_1=0$ and are omitted in \unnu.)
 In fact, one can think of $\tilde \CF_M(u)$
as the prepotential of the effective theory of dual photons and monopoles
near $u=1$.
In the other terms,
$\Tr R\wedge R$ and $\Tr R\wedge \tilde R$ are differential forms whose
integrals
are multiples of the signature $\sigma$
and Euler characteristic $\chi$ of $X$.
Upon exponentiation in the path integral, these interactions give
factors of the general form $A^\chi B^\sigma$ familiar from section 2.

The three couplings indicated explicitly in \unnu\ give rise in the path
integral for a given ${\rm Spin}^c$ structure $\lambda$ to factors
\eqn\homigo{  C(u)^{\lambda^2/2}P(u)^{\sigma/8} L(u)^{\chi/4} ,}
where the functions $C,P,$ and $L$ are essentially exponentials of $c,p,$ and
$\ell$.
We have seen such factors before; the first corresponds to a factor in the
lattice
theta function on the $u$-plane, and the other two are clearly analogous to the
functions in equations \grvcpl\ and \dono.

One might ask why the functions $c,p$, and $\ell$, or $C,P$, and $L$, do
not
precisely equal the analogous functions in the $u$-plane calculation.
The answer is that in the $u$-plane analysis, one deals with an effective
action in which the monopole hypermultiplet has been integrated out, while
in the present discussion we are using an effective Lagrangian that includes
the monopole fields.  In particular, the functions $C,P$, and  $L$ should be
completely non-singular (and non-zero) at $u=1$, as the singularities
in the $u$-plane description come from integrating out the massless
hypermultiplet.
It is quite conceivable that there is a simple recipe to relate the couplings
with the hypermultiplets  present to couplings with hypermultiplets
integrated out,
but in the present paper we will simply determine the functions $C,P$, and  $L$
by comparison to the wall-crossing formula.

An analogous issue arises in the mapping from the microscopic observable
$\exp(p\CO+I(S))$ to an effective interaction in the theory at $u=1$.
The general considerations are as in the discussion on the $u$-plane.
The microscopic operator $\exp(p\CO)$ simply maps to $\exp(2pu)$ in the
$u$-plane description.
The operator $I(S)$ maps to
$  \int_S\left({i \over  4 \pi} {du\over da_D}(F_-+D_+) - {i \over
32\sqrt{2}\pi} {d^2u\over da_D^2}
\psi\wedge\psi\right).$  This computation is in fact precisely as it was on the
$u$-plane, except now using the vector multiplet that contains $a_D, A_D$.
Because of the existence of a fermi field $\chi$ with $\delta\chi=i(F_+-D_+)$
and
the fact that the $u=1$ computation is done by localization on supersymmetric
fields,
we actually can replace $D_+$ here by $F_+$, so that $I(S)$ maps to the
familiar topological field theory expression
$  \int_S\left({i \over  4 \pi} {du\over da_D}F- {i \over 32\sqrt{2}\pi}
{d^2u\over da_D^2}
\psi\wedge\psi\right)$;
also, the terms involving $\psi$ can be dropped if $b_1(X)=0$.  In reducing
$\exp(I(S))$ to the $u=1$ description, there is a contact term, governed
by the same general logic that applied on the $u$-plane.
So we get (assuming $b_1(X)=0$)
\eqn\torro{\exp(p\CO +I(S))\to \exp\left(
2pu + {i\over  4\pi} \int_S{du\over da_D}F
+S^2T^*(u)\right).}
{\it A priori} the function $T^*(u)$ might differ from the analogous function
$T(u)$ on the $u$-plane, since $T(u)$ is obtained after integrating out the
hypermultiplet.  However, we will see that $T^*=T$ -- this particular coupling
is unchanged by integrating out the monopoles.  (Therefore $T$ has
no singularity at $u=1$ or $-1$, an assumption that was made in determining
$T$.)

\bigskip\noindent{\it Form Of The Path Integral}

We want to analyze the contribution from the monopole vacuum at $u=1$
to the path integral that defines the generating function $\langle
e^{p\CO+I(S)}
\rangle_\xi$ (where as in the introduction, we sum over all classes of
$SO(3)$ bundles $E$ of fixed $\xi=w_2(E)$).
The path integral in this vacuum  includes a sum over ${\rm Spin}^c$
structures
$\lambda$, that is over the ``line bundle'' of the dual photon $A_D$.
We will write $\langle e^{p\CO+I(S)}\rangle_{\xi,u=1}$ for the contribution
of the $u=1$ vacuum to the generating function, and $\langle e^{p\CO+I(S)}
\rangle_{\xi,
1,\lambda}$ for the contribution in the $u=1$ vacuum
from a particular ${\rm Spin}^c$ structure
$\lambda$.  Thus
\eqn\momo{\langle e^{p\CO+I(S)}
\rangle_{\xi,u=1}=\sum_\lambda\langle e^{p\CO+I(S)}\rangle
_{\xi,1,\lambda}.}
Our goal here is to work out a formula for $\langle e^{p\CO+I(S)}
\rangle_{\xi,1,\lambda}$
in terms of the invariant $SW(\lambda)$.

The requisite path integral has several factors.  We must use the formula
\torro\ by which $e^{p\CO+I(S)}$ is translated to an effective operator in the
$u=1$ vacuum, the factors in \homigo, and two additional factors. The first  is
a factor of 2, because in defining the Donaldson invariants (and in
particular $\langle e^{p\CO+I(S)}\rangle_\xi$), it is not customary to divide
by the order of the center of $SU(2)$.  Also,
upon making the duality transformation to the natural description
at $u=1$, one gets an analog of the phase factor \joxoz.
This phase factor is
$e^{2i\pi(\lambda_0\cdot\lambda+\lambda_0^2)}$, where $\lambda_0$ is
the fixed element  of $\half w_2(E)+\Gamma$ that entered in \joxoz.
(A path integral explanation of the $\lambda$ dependence of this
factor is in \abelS.)

After multiplying these factors together, we get a function of $a_D$ that
must be integrated over ${\cal M}_\lambda$ to get the contribution of the
${\rm Spin}^c$ structure $\lambda$ to the Donaldson invariants.
So we have
\eqn\urko{\eqalign{
\langle e^{p\CO+I(S)}\rangle_{\xi, 1,\lambda}= & \int_{{\cal M}_\lambda}
2e^{2i\pi(\lambda_0\cdot\lambda+\lambda_0^2)}\exp\left(
2pu + {i\over  4\pi} \int_S{du\over da_D}F
+S^2T^*(u)\right)\cr & \cdot
 C(u)^{\lambda^2/2}P(u)^{\sigma/8} L(u)^{\chi/4} .\cr}}
What it means to integrate over ${\cal M}_\lambda$ is that one must expand
in powers of $a_D$ -- which represents a two-dimensional cohomology class on
${\cal M}_\lambda$ -- near $a_D=0$, and extract the coefficient of
$a_D^{n}$, where $n=d_\lambda/2=-(2\chi+3\sigma)/8+\lambda^2/2$.
Then we integrate over ${\cal M}_\lambda$, using
$\int_{{\cal M}_\lambda}
a_D^{n}=SW(\lambda)$.
We can write this as  a residue,
\eqn\purko{\eqalign{
\langle e^{p\CO+I(S)}\rangle_{\xi,1,\lambda}&=SW(\lambda)\cdot
{\rm Res}_{a_D=0}
{da_D\over a_D^{1-(2\chi+3\sigma)/8+\lambda^2/2}}\cdot
2e^{2i\pi(\lambda_0\cdot\lambda+\lambda_0^2)}\cr &\exp\left(
2pu + {i\over  4\pi} \int_S{du\over da_D}F+S^2T^*(u)\right) \cdot
 C(u)^{\lambda^2/2}P(u)^{\sigma/8} L(u)^{\chi/4} .\cr}}
Note that we have not been careful to normalize the operator $a_D$
topologically;
it is not necessary to do so as a change in the normalization could be absorbed
in a rescaling of the yet-undetermined functions $C,P$, and  $L$.

\bigskip\noindent{\it Determination Of $C,P$, and  $L$}

To determine those factors, we will compare to the wall crossing formulas in
the
special case $b_2^+=1$,  that is $\chi+\sigma=4$.
Assuming that $\lambda$ is such that $n\geq 0$ (otherwise  \purko\ vanishes),
SW wall crossing occurs at walls at which $\lambda_+=0$.  It is known
topologically that at such a wall, $SW(\lambda)$ changes by $\pm 1$ (when
$b_1=0$).  The sign
depends on the direction in which one crosses the wall, and it will not
be necessary to keep track of it in order to determine the unknown functions.

The change in \purko\ is hence
\eqn\opurko{\eqalign{\Delta\langle
e^{p\CO+I(S)}&\rangle_{\xi, 1,\lambda}=\pm
{\rm Res}_{a_D=0}
{da_D\over a_D^{-\sigma/8+\lambda^2/2}}\cdot
2e^{2i\pi(\lambda_0\cdot\lambda+\lambda_0^2)}
\cr & \exp\left(2pu + {i\over  4\pi} \int_S{du\over
da_D}F+S^2T^*(u)\right) \cdot
 C(u)^{\lambda^2/2}P(u)^{\sigma/8} L(u)^{1-\sigma/4}.\cr}}

On the other hand, the
contribution of $u=1$ to the wall-crossing formula
for the $u$-plane integral was given in equation
\mwcrssing.
In order to compare these formulae note first that
$da_D/du = -h_M(\tau_D)={i \over  2} \vartheta_3\vartheta_4(\tau_D)$.
\foot{There is a subtle minus sign here related to the
fact that $h$ is a form of weight 1.}  Similarly
we must have $T^*(u)=T(u)$, since the
$q_D$ expansion of $T(u)$ around the monopole
cusp $\tau=0$ is exactly $T_M(q_D)$ defined in \yooemm.
\foot{If we follow our notation to its logical conclusion
then we should speak of a $q_M$ expansion and
$a_M$, etc. However, we do not do this since
the notation $a_D$ is standard.}
In a similar way   we find that the unknown functions
$C,P,$ and $L$ can be uniquely determined in order for the formulas to
agree:
\eqn\polo{\eqalign{
C & = {a_D \over  q_D}       \cr
 P & = - {4 \vartheta_2(\tau_D)^8 \over    h_M^4}  a_D^{-1} \cr
 L & = -{2i  \over h_M^2} \cr}}
where we have used the identity:
\eqn\dervadee{
q_D{da_D \over  d q_D} = {1 \over  16i} {\vartheta_2^8 \over \vartheta_3
\vartheta_4} .
}
In particular, the prepotential of the theory is obtained
from
\eqn\smoothprep{
{d^2 \over  d a_D^2} \tilde \CF_M(a_D) = \tau_D -{1 \over  2\pi i} \log a_D.
}

Substituting  \polo\ in \purko\ we  get a  formula for the contribution of a
${\rm Spin}^c$ structure
$\lambda$ at $u=1$ to the Donaldson invariants:
\eqn\lilpo{
\eqalign{
\langle e^{p\CO+I(S)}\rangle_{\xi,1,\lambda}
&
= {SW(\lambda) \over 16}
\cdot e^{2i \pi (\lambda_0\cdot\lambda+\lambda_0^2) }
 \cr
\cdot {\rm Res}_{q_D=0} \Biggl[
{dq_D\over q_D}
q_D^{-\lambda^2/2}
&
 {\vartheta_2^{8+\sigma} \over  a_D h_M }
\biggl(2i {a_D \over  h_M^2 }\biggr)^{(\chi+\sigma)/4} \qquad\qquad\cr
&
\exp\bigl[ 2 p u_M -
i(\lambda,S)/h_M + S^2 T_M \bigr] \Biggr] \cr}
}
In order to complete the result we must say how to
expand $a_D$ in $q_D$. This is a simple exercise
in elliptic functions. We know from
 \refs{\swi,\swii}  that $a_D(u)$ is given by
$a_D(u) =  ({1-u \over 2i }) {}_2F_1(\half,\half;2;{1-u\over 2})$.
This may be expressed in terms of complete
elliptic integrals which themselves may be expressed in
terms of modular functions. The result is that:
\eqn\ayedee{
a_D = -{i \over  6} \biggl({2 {E_2}-\vartheta_3^4-\vartheta_4^4 \over
\vartheta_3\vartheta_4}
\biggr)
}
as a function of $q_D$.

The analogous contribution from $u=-1$ follows almost immediately as the theory
when formulated on flat $\IR^4$
has a symmetry under $u\leftrightarrow -u$.
One need only replace $u$ by $-u$, and $a_D$ by $a+a_D$
(which is the appropriate
local parameter near $u=-1$), to get the formula analogous to \lilpo\ for
the contribution of the vacuum at $u=-1$
to the Donaldson invariants.
As in eqn. (2.66) of \kahler,  this replacement,
apart from $u\to -u$, introduces a multiplicative
factor of $i^{(\chi+\sigma)/4-\xi^2}$ that reflects the fact that the
transformation $u\to -u$, though a symmetry on flat $\IR^4$, has
an anomaly on a four-manifold.

As a simple but important example of the above procedure, one
easily recovers equation  (2.17) of \monopole\
in the simple type case when $d_\lambda=0$.
In this case we need only take the
leading terms in the $q$-expansions in
\lilpo. Using $u_M=1+\cdots$,
$T_M=1/2 + \cdots $,
$h_M = 1/(2i) + \cdots $ and
$a_D = 16i q_D + \cdots$
we find that \lilpo\ reduces to
\eqn\swwcii{
2^{1+{7\chi\over 4}+{11\sigma\over 4}} e^{2p + S^2/2} e^{2(S,\lambda)}
e^{2i \pi (\lambda_0\cdot\lambda+\lambda_0^2)  }
}
Identifying
$2\lambda $ with the variable called  $x$ in \monopole, equation (2.17), we get
perfect agreement.\foot{We recall from the footnote just before eqn.
\lateth\ that, to agree with the mathematical literature, we have
modified the orientation convention of \monopole\ by
a factor $(-1)^{\half(w_2(E)^2+w_2(E)\cdot w_2(X))}$.}

More generally, the above formula for $\langle e^{p{\cal O}+I(S)}\rangle_{\xi,
1,\lambda}$ implies that four-manifolds of $b_1=0$, $b_2^+>1$ (for which the
$u$-plane integral vanishes, so everything comes from the
SW contributions at $u=\pm 1$) are of simple type in the generalized
sense that $({\p^2\over\p p^2}-4)^r \langle e^{p{\cal O}+I(S)}\rangle_{\xi,
1,\lambda}=0$ for some $r$.
One simply takes $r$ to exceed
the maximum value of $n=d_\lambda/2$ that arises for any $\lambda $ with
$SW(\lambda)\not=0$.  (There is such a maximum, since it is known
that $SW(\lambda)=0$ for all but finitely many $\lambda$.)
The formula \lilpo\ entails an expansion in powers of $(u-1)$ (or
$a_D$) in which one only ``sees'' terms of order at most $(u-1)^n$, hence
\eqn\genspltp{
({\p^2\over\p p^2}-4)^r Z_D = 0
}
if $r$ exceeds the maximum possible value of $n$.
Hence the manifold is of generalized simple type
in the sense formulated by Kronheimer and Mrowka.

\newsec{Evaluation in certain chambers}

As explained in the introduction, the wall-crossing,
blowup, and vanishing properties of
$Z_u$ completely determine its
value for all $X$ of $b_1=0, b_2^+=1$. However,
as explained in the introduction,
it is desirable to have a more effective formula.

We derive such a formula here in the case
$b_->1$ and $w_2(E)=0$. We follow the general
calculation of Borcherds
\borcherds\ (which in turn is an
improvement on calculations done in the
literature on quantum string corrections,
such as \refs{\dkl,\hm}).

We let $I^{p,q}$ be the lattice
with quadratic form $\sum_{i=1}^px_i^2-\sum_{j=1}^qy_j^2$.
The lattice $H^2(X,\IZ)$ is isomorphic to $I^{1,b_2-1}$.  If
we let $II^{1,1}$ be the even unimodular rank two
lattice (often called $H$) with intersection
form $\left(\matrix{ 0 & 1 \cr 1 & 0 \cr}\right)$,
then $I^{1,b_2-1}$ is isomorphic to $II^{1,1}\oplus I^{0,b_2-2}$.
Our computation will depend on a choice of such a decomposition
(and for each such decomposition, we will compute the $u$-plane
integral in a certain chamber).  Fixing such a decomposition,
we choose a basis of primitive null vectors $z,z'$, with $(z,z')=1$,
which generate the summand $II^{1,1}$ of $H^2(X,\IZ)$.  The second
summand $I^{0,b_2-1}$ will be called $K$.
$K$ can be identified  as
\eqn\nrj{K=z^\perp/\langle z\rangle,}
and with this definition depends only on the choice of a primitive
null vector $z$ and not on $z'$.

An example of the situation considered here is that $X$ is a rational
ruled surface, and $z$ is the class of the fiber.  The situation we
are considering is thus very close to that of section 5, except
that $w_2(E)$ is zero.

\subsec{The Lattice $K$ and the Reduction Formula}

Our goal is to reduce the computation of the $u$-plane integrals
to one involving  theta functions for the lattice $K$.

We let $P_\pm$ be the orthogonal projections of $H^2(X,\IZ)$ to self-dual
and anti-selfdual parts, and
and let $\tilde P_\pm$ be the orthogonal projections
onto the orthogonal complement of $z_\pm=P_\pm(z) $ within the
self-dual and anti-self-dual subspaces of $H^2(X,\IZ)$.
In particular, $z_+ = (z,\omega) \omega$, and $\tilde P_+=0$
since the self-dual subspace of $H^2(X,\IZ)$ is one-dimensional,
generated by $z_+$.  We have
\eqn\prj{
\eqalign{
\lambda_+ & ={ (\lambda_+, z_+)\over  z_+^2}  z_+ = (\lambda,\omega) \omega \cr
\lambda_- & ={ (\lambda_-, z_-)\over  z_-^2}  z_-  + \tilde P_-(\lambda) \cr}
}
and the second line is an orthogonal sum in $\IR^{0,b_-}$:
$(z_-,\tilde P_-(\lambda)) =0$.

The reduction formula is now obtained by writing
\eqn\reducform{
\eqalign{
\lambda
& = \lambda^K + c z' + n z\cr}
}
with $c,n\in\IZ$ and $\lambda^K\in K$,
and then doing a Poisson summation formula on $n$.
Using the isomorphism
$H^2(X,{\bf Z}) \cong I^{1,b_-} \cong II^{1,1} \oplus I^{0, b_- - 1} $
we can take
\eqn\vctrchc{
\eqalign{
z& = (1,0;\vec 0)\cr
z'& = (0,1;\vec 0)\cr
w_2(X) & = (0,0; 1,\cdots , 1) \cr}
}
Moreover, since
$w_2(E)=0$ we have
$(\alpha_I, z) = (\beta_I, z) = 0 $ for all $I$ ($\alpha_I$ and $\beta_I$
were introduced in equation \thetcoset).

The net result of these manipulations is that the
integral $\CG(\rho)$ of \brght\   can be written as:
\eqn\cdint{
\eqalign{
{1 \over
\sqrt{2  z_+^2}} \int_{\CF }
{dx dy \over  y^{2}} \sum_I
\hat f_{I}
&
\sum_{c, d\in \IZ}
\exp\biggl[-{\pi \over  2 y z_+^2}\vert c \tau
+ d \vert^2-{\pi \over  y z_+^2} (\bar \xi_+^I, z_+)(\bar \xi_-^I,z_-) \biggr]
\qquad
\cr
\exp\Biggl[ - {\pi (\bar \xi_+^I, z_+) \over    z_+^2}
\bigl({ c   \tau + d \over  y}\bigr)
&- {\pi (\bar \xi_-^I, z_-) \over    z_+^2}
\bigl({ c   \bar \tau + d \over  y}\bigr)
+ 2 \pi i {(\alpha_+^I,z_+)\over  z_+^2} c - i \pi (\mu,\alpha^I) c \Biggr]
\cr
&
\bar \Theta_{(H^2+\beta^I)\cap z^\perp/z}
(\tau, \mu d + \alpha^I, - \mu c; \tilde P(\xi^I))
\cr}
}
where the vector $\xi$ is defined in \defofx.
Following Borcherds, we have introduced a vector
$\mu\in K$ as follows:
\eqn\dfnmu{
\mu \equiv -z' + {z_+ \over  2 z_+^2} + {z_- \over  2 z_-^2}
= - z' + {z_+ - z_- \over  2 z_+^2}.
}
This vector satisfies $(\mu,z)=0$ and thus descends to
 a vector in $K$. Note that this vector is
{\it metric dependent}.

We now apply the unfolding technique to
the integral \cdint.  One looks at the action of $SL(2,\IZ)$ on
$c$ and $d$.
The degenerate orbit with $c=d=0$ gets special treatment.
That contribution to the theta function is modular-invariant by itself,
and gives the integral over the fundamental domain of a holomorphic
form divided by a power of $y$.  Such an integral can be done in a standard
way by integrating by parts, picking up a contribution at infinity.
Orbits with $c$ and $d$ not both zero can be transformed by $SL(2,\IZ)$
to have $d=0$, giving an integral over a strip $0\leq x\leq 1$ in the
upper half plane (rather than a modular region), together with
a sum on $c$ from
$-\infty$ to $+\infty$, omitting zero.
The integral over the strip is  straightforward, if tedious.
We give some details of the derivation in appendix C.

In order to write the final result, we define:
\eqn\combexp{
S^K   \equiv S - (S,z)z'
}
and
\eqn\defpsieye{
\psi_I \equiv    { 1  \over  h_I}
\biggl[ (\lambda
+ \beta_I, S^K ) - [(\lambda
+ \beta_I, \mu)](S, z)\biggr]
}
where $[\cdot ]$ is the greatest integer function.
The final result is then
(recall the notation \cosetnot ):
\eqn\compare{
\eqalign{
Z_u =& -  4 \sqrt{2} \pi i   \Biggl\{
4  \Biggl[      {   f_\infty h_\infty \over  1- e^{- i(S,z)/h_\infty  } }
\tilde \Theta_{K,\infty}^{(\mu)}(S)  \Biggr]_{q^0} \cr &
+
\Biggl[      {   f_M h_M \over  1- e^{- i(S,z)/h_M  } } \tilde
\Theta_{K,M}^{(\mu)}(S)  \Biggr]_{q^0}
+
\Biggl[      {   f_D h_D \over  1- e^{- i(S,z)/h_D  } } \tilde
\Theta_{K,D}^{(\mu)}(S)  \Biggr]_{q^0}\Biggr\}
\cr}
}
where $\tilde \Theta$ is a theta-function-like object
\eqn\thetalike{
\eqalign{
\tilde \Theta_{K,I}^{(\mu)}(S)   =
\sum_{\lambda\in K}
\exp\Biggl\{
& - i \pi \tau (\lambda
+ \beta_I)^2 + 2 \pi i (\lambda + \beta_I , \alpha_I) -    i  (\lambda +
\beta_I, S^K )/h_I
 \Biggr\} \cdot \cr
&  \qquad \cdot \exp\bigl\{  i {(S, z) \over h_I} [(\lambda + \beta_I, \mu)]
\bigr\}
\cr}
}
and $f_I$ is the holomorphic version of the
function $\hat f_I$ introduced in \effhat.
(That is, we replace $\hat T \rightarrow T $ in
\effhat.)

There are several remarks which should be made
about this result.

\item{1.} In order to apply the unfolding method we must
regard the expressions as power series in $p, S$.
The decay from the $\Theta$ function
$\sim \exp\bigl[-{\pi \over  2 z_+^2}  {1 \over  y} \bigr]$
in \cdint\
must not be overwhelmed by the ``tachyon'' divergence
of the modular functions.
At order $p^\ell S^r$ the modular functions diverge like:
\eqn\moddivi{
\eqalign{
 \quad & \quad \exp\bigl[ -{2 \pi \over  y} (8 + \sigma)\bigr] \cr}
}
in the regions $(\infty,0), (\infty, 2)$ and like
\eqn\moddivii{
  \exp\bigl[{2 \pi \over  8 y} (r+2 \ell +3)  \bigr] }
in the regions $(\infty,1), (\infty, 3) , M, D $.
Therefore, for $b_-\leq 9$ we have the condition:
\eqn\ufldcon{
z_+^2 < {2 \over  r + 2 \ell + 3}  ,
}
and for $b_->9$ we have an extra condition:
\eqn\extrcnd{
z_+^2 < { 1 \over  4 (b_- - 9)}
}
for the validity of \compare.

\item{2.} There are two sources of chamber dependence
in the answer. First, the expression is only valid in an appropriate set of
chambers defined above. Second, the   factor $[(\lambda + \beta_I, \mu)]$
 prevents
 $\tilde \Theta_{K,I}^{(\mu)}(S)$ from being a lattice theta function.
Note that the vector $\mu$ depends continuously
on the metric. However, $\mu$ only enters through the
greatest integer function, so the expression is metric-independent within   a
chamber.

\item{3.} The result \compare\ bears some similarity
with Conjecture 4.12 of \gottzag.

\subsec{Example: $X=\IP^1 \times \IP^1$}

As a special case of the above formula we
write the invariants for $X=\IP^1 \times \IP^1$.
In this case $K=0$ and the expressions
simplify somewhat. Moreover, only the semiclassical
cusp contributes (this is related to the fact that
the SW invariants vanish, since $X$ admits a
metric of positive scalar curvature).

The expression \compare\ thus simplifies to:
\eqn\sttoag{
( - 8) \sqrt{2} \pi i \Biggl[ f h
\coth\biggl( i { (S,z)\over 2 h  } \biggr)
\Biggr]_{q^0}
}
This agrees with the expression found in
\gottzag.
\foot{Except for a factor of $4$. Since the
answers below would not be integral in the
stable range if divided by $4$ we suspect
that our normalization is the correct one.}

At first order in $S$ we have:
\eqn\stoops{
 Z_D = -   (S,z) \bigl[ 1 +{ 17 \over  2^4} {p^2\over  2!}
 + {71 \over  2^5} {p^4\over  4!}
+ {23505 \over  2^{12}} {p^6 \over  6!}  + \cdots \bigr]
}

Similarly, we can easily extract the
 first few Donaldson polynomials in $S$.
To do this, recall the relation between the
Donaldson polynomials and the generating function
of equations \degree\ and \genfun.
Using this relation, we have:
\eqn\dnplystoo{
\eqalign{
\CD_2 & = - (S,z)  \cr
\CD_{10} & = 5 S^4 (S,z) - {5 \over  2} S^2 (S,z)^3 + (S,z)^5\cr
\CD_{18} & = 252 S^6 (S,z)^3 - 216 S^4 (S,z)^5 + 108 S^2 (S,z)^7-40 (S,z)^9 \cr
\CD_{26}& = 102960 S^8 (S,z)^5   - 108108 S^6 (S,z)^7\cr
&
+ 63180 S^4 (S,z)^9 - 26949
S^2 (S,z)^{11} + 9345 (S,z)^{13}\cr}
}

\newsec{Donaldson invariants of the projective plane}

In this section we consider the Donaldson invariants
for $\IP^2$. Since the SW invariants vanish (as
$\IP^2$ admits a metric of positive scalar curvature),
the Donaldson invariants coincide with the
$u$-plane integral. $\IP^2$ is therefore, in a sense, as far as possible
from being of simple type.

For simplicity we focus on the
case $w_2(E)=0$.  Under this condition, the integral that must be
evaluated is:
\eqn\cptoo{    {(S,\omega) \over  32 \sqrt{2} \pi}
\int_{\gof \backslash\CH} {dx dy \over  y^{3/2}}
{\vartheta_4^{9} \over  (\half \vartheta_2 \vartheta_3)^4 }
 \exp\biggl\{ 2p u -{1 \over  24} S^2
\bigl[   {\hat E_2\over  h(\tau)^2}  - 8 u \bigr] \biggr\}
\bar \vartheta_4
}

If we try to do this integral using the unfolding
technique on, say, the lattice theta function
$\vartheta_4 \bar \vartheta_4$, we find that we are
always outside the range of validity of this method.
The reason is
that there is no parameter $z_+^2$ to vary; indeed,
it is effectively always equal to one,  hence
always outside the domain of validity
defined by \ufldcon. Therefore, we need another approach.
One approach is to write the integral as a
total divergence and pick up the constant term
at infinity. This can be done using a
nonholomorphic modular form of
weight $(3/2,0)$ introduced by Zagier.  Similar
integrals have been  done using this form by Borcherds
\borcherds.

\subsec{Zagier's form and related forms}

Zagier's nonholomorphic
modular form \refs{\zagi,\zagii}
of weight $(3/2,0)$  for $\Gamma_0(4)$ is given by
\eqn\zagfrm{
\eqalign{
{\bf G}(\tau,y) & = \sum_{n\geq 0} \CH(n)  q^n +
\sum_{f=-\infty}^\infty q^{-f^2} {1 \over  16 \pi y^{1/2}}
\int_1^\infty e^{- 4 \pi f^2 u y} {du \over  u^{3/2}}\cr
& = \sum_{n\geq 0}  \CH(n) q^n
+   y^{-1/2} \sum_{f=-\infty}^\infty q^{-f^2} \beta(4 \pi f^2 y)  \cr}
}
where
\eqn\defbet{
\beta(t)   = {1 \over  16 \pi} \int_1^\infty e^{-ut} {d u \over  u^{3/2}}
}
We define Fourier coefficients by:
\eqn\dffcffs{
{\bf G}(\tau,y) \equiv
  \sum_{n\in \IZ} c(n,y) e^{2 \pi i n x}
}
In \zagfrm, the  $\CH(n)$ are Hurwitz class numbers, which are closely
related to the class numbers of imaginary quadratic fields.  The
$\CH(n)$ for small $n$ are given by
\eqn\hurwitz{
\eqalign{
\CH(\tau) & = \sum_{n\geq 0} \CH(n) q^n \cr
& = -1/12 + q^3/3 + q^4/2 + q^7 + q^8 + q^{11}+
(4/3) q^{12} +\cr
& + 2 q^{15} + 3/2 q^{16} + q^{19} + 2 q^{20} + 3 q^{23} + 2 q^{24}+\cr
&+ (4/3) q^{27} + 2 q^{28} + 3 q^{31} + 3 q^{32} + 2 q^{35} \cr& +
 (5/2) q^{36} + 4
q^{39} + 2 q^{40} +   \cdots \cr}
}
The form ${\bf G}$ satisfies the equation:
\eqn\zagfrmi{
{ \p \over  \p \bar \tau} {\bf G} = {1 \over  32 \pi i}  { 1 \over  y^{3/2}}
\overline{ \vartheta_3(2 \tau)}
}

{}From Zagier's form we construct a two-dimensional
representation of the modular group:
\eqn\twodee{
\eqalign{
{\bf G_1}& = \sum_{n\in \IZ} c(4n, y/4) e^{2 \pi i nx} \cr
& = \sum_{n\geq 0}  \CH(4n) q^n
+   2y^{-1/2} \sum_{f=-\infty}^\infty q^{-f^2} \beta(4 \pi f^2 y)   \cr
{\bf G_2}& = \sum_{n\in \IZ} c(4n-1, y/4) e^{2 \pi i (n-1/4)x} \cr
&   = \sum_{n> 0}  \CH(4n-1) q^{n-1/4}
+   2y^{-1/2} \sum_{f=-\infty}^\infty q^{-(f+\half)^2} \beta(4 \pi (f+\half)^2
y)   \cr}
}
These form a weight $(3/2,0)$ representation of the
modular group with (\ez, Theorem 5.4, \borcherds):
\foot{This means the functions transform like the
third power of a $\vartheta$ function. We use the
principle branch of the logarithm, and define $\sqrt{z}$ to have its
argument in $(-\pi/2, \pi/2]$ for $z$ in the complex
plane cut along $(-\infty,0]$.}
\eqn\repmd{
\eqalign{
\pmatrix{ {\bf G_1}(\tau+1) \cr {\bf G_2}( \tau +1) \cr}
& = \pmatrix{ {\bf G_1}(\tau ) \cr -i {\bf G_2}( \tau  ) \cr} \cr
\pmatrix{ {\bf G_1}(-1/\tau) \cr {\bf G_2}(-1/\tau) \cr}
& = {1 + i \over  2} \tau^{3/2} \pmatrix{1& 1 \cr 1& -1 \cr}
\pmatrix{ {\bf G_1}( \tau) \cr {\bf G_2}( \tau) \cr} \cr
& =  -{1 \over  \sqrt{2}} (-i\tau)^{3/2} \pmatrix{1& 1 \cr 1& -1 \cr}
\pmatrix{ {\bf G_1}( \tau) \cr {\bf G_2}( \tau) \cr} \cr}
}

Note that
\eqn\rltzfs{
\eqalign{
{\bf G}(\tau) & = {\bf G_1}(4 \tau) + {\bf G_2}(4 \tau) \cr
 { \p \over  \p \bar \tau} {\bf G_1} & = {1 \over  16 \pi i}  { 1 \over
y^{3/2}} \overline{ \vartheta_3(2 \tau)} \cr
{ \p \over  \p \bar \tau} {\bf G_2} & = {1 \over  16 \pi i}  { 1 \over
y^{3/2}} \overline{ \vartheta_2(2 \tau)} \cr}
}

We would like to construct a form for $\Gamma^0(4)$
such that we can integrate by parts in \cptoo. Such a
form can be constructed by noting that
${\bf G}((\tau+1)/2)$ will be modular for $\Gamma(2)$.
Now $\Gamma(2) \cap \Gamma^0(4)$ is index
two in $\Gamma^0(4)$, a coset representative being
$\tau \rightarrow {\tau \over  \tau+1}$. Adding the
transform we obtain the desired form:
\eqn\desfrm{
\eqalign{
Q_{(\infty,0)} & \equiv {\bf G}({\tau+1 \over  2}) +
\half {\bf G_1}({\tau+1 \over  2}) \cr}}
Thus, $Q_{(\infty,0)}$ is modular for $\Gamma^0(4)$ and obeys
\eqn\zaagfrmi{
{ \p \over  \p \bar \tau} {Q_{(\infty,0)}} =
{1 \over  8\sqrt 2 \pi i}  { 1 \over  y^{3/2}}
\overline{ \vartheta_4(\tau)}
}

The sum over cosets will now involve the functions:
\eqn\csfncs{
\eqalign{
Q_{(\infty,0)} = Q_{(\infty,2)} & = {\bf G_1}(2 \tau) - {\bf G_2}(2 \tau) +
\half
{\bf G_1}({\tau+1 \over  2})\cr
Q_{(\infty,1)}= Q_{(\infty,3)} & = {\bf G_1}(2 \tau) + {\bf G_2}(2 \tau) +
\half
{\bf G_1}({\tau  \over  2})\cr
Q_{M}= Q_D & = \half \bigl[{\bf G_2}({\tau  \over  2})
+ \kappa {\bf G_2}({\tau+1 \over  2})\bigr] \cr}
}
where $\kappa=e^{2\pi i /8}$.

The integral in \cptoo\ now can be written:
\eqn\invtintg{
\CI = 8 \pi i \sqrt{2} (S,\omega)
\int_{\bar \Gamma\backslash \CH} dx dy
\sum_I \hat f_I {\p \over  \p \bar \tau} Q_I
}

Like Zagier's functions the $Q_I$  can be written as
\eqn\rpddec{
Q_I(\tau,y) = \CH_I(\tau) + {cnst. \over  y^{1/2}} + \CF_I(\tau,y)
}
where $\CF_I(\tau,y)$ has exponential decay
for $y \rightarrow \infty$ and only negative Fourier components in $x$
(which will prevent it from contributing in the computation below)
and $ \CH_I(\tau)$ is
expressed in terms of class numbers
analogous to \hurwitz.

We would like to integrate by parts and use the rule:
\eqn\ibpi{
\int_{\CF} dx dy {\p \over  \p \bar \tau}  F(x,y)
= + { i \over  2} \lim_{\Lambda\rightarrow\infty}
\int_{-\half}^{+\half} dx F(x,\Lambda)
}
However, we need to take into account the nonholomorphic
dependence in $\hat f$. Thus, we
need to generalize Zagier's form, that is, we need
(nonholomorphic)
modular forms ${\bf G}^{(\ell)}$ for $\Gamma_0(4)$
of weight
$({3 \over  2} + 2\ell, 0)$ such that
\eqn\zgfrmii{
{\p \over  \p \bar \tau} {\bf G}^{(\ell)}
={1 \over  32 \pi i } {1 \over  y^{3/2}}
 \overline{ \vartheta_3(2 \tau) } \hat E_2^\ell
}

This may be done with the aid of the following.

\bigskip
{\bf Lemma}. Suppose $g(\tau,\bar \tau)$ is
modular of weight $(3/2,0)$ for some congruence
subgroup $\Gamma'$. Suppose moreover
that
\eqn\lemmi{
{\p \over  \p \bar \tau} g(\tau,\bar \tau)= {h(\bar \tau) \over  y^{3/2}}
}
Then
\eqn\lemmii{
\eqalign{
\CE^{\ell}[g] & = \sum_{j=0}^\ell
 {\ell \choose j} {\Gamma(3/2) \over  \Gamma(3/2 + j)}
\bigl({6i \over  \pi}\bigr)^{j} E_2^{\ell-j}
\bigl( { d \over  d \tau} \bigr)^j g \cr}
}
is modular for $\Gamma'$ of weight $(2 \ell + {3 \over  2}, 0)$
and satisfies:
\eqn\lemmiii{
{\p \over  \p \bar \tau} \CE^{\ell}[g] = {h(\bar \tau) \over  y^{3/2}}
\hat E_2^{\ell}
}

{\it Proof}: It is straightforward to show \lemmiii\
by differentiating and using the fact that $h(\bar \tau)$
does not depend on $\tau$. It is not obvious that
the expression in \lemmii\ transforms well under
modular transformations.  However,  \lemmii\ may be
  expressed in terms of
Cohen's operator   \cohen, which is
essentially just a product of covariant
derivatives normal-ordered.
Cohen's operator is:
\eqn\cohop{
\CF_r[f] = \sum_{j=0}^r {r \choose j}
{\Gamma(k+r) \over  \Gamma(k + j)}
\bigl( {1 \over  2iy}\bigr)^{r-j} \bigl({d \over  d \tau}\bigr)^j f
}
If  $f$ is modular of weight $(k,0)$ then
$\CF_r[f] $ is modular of weight $(k+2r,0)$.
It is straightforward to check that
\eqn\zgfrmiii{
\CE^{\ell}[g] = \sum_{t=0}^\ell
{\ell \choose t} {\Gamma(3/2) \over  \Gamma(3/2 + \ell - t)}
\bigl({6i \over  \pi}\bigr)^{\ell-t} \hat E_2^t
\CF_{\ell - t} [g ]
}
and thus $\CE^{\ell}[g] $ is modular. $\spadesuit$

\subsec{Answer for $\IP^2$}

It is now straightforward to evaluate the
integral for $\IP^2$.
One expands the exponential $\exp[ -S^2 \hat E_2/(24 h^2) ]$
in powers of $S^2$ and expresses each term as a
total derivative using the lemma. Then integrating by
parts, and keeping the zeroth Fourier coefficient, one
obtains a double sum. This is easily rewritten so that
one can re-exponentiate
one infinite sum to get a factor
$\exp [-S^2  E_2/(24 h^2) ]$.
The result of all these manipulations is the formula:
\eqn\allord{
Z_D= -4 \sqrt{2} \pi  (S, \omega)
 \sum_{j=0}^\infty {\Gamma(3/2) \over  j! \Gamma(3/2 + j) }
\biggl({S^2 \over  2}\biggr)^j \Biggl[ \sum_I {f_I \over  h_I^{2j} }\bigl(q {d
\over  d q}\bigr)^j
  \CH_I(\tau)      \Biggr]_{q^0}  }

Examining the sum over cosets we find
 the entire contribution comes from
the semiclassical cusp.
Consequently, the $SU(2)$ Donaldson polynomials for $\IP^2$
are given by:
\eqn\dcptwo{
\eqalign{
\CD_{w_2(E)=0}(S,p ) & = \sum_{x,t} S^{2x+1}
p^t d_{2x +1, t} \cr
d_{2x +1, t}  = -2^{t-1}  (2x+1)!
& \sum_{j=0}^x {1 \over  (x-j)!}
\cr
{\Gamma(3/2) \over  j! \Gamma(3/2 + j) } \bigl(\half\bigr)^j
&
\cdot
\Biggl[ {\vartheta_4^9 \over  h^{4 + 2j} } T^{x-j} u^t \CH_{(\infty,0)}(q,j)
\Biggr]_{q^0}\cr
\CH_{(\infty,0)}(q,j) &  \equiv
\bigl(q {d \over  d q}\bigr)^j \CH_{(\infty,0)}(q)\cr
\CH_{(\infty,0)}  =
\sum_{n\geq 0}
& H(4n) q^{2n} \cr
+ \half \sum_{n\geq 0} H(4n) q^{n/2}(-1)^n  -
&  \sum_{n> 0} H(4n-1)
q^{2n-1/2} . \cr}
}

Expanding,  we
find that the first four polynomials are:
\eqn\ourdnld{
\eqalign{
\CD_{2} & = - {3 \over  2}  S \cr
\CD_{10} & = S^5 - p S^3 - {13 \over  8} p^2 S \cr
\CD_{18} & =  3 S^9 + {15 \over  4} S^7 p - {11 \over  16} S^5 p^2 -{141 \over
64} S^3 p^3 - {879 \over  256} S p^4 \cr
\CD_{26} & =  54 S^{13} + 24 S^{11} p + {159 \over  8} S^9 p^2\cr
&  + {51 \over  16}
S^7 p^3 - {459 \over  128} S^5 p^4 - {1515 \over 256} S^3 p^5 - {36675 \over
4096} S p^6 \cr}
}
in agreement with the results obtained
previously in \refs{\eg,\gottsche}.

\subsec{Class number relations}

In \gottsche, G\"ottsche gave a closed expression for the
Donaldson invariants for $\IP^2$. His expression differs
from the above, and comparing the two implies relations on
class numbers similar to the famous relations of
Kronecker, Weber, and Zagier. See, for instance,
\refs{\zagi\zagii}.

The answer obtained above in terms of class numbers is:
\eqn\allordii{
{\it Res}_{q=0}\left({dq\over q}\cdot(-\half)
 \sum_{j=0}^\infty {\Gamma(3/2) \over  j! \Gamma(3/2 + j) }
 {S^{2j+1} \over  2^j }  \Biggl[
e^{2 p u + S^2 T} {\vartheta_4^9\over  h^{4 + 2j} }\CH_{(\infty,0)}(q,j)
\Biggr]\right).  }

In \gottsche, G\"ottsche first
blows up $\IP^2$ to $\IF_1$ and (in
the notation
of section 5) considers
an $SO(3)$ bundle $\tilde E$ with
$w_2(\tilde E)=F$ in the chamber
$\omega\cdot B =0$, indeed, $\omega=H$.
This is on the wrong side of the K\"ahler cone
to apply the vanishing theorem. He then adds
all the wall crossings in the K\"ahler cone to
go to $\omega = F$ where the vanishing theorem
applies. The walls are $W_\lambda$ for: $\pm \lambda = (n_2+\half)B - n_1 H$
with $n_1>0, n_2\geq n_1$. The answer obtained for the generating
function of $\IP^2$ Donaldson invariants is
\eqn\allordiii{{\rm Res}_{q=0}\left({dq\over q}
\cdot(-\half)
 \sum_{j=0}^\infty {(-1)^j S^{2j+1} \over (2j+1)! }
   \Biggl[
e^{2 p u + S^2 T} {\vartheta_4^8 \over  h^4 }
V(q,j)  \Biggr]\right) }
where
\eqn\allordiv{
V(q,j)  = \sum_{n_1>0,n_2\geq n_1} (-1)^{n_1 + n_2} (2n_2+1)n_1^{2j+1}
{q^{\half(n_2(n_2+1)-n_1^2 )+ 1/8} \over  h^{2j+1} }
}

It is tempting to try to cancel the common factors
$e^{2 p u + S^2 T} {\vartheta_4^8\over  h^4 }$
in \allordii\ and \allordiii\
and then equate the power series in $S,q$. This
gives false formulae because the function
$e^{2 p u + S^2 T} {\vartheta_4^8\over  h^4 }$
is not sufficiently generic.
A safe way to proceed is to note that the function $e^{2pu}$, when
expanded in powers of $p$, generates arbitrary non-negative powers of $u$.
Hence, it is helpful to change variables from $q$ to $u$, and replace the
extraction of residues at $q=0$ with residues at $u=\infty$, using
the relation
\eqn\uu{{dq\over q}={du\over u}\cdot {u\over q}{dq\over du}.}
  Then, equality of residues at $u=\infty$ for all $p$ means that
  the functions multiplying $e^{2pu}$ in the two expressions have
  the same non-positive terms in their Laurent expansions near $u=\infty$.
Thus:
\eqn\clssidenty{
\eqalign{
  \Biggl[  \biggl({d q\over  du} {u\over  q} \biggr)
e^{  S^2 T} {\vartheta_4^8\over  h^{4  } }
&
\sum_{j=0}^\infty {\Gamma(3/2) \over  j! \Gamma(3/2 + j) }
 {S^{2j+1} \over   2^j h^{2j} } \vartheta_4  \CH_{(\infty,0)}(q,j)
\Biggr]_{\leq 0  }
= \cr
\Biggl[   \biggl({d q\over  du} {u\over  q} \biggr)
e^{  S^2 T} {\vartheta_4^8\over  h^4 }
&
\sum_{j=0}^\infty {(-1)^j S^{2j+1} \over (2j+1)! }
V(q,j)  \Biggr]_{\leq 0 } \cr}
}
We need care here, since
   $T(u)$ and other functions appearing here have series expansions
in inverse powers of $u$, but they are multiplying
expressions with series in positive powers of $u$,
so we cannot cancel these factors.

Nevertheless, \clssidenty\ does imply
some very interesting results on class numbers.
As an example of these relations we take the
term at first order in $S$.
Equation \clssidenty\ means that the functions on the left and right
differ by an entire function of $u$.  Taking only the linear term in
$S$, we get on the left a function that is constant at $u=\infty$
and on the right a function that vanishes there, so the functions
on the left and right actually differ by a constant.
Using this information together with the relation
\eqn\simprel{
{1 \over  u} q{du \over  dq} = -{1 \over  8 u}{\vartheta_4^8\over (\vartheta_2
\vartheta_3)^2}
}
we obtain an explicit formula for class numbers:
\eqn\nwclass{
\CH_{(\infty,0)} = \sum_{n_1>0,n_2\geq n_1} (-1)^{n_1 + n_2} (2n_2+1)n_1
{q^{\half(n_2(n_2+1)-n_1^2 )+ 1/8}
\over  \eta^3 } -{\vartheta_2^4 + \vartheta_3^4 \over  8 \vartheta_4}
}
We have checked this numerically to order
$q^{11/2}$. It would be very interesting to see
if this relation leads to improved estimates on the
asymptotic behavior of class numbers.

\newsec{Extension to $\pi_1(X) \not=0$ }

We indicate here how some of the above
results generalize to Donaldson theory on a
four-manifold $X$ that is not simply-connected and which in general
has $b_1\not= 0$.  For simplicity, we assume that there is no two-torsion
in $H^2(X,\IZ)$, so that $w_2(E)$ has an integral lift and some of
our formulas which involve dividing by two still make sense.

When $b_1(X) \not=0$, we
 can add new operators to the microscopic correlation
function, namely $\int_\gamma \Tr \phi \psi = \int_\gamma Ku $ for
a one-cycle $\gamma$ and
$\int_L K^3 u$   for
a three-cycle $L$. To compute correlation functions of
 such operators, one will have to determine several
new contact term corrections. For simplicity we
omit these operators.

On general grounds the photon partition function always
involves a sum over all line bundles.
Line bundles are classified topologically by
$c_1 \in H^2(X;\IZ)$. Moreover, on each class we
must integrate over the moduli space of harmonic connections.
This moduli space is a torus of dimension $b_1(X)$.

Proceeding as in section three, we reduce the
  evaluation
of the $u$-plane contribution to
the path integral to the finite-dimensional integral:
\eqn\coulbrnch{
\eqalign{
Z_u =
2 \int [da \,d\bar a \,d \eta \,d \chi]
  \int_{Pic(X)}  d\psi
\int dD
&
A^\chi B^\sigma y^{-1/2}  \cr
\exp\Biggl[ {1 \over  8 \pi}
\int ({\rm Im }\tau) D \wedge *D \biggr]
\exp\biggl[ - i \pi
\bar \tau \lambda_+^2 - i \pi \tau \lambda_-^2
&
+   \pi i (\lambda,   w_2(X) ) \biggr]
\cr
\exp\Biggl[
 - {i \sqrt{2}   \over  16 \pi } \int {d \bar \tau \over  d \bar a} \eta
\chi\wedge
(D_+ + 4\pi\lambda_+ )
 + {i \sqrt{2}   \over  2^7 \pi } \int {d \tau \over  da}
(\psi\wedge \psi) \wedge
&
( 4\pi\lambda_-  +
  D_+)    \cr
 + {1 \over  3 \cdot 2^{11} \pi i }  \int    {d^2 \tau \over  da^2} \psi\wedge
\psi
\wedge \psi\wedge \psi
 + 2 p u
+ {i \over  \sqrt{2} \pi}
\int_S K^2 u +
&
S^2 T(u) \Biggr]\cr}
}
Here $\int_{Pic(X)}$ is the sum/integral over all pairs consisting
of a complex line bundle on $X$ together with a harmonic connection.
These bundles
have first Chern class $2 \lambda \in H^2(X;\IZ)$ which is congruent
to $w_2(E)$ modulo two.
In other words, $\int_{Pic(X)}$ is a sum over $H^2(X,\IZ)$ times an
integral over a $b_1(X)$-dimensional torus. The $\psi$ zero modes
are tangent to $Pic(X)$, so the integration over them is naturally understood
as the integral of a differential form on $Pic(X)$.  As for the $\eta$ and
$\chi$ zero modes, they are normalized as in section 2.3 to represent
fixed cohomology classes (which means in practice that the zero mode
wave function for $\eta$ is $\eta_0=1$, and for $\chi$ we use a basis
of orthonormal self-dual harmonic two-forms).
   The integral over $D$ is
over the space of self-dual two-forms.  The
``integral'' over $D$ is really a Gaussian integral with
 an exponent of the wrong sign and
is to be interpreted algebraically.  (Also, the $D$ determinant
should be ignored; this is part of the supersymmetric cancellation
of all the bose and fermi determinants.)

The integrals in \coulbrnch\
are    most readily done by first
doing the formal Gaussian integral on the auxiliary field $D$
and then soaking up the fermion zero modes.
The resulting expressions may be
seen to be modular invariant if we treat
 $\psi$ as weight $(1,0)$. Indeed, making the
simple redefinition
\eqn\shftess{
\tilde S \equiv S
 - { \sqrt{2} \over  32} {d \tau \over  du} \psi\wedge \psi
}
we get  an expression with the same $F,D,\chi,\eta$
dependence as in the simply connected case. Therefore
we can say without further ado that the
$u$-plane integral is:
\eqn\newcpione{
Z_u = \int_{\gof \backslash \CH}
{dx dy \over  y^{1/2}}
\mu(\tau) \int_{Pic(X)} d \psi e^{2 p u + \tilde S^2 \hat T(u)+
(\tilde S, \psi^2) H(u) + \psi^4 K(u) } \Psi(\tilde S)
}
where
\eqn\nwcnvpione{
\eqalign{
\mu(\tau) & = - {\sqrt{2} \over  2} {da \over  d \tau} A^\chi B^\sigma\cr
\Psi(\tilde S) & = \exp(2i\pi \lambda_0^2)
\exp\bigl[  - { 1 \over  8 \pi y}({d   u \over  d  a})^2
\tilde S_-^2 \bigr]\cr &
\sum_{\lambda\in H^2+ \half w_2(E) }
\exp\biggl[ - i \pi
\bar\tau (\lambda_+)^2 - i \pi   \tau(\lambda_-  )^2
+ \pi i (\lambda -\lambda_0 ,  w_2(X)) \biggr]  \cr
&
\exp\bigl[- i   { d    u \over  d   a} (\tilde S_-,\lambda_-) \bigr]  \biggl[
(\lambda_+ ,\omega) +{i \over  4 \pi y} {d u\over  d a} (\tilde S_+,\omega )
   \biggr] \cr
H(u) & = {i \sqrt{2} \over  64 \pi} \biggl( {d^2 u \over  da^2} - 4\pi i
{d \tau \over  du} T(u) \biggr) \cr
K(u) & = - {i \over  3\cdot 2^{11} \pi} \biggl({d^2 \tau \over  da^2} - 6 { d
\tau \over  du} {d^2 u \over  da^2} + 12\pi i ({d \tau \over  du})^2 T(u)
\biggr) \cr}
}
It is easy to check that, although $T(u)$ transforms by a shift under
modular transformations,
$H(u)$ and $K(u)$ transform covariantly with weights $(-2,0),(-4,0)$
respectively.
Indeed, using
\eqn\aeetwo{
a  = {1 \over  6}
\biggl(  {2 E_2+ \vartheta_2^4 + \vartheta_3^4   \over  \vartheta_2 \vartheta_3
 }  \biggr)
}
one can derive the explicit $q$-expansions:
\eqn\achkay{
\eqalign{
H(u) & = { i \sqrt{2} \over  16\pi} {\vartheta_2^4 + \vartheta_3^4\over
\vartheta_4^8} \cr
K(u) & = -{7 \over  3\cdot 2^7 \pi^2} {(\vartheta_2 \vartheta_3)^2(
\vartheta_2^4 + \vartheta_3^4)\over  \vartheta_4^{16}} \cr}
}
This allows us to  extend immediately all the above
results on wall-crossing, blowup, explicit evaluations, and the like, to the
case $b_1(X)>0$, since the integral
has the same form.
In fact, the above formulae can be further simplified, since for
any four elements $\psi_1, \dots, \psi_4$ of $H^1(X;\IR)$ we
have $\psi_1 \wedge \cdots \wedge \psi_4 = 0$ when
$b_{2}^+<3$. Hence we may drop the $\psi^4$ terms.
(We have given the formulae for $K$ since they are likely
to be useful in related contexts.)

The discussion of section 7 can also be extended
to the nonsimply-connected case. Define
\eqn\monesstilde{
\tilde S_M \equiv S
- { \sqrt{2} \over  32} {d \tau_D \over  du} \psi\wedge \psi
}
Then, using \smoothprep\ one finds the generalization
of \urko\  to be
\eqn\newurko{\eqalign{
\langle e^{p\CO+I(S)}\rangle_{\xi,u=1}=
\int_{Pic(X)} d \psi \int_{{\cal M}_\lambda}
&
2e^{2i\pi(\lambda_0\cdot\lambda+\lambda_0^2)}
C(u)^{\lambda^2/2}P(u)^{\sigma/8} L(u)^{\chi/4} \cr
\exp\left( H_M (\tilde S_M, \psi^2)
+  {\sqrt{2} \over  2^6 \pi^2} {1 \over  a_D} (\psi^2,\lambda)  \right)
& \exp\left(
2pu + i  {du\over da_D}(\tilde S_M, \lambda)
+\tilde S_M^2 T_M(u)\right)\cr
}}
where  the line bundles in $Pic(X) $ now have
$2\lambda\in H^2(X;\IZ)$ congruent to $w_2(X)$
modulo two,
and we have dropped two  $\psi^4$ terms since $b_{2}^+=1$.

We can now turn the reasoning of section seven around
and use \newcpione\ and  \newurko\  to give a
new derivation of the generalized  wall-crossing formulae
for Seiberg-Witten invariants given in  \refs{\lilu,\okonek}.

\newsec{Incorporation Of Matter}

${\cal N}=2$ supersymmetric gauge theories in four dimensions can
be generalized to include hypermultiplets in some representation of the
gauge group. Insuring that the beta
function should be zero or negative places a restriction on the possible
representations.  For the case
that the gauge group is $SU(2)$, the possibilities are that the
hypermultiplets consist of $2N_f$ copies of the two-dimensional representation
of $SU(2)$, for $N_f\leq 4$,\foot{The number of copies of the
two-dimensional representation must be even; otherwise the quantum theory
is inconsistent because of a global anomaly.  A single copy of the
two-dimensional representation gives what is sometimes called a
half-hypermultiplet.
A pair of half-hypermultiplets is sometimes called
a quark or a quark flavor, a terminology we will sometimes use below.}
or a single copy of the adjoint representation.  These theories allow
hypermultiplet bare masses and
all have the $SU(2)_R$ group of $R$ symmetries.  An additional
$U(1)_R$ symmetry group is generally anomalous and is also explicitly
violated by hypermultiplet bare masses.  The theory with the adjoint
hypermultiplet actually has more symmetry (${\cal N}=4$ supersymmetry
and $SO(6)_R$ in the absence of a hypermultiplet bare mass, broken
to ${\cal N}=2$ and $SO(4)_R$ if there is a bare mass); we will
call it the ${\cal N}=4$ theory (though we are mainly interested
in the case in
which the ${\cal N}=4$ is softly broken to ${\cal N}=2$ by the bare
mass).

After including hypermultiplets, the Coulomb branch of vacua is still
parametrized by a copy of the $u$-plane (where $u$ is
related to $\langle \Tr \phi^2 \rangle$ in the underlying theory),
but now the $u$-plane parametrizes a different family
of elliptic curves.  The appropriate
families (which depend on the
hypermultiplet bare masses) were determined in \swii.
They have the form:
\eqn\curvei{
y^2 = x^3 + a_2 x^2 + a_4 x + a_6
}
where $a_2, a_4,a_6$ are polynomials in $u$ and in the
masses  $m_i$. They are also polynomials in the
scale  $\Lambda$ of the theory for $N_f < 4$,  or
of certain modular functions $e_i(\tau_0)$ for $N_f=4$ or
for ${\cal N}=4$.
Here $\tau_0$ will refer to the
coupling as measured at $u=\infty$ in the $N_f=4$ or ${\cal N}=4$ theory.
In this paper we have put $\Lambda=1$ for $N_f=0$.

Any of these theories can be twisted to obtain a topological field
theory.  We will consider here only the standard twist, which as reviewed
in section 2 is obtained
by decomposing the four-dimensional rotation group
as $Spin(4)=SU(2)_-\times SU(2)_+$ and then picking a diagonal
subgroup of $SU(2)_+\times SU(2)_R$.  Some additional twists are possible,
using the $ Spin(2N_f)$ global symmetry in the case of
matter multiplets in the two-dimensional representation and picking
a homomorphism of $SU(2)_+$ to $Spin(2N_f)$, or using the extended symmetry
of the ${\cal N}=4$ theory.  (An alternative twist in the
latter case, related to the Euler characteristic of instanton
moduli space, was explored in \vw.)

The invariants of  smooth four-manifolds associated with these twisted
theories with hypermultiplets could be computed at short distances in terms
of the underlying $SU(2)$ gauge theory with matter.  Such an analysis would
proceed roughly along the lines in \refs{\vw,\marino}
and will not be explored here.
Our goal will be to compute at long distances in terms of the physical
vacua.  We will consider mainly the case that the hypermultiplet
bare masses are generic, so that there is no Higgs branch of vacua, and
the entire contribution
 comes from the Coulomb branch.  On the Coulomb branch, there is
a finite set of exceptional points at which a massless charged hypermultiplet
appears.  There will be SW contributions from these points, which can be
analyzed rather as in section 7.  There is also a continuous
$u$-plane integral, similar to the ones we have already studied
but with some differences that we will analyze.

We will obtain the analogue of \hoggo\ for the
theories with hypermultiplets. It is given by
equation $(11.8)$ below. Then we will discuss special
properties of the integrals and of the various models.

\subsec{The Measure Factor And The Contact Term}

First we analyze the measure factor $A^\chi B^\sigma$ for
the $u$-plane integral with hypermultiplets.

The factor $B^\sigma$, for the theory without hypermultiplets, was
determined in \abelS\ as follows.  We will express
the argument in  a way that carries over immediately to the general
case.  This factor has neither zeroes
nor poles on the $u$-plane except at points $u_i$ at which there
is a massless charged hypermultiplet.  Integrating out the light
hypermultiplet produces a singularity $B^\sigma\sim (u-u_i)^{\sigma/8}$.
Hence,
if we set\foot{We recall from \swii\
that for the theory with doublet hypermultiplets, one uses a formalism
that generalizes the $\Gamma^0(4)$ formalism of the pure gauge theory.
In this description, the discriminant of the family of elliptic curves
is up to a constant multiple the function $\Delta$ defined in the next
equation.  For the ${\cal N}=4$ theory, it is more
convenient to use
instead a formalism related to $\Gamma(2)$, and then the discriminant
is the square of what we are here calling $\Delta$.}
\eqn\ikoo{\Delta=\prod_i(u-u_i)}
 then up to a constant multiplicative factor one has
\eqn\hogbo{B^\sigma=\Delta^{\sigma/8}.}
The case treated in \abelS\ was the case $\Delta=u^2-1$.
There is no way to determine an overall multiplicative factor  in \hogbo\
except by comparing to a precise definition of the theory at short distances;
we have done so earlier in this paper for the pure gauge theory, but
will not do so for the theories with hypermultiplets.

Just as in \abelS, this formula can be checked by looking at the behavior
near $u=\infty$.  Near $\infty$, the function $B^\sigma$ should behave
as a power of $u$,
 in a way that  reproduces the anomaly of those of the elementary
fields that are massive at large $u$ and have been integrated out to
produce the $B$ function.  (Possible hypermultiplet bare masses which
break the $U(1)_R$ symmetry explicitly are irrelevant at large $u$.)
The charged components of the vector multiplet give an anomaly that
corresponds to a behavior at infinity $B^\sigma\sim u^{\sigma/4}$,
as in eqn.  (3.5) of \abelS.  Including the contributions of charged
components of the hypermultiplets, the   behavior at infinity
should be $B^\sigma\sim u^{(2+N_f)\sigma/8}$ for the theory with
$2N_f$ doublets, or $u^{3\sigma/8}$ for the theory with an adjoint
hypermultiplet.  This agrees with \hogbo, since \swii\ $\Delta$ is
a polynomial of degree $2+N_f$ or of degree 3 in the two cases.

Now we consider the ``function'' $A^\chi$.  This quantity was determined
in \abelS\ from the following properties:

(1) It is actually not a function in the ordinary sense, but transforms
under duality transformations in the low energy theory like a
holomorphic modular form of weight $-\chi/2$.

(2) In the appropriate local description of the low energy theory,
it has neither zeroes nor poles at zeroes of $\Delta$ or elsewhere
on the $u$-plane.

(3) It behaves near $u=\infty$ as $u^{\chi/4}$.

All of these properties are unchanged by the incorporation of
hypermultiplets.  In fact, the anomalies of the elementary
hypermultiplets involve
only $\sigma$ and not $\chi$, so incorporation of such hypermultiplets
does not modify assertion (3).  Likewise, massless charged hypermultiplets
of the low energy theory have an anomaly that is independent of $\chi$,
which is the reason for the assertion in (2) that $A^\chi$ is regular
and non-zero even at zeroes of $\Delta$.
Finally, because the hypermultiplet kinetic
energy has no explicit $\tau$-dependence, the analysis of the modular
weights proceeds just as in \abelS, leading to assertion (1).

One would therefore expect that in some sense $A^\chi$ would be the
same as in the theory without hypermultiplets.  That is so, but some care
is required.  The result in \abelS\   was $A^\chi=((u^2-1)d\tau/du)^{\chi/4}$.

The most obvious thing to do is to replace $u^2-1$ by $\Delta$.
However, it is not true that $A^\chi$ is equal to
$(\Delta\, d\tau/du)^{\chi/4}$.
This fails to obey property (2), which fails at zeroes of $d\tau/du$.
(For the theory without elementary hypermultiplets, there are no such
zeroes, as shown on p. 398 of \abelS\ by an argument that does not
carry over when hypermultiplets are included.)
But we can proceed as follows.  For the theory
without hypermultiplets, there is an identity
%%%
\matone:
%%%
\eqn\ikkop{(u^2-1){d\tau\over du}={i \over  4 \pi}
\left({du\over da}\right)^2.}
Hence the result in \abelS\ could have been written
\eqn\mogbo{A^\chi=\left({du\over da}\right)^{{\chi\over 2}}.}
This expression obeys properties (1)-(3), irrespective of the
presence of matter hypermultiplets.  Property (1) is verified using
the fact that $du/da_D=(da/da_D)du/da=(1/ \tau)du/da$.  As regards
(2), the absence of zeroes or poles of $A^\chi$ away from zeroes
of $\Delta$ follows from the fact that $da/du$ is a period of a holomorphic
differential and so is never zero.  Regularity at zeroes of $\Delta$
follows from the fact that, in the appropriate local description, $a$
is a good local coordinate at such a zero so $du/da \not= 0$.
Finally, (3) is a consequence of the fact that near infinity
$a\sim \sqrt u$.

To summarize our results so far, the measure factor is
\eqn\jogbo{A^\chi  B^\sigma =\left({du\over da}\right)^{\chi/2}
\Delta^{\sigma/8}}
up to multiplicative constants that depend on a precise microscopic
definition of the theory.

The other somewhat similar function that must be determined
is the contact function $T(u)$ that arises in the product of
two-observables.
The derivation in section 2.2, which led in equation \jokko\ to the general
structure
\eqn\pillop{T=-{1\over 24}  E_2(\tau)\left(du\over da\right)^2 + H(u)
}
carries over here.  We recall that $H$ is here an ordinary holomorphic
function of $u$.
Moreover the determination of the function
$H$ that was given in section 2.2 for the theory without hypermultiplets
carries over with only small modifications to the general case.
One modification is that in general the theory with hypermultiplets
has no symmetry under $u\to -u$.  An examination of the determination  of
 $H$ for $N_f=0$ shows that the same result -- that is, $H(u)=u/3$ --
follows without any assumption of
this symmetry if $T$ vanishes for $u\to\infty$.  This is so by asymptotic
freedom for doublet hypermultiplets with $N_f=1,2,3$, so the form
of $T(u)$ obtained previously for $N_f=0$ carries over to these cases.
  For $N_f=4$, or for ${\cal N}=4$,
instead of asymptotic freedom one has conformal invariance near $u=\infty$.
In those cases, instead of vanishing near infinity, $T$ might approach
a constant (independent of $u$, but depending on the bare coupling
constant or more precisely on the coupling
constant $\tau_0$ measured at $u=\infty$).
Thus for $N_f=4$ or ${\cal N}=4$, we have
$H(u) = E_2(\tau_0) \cdot u/3$ up to a possible additive constant.  The
coefficient of $u$ has been chosen to cancel a pole in $T$ at $u=\infty$.
Since topological invariance would not be spoiled by adding a constant to
$T$,
the constant term in $T(u)$ can only be determined in these examples
by comparing to a microscopic definition of the theories, or possibly
by using $S$-duality and holomorphy in $\tau$.

\subsec{Expression for the $u$-plane integrals }

We now consider the twisted theories with hypermultiplets
on a four-manifold $X$ with $b_1=0$, $b_2^+=1$.
For the theories with doublet hypermultiplets, one must set
$\xi=w_2(E)=w_2(X)$. The reason is that
the hypermultiplets, being doublets,
 transform non-trivially under
the center
of the gauge group
$SU(2)$ and, being spinors,
also transform under the center of the cover of  the Lorentz group.  (For some
alternative twists that use suitable homomorphisms
of $SU(2)_+$ to $Spin(2N_f)$, for $N_f=2,4$, this restriction would be
modified.)
For the case of the adjoint hypermultiplet, the
hypermultiplets transform trivially under the center of
$SU(2)$ but are still spinors with respect to the Lorentz
group. Thus these theories only make sense
for $w_2(X) =0$, i.e., for spin manifolds.

Apart from
factors examined in the last subsection, the derivation of the $u$-plane
integrand in these theories
proceeds rather as in section 3.  In particular, the definition of the
photon path integral $Z$ of eqn. \momobo\ is unchanged.
One important difference, which leads to some complications in actually
performing integrals, is that the $u$-plane is generically not a modular
curve, and hence one cannot conveniently map the integration region
to a fundamental domain in the upper half $\tau$ plane, as we did in the theory
without elementary hypermultiplets.

Putting together all the above remarks, we conclude
that  the $u$-plane integral for all values of $N_f$ is
given by:
\eqn\hoggomatter{
Z_u(p,S;m_i,\tau_0) = \int_{\IC}
{du  d\bar u \over  y^{1/2}}
\mu(\tau) e^{2 p u + S^2 \hat T(u)} \Psi
}
(For $N_f < 4$ we replace $\tau_0 \rightarrow \Lambda$ on the
left hand side.)  The function $\Psi$ is exactly the
same as \newtheta. The measure is now:
\eqn\measmatt{
 \mu(\tau)   = 2
 \alpha^\chi \beta^\sigma  { d \bar \tau \over  d \bar u}
\bigl({d a \over  du } \bigr)^{1- \half \chi  } \Delta^{\sigma/8}
}
Here $\alpha,\beta $ are functions
of $m_i,\Lambda$ for $N_f <4$ and functions of
$m_i, \tau_0$ for $N_f=4$. It is possible that
they can be determined
by constraints of symmetry, holomorphy, and RG
flow. (We hope to return to this in future work.)
Of course, the definition of
$Z_u$ also depends implicitly on
a choice of metric through a choice of
period point $\omega$. We will study the
dependence on $\omega$ presently.

As for $N_f=0$, the integral \hoggomatter\ requires
precise definition. There are singularities in the integral
near the zeroes of $\Delta$ and near $u=\infty$.
Near the zeroes of $\Delta$
one can
express the integrand in terms of the
appropriate $\tau$ parameter and use the definition
discussed in section 3.2. This also works at
$u=\infty$ for $N_f<4$.

For $N_f=4$ (or similarly ${\cal N}=4$) the $\tau$ parameter behaves at
$u \rightarrow \infty$ like:
\eqn\jayexp{
\tau(u;m_i, \tau_0) =
\tau_0 + \CO(1/u)
}
We also have:
\eqn\perlim{
{da \over  du } \rightarrow {1 \over  \sqrt{8u}}(1 +
 \CO({1\over  u}) )
}
Finally, $\Delta$ is a sixth order polynomial in
$u$.
Hence the measure behaves as
\eqn\limmeasure{
F(m_i,\tau_0) du d\bar u {1 \over  \bar u^2} u^{-\half + \chi/4 + 3\sigma/4}
\biggl(1 + \CO({1 \over  u} , {1 \over  \bar u})
\biggr)
}
for some function $F$.
The series in ${1 \over  \bar u}$ comes from the
expansion of $y=\Im \tau$ and $d \bar \tau/d \bar u$.
An operator insertion of ghost number $Q$
modifies the integrand at $u \rightarrow \infty$
by an  insertion of $u^{Q/4} $. This is always
holomorphic. Therefore, in
  order to define the integral at infinity we first
 integrate over the region $\vert u \vert<R$ and then take $R\to\infty$.
\foot{In principle other regularizations are possible.
For instance, one could  use the coordinate
$a$ at infinity and make a similar definition.  We expect that
any difference between   the two answers could be interpreted
in terms of a     redefinition of the observables, along the
lines suggested in section 3.2.}
The integrand of
\hoggomatter\ has an expansion in terms
$\sum_{\mu,\nu}u^\mu\bar u^\nu$, where the largest possible value of $\nu$
is $-2$.  Hence, upon integrating over a large circle in the $u$-plane
with $\vert u \vert$ fixed and then
integrating over $\vert u \vert$, the dangerous terms
vanish and one gets a convergent result.

\subsec{Topological invariance of the integrals}

We will now investigate the topological
invariance of the integral \hoggomatter.
Actually, anticipating that wall-crossing at zeroes of $\Delta$ will cancel
similar behavior of SW contributions, we will focus on the behavior near
$u=\infty$.
Given the
delicate convergence of
the integrals at large $u$,  the topological
invariance is not obvious. We will discover
that for $N_f<4$, the integrals are locally constant as a function
of the metric and have a wall-crossing behavior similar to that of
Donaldson theory.
In marked contrast, for $N_f=4$
(and $\CN=4$) the integrals
have no wall-crossing, but instead have
a {\it continuous} dependence on the
period point $\omega$. Nevertheless,
 the $N_f=4$ theory
does have a truly topological subsector.
This is defined by correlation functions
of observables
satisfying an upper bound on the ghost number
which we derive below. For these observables
the integrals are absolutely convergent at
infinity and have no metric dependence at all, even for $b_2^+=1$.

Let us consider a family of period points
$\omega(t)$ and study
$Z_u(\omega(t))$. The variation of the path
integral
with respect to the metric is given by the
one-point function of the energy momentum
tensor $T$. Since we have a topological field
theory, the energy momentum tensor is
BRST exact; that is,
$T=\{ \bar \CQ, L \}$ for
some local  and duality invariant quantity
$L$. When the path integral is
reduced to an integral on the $u$-plane, the
BRST exact integrand is expected to become a total
derivative in $\bar u$. We will now exhibit this behavior
directly from the expression \hoggomatter.

The nonholomorphic
and metric dependent factors in the integrand
of \hoggomatter\
are all contained in the expression:
\eqn\topdepi{
\eqalign{
\tilde \Psi \equiv  {d \bar \tau \over  d \bar u}
 \exp\bigl[ { 1 \over  8 \pi y}
&
({d   u \over  d  a})^2 S_+^2 \bigr]
e^{2\pi i \lambda_0^2}
\sum_{\lambda\in H^2+ \half w_2(X) }(-1)^{(\lambda-\lambda_0)\cdot w_2(X)}\cr
\biggl[ {(\lambda ,\omega) \over  y^{1/2}} +{i \over  4 \pi y^{3/2}} {d u\over
d a} (S,\omega )
   \biggr] \cdot  &
\exp\biggl[ - i \pi \bar\tau (\lambda_+)^2-i \pi   \tau(\lambda_-  )^2-i
{ d    u \over  d   a} (S,\lambda_-) \biggr]  \cr }
}
The derivative of
$\tilde \Psi(\omega(t))$ with respect to $t$ can
be rewritten as a total derivative $d/d\bar u$:
\eqn\oopsilon{
{d \over  dt} \tilde \Psi(\omega(t)) =
{d \over  d \bar u} \Upsilon}
The explicit formula is:
\eqn\oopsii{
\eqalign{
\Upsilon =    -4i
\exp\bigl[ { 1 \over  8 \pi y}({d   u \over  d  a})^2 S_+^2 \bigr]
e^{2\pi i \lambda_0^2}
&
\sum_{\lambda\in H^2+ \half w_2(X) }(-1)^{(\lambda-\lambda_0)\cdot w_2(X)}
\cr
\biggl[ (\lambda ,\dot \omega)    y^{1/2}  +{y^{-1/2} \over  4 \pi i} {d u\over
 d a} (S,\dot \omega )
   \biggr] \cdot
&
\exp\biggl[ - i \pi \bar\tau (\lambda_+)^2
- i \pi   \tau(\lambda_-  )^2-i
{ d    u \over  d   a} (S,\lambda_-) \biggr]  \cr }
}
An important feature of $\Upsilon$ is that
it transforms well under modular transformations
when combined with the contact term
$\exp[S^2 T(u)]$. Thus, one can integrate
by parts. It is possible to
 write similar expressions directly for
the integrand of \hoggomatter; however, these
expressions are in general not useful because
they do not have good modular transformations
(they are quite analogous to the second term
in \zagfrm).

Using \oopsilon, we can perform
 the integration by
parts, and find that
the continuous variation of
the Coulomb branch integral is
\eqn\contvari{
{d \over  dt} Z_u(\omega(t)) = -i \alpha^\chi
\beta^\sigma
\lim_{R \rightarrow \infty}
\oint_{\vert u \vert = R} du
   \bigl({d a \over  du } \bigr)^{1- \half \chi  } \Delta^{\sigma/8}
e^{ 2 p u + S^2 T(u)} \Upsilon
}
Now, if $N_f<4$,
 $\tau(u) \rightarrow i \infty$ for $u \rightarrow \infty$,
indeed, $q\sim 1/u^{4-N_f}$.  Hence the terms in
the lattice theta function decay at infinity.
 If we do not cross
a wall, then $\lambda_+^2>0$, so to cancel the phase
integral on $u$ the integrand must go like
$u^{-1} (u \bar u)^{-\nu}$ for $\nu>0$, and hence
the variation of the integral vanishes.
 Thus, except for the wall-crossing,
the integral is topologically invariant for
$N_f<4$.

The situation is quite different for $N_f=4$.
Since $\tau$ approaches a constant $\tau_0$,
there is no  suppression from the lattice
theta function. The integrand pertaining
to the general correlation function
of high order involves a sum of terms
including  a
term $\sim {du \over u}$. Thus, there is
{\it continuous} variation of the integral as a function
of $\omega$!

While for $N_f=4$ we have thus lost topological  invariance
for the general correlator at $b_2^+=1$,
there is a special subclass of correlators
which are completely invariant, that is,  have no continuous variation or
wall-crossing, and are thereby true
topological invariants. These are the
correlators for which the integral at
$u=\infty$ is absolutely convergent.
The variation \contvari\ of  a correlator involving
$p^\ell S^r$ has an integrand which behaves
for large $u$ like
\eqn\contvarii{
\oint du\,\,
 u^{(\sigma+1)/2} p^\ell u^\ell S^r u^{r/2} \biggl[
\Upsilon_0 + \CO(1/u, 1/\bar
u)\biggr],
}
Here $\Upsilon_0$ is the limiting value of
\oopsii\ at the appropriate order in $S$. It is
generically nonzero. The term $u^{ (\sigma+1)/2}$
 comes from the
measure and we have used $\chi + \sigma=4$.
If the leading power of $u$ is less than $-1$
then we cannot have any variation of the correlator,
either continuous or discontinuous.
Hence,  correlators
of ghost number $Q=4 \ell + 2 r$ are true topological (or rather
differentiable)
invariants, even for manifolds of $b_2^+=1$ (and $b_1=0$), for
\eqn\ghostbound{
6 +2 \sigma + Q < 0.
}

\subsec{Other Properties Of The $u$-plane Integrals}

Now we will discuss other general properties of the $u$-plane integrals.
Many  results we found for Donaldson theory
 generalize nicely, but there are some changes.

The
vanishing  in certain chambers
of the $u$-plane integral, discussed in section 5,
 does not have a precise analog for $N_f>0$
because of the restriction $w_2(E)=w_2(X)$, which prevents one from
considering the appropriate bundles.  This vanishing does have an analog
for ${\cal N}=4$,
which can be obtained in the same way.  Since the
twisted ${\cal N}=4$ theory is restricted to four-manifolds $X$ with
 $w_2(X)=0$, the only practical case of the vanishing is $\IP^1\times \IP^1$.

The blowup formula analyzed in section 6 generalizes as follows.
The derivation of the blowup formula is exactly the
same as before, but for $N_f>0$, we must choose the case
$w_2(\tilde E) \cdot B =1 ~\mod ~2$ because of the
restriction $w_2(E)=w_2(X)$.
Following through the steps
\ottop\ to \blwiiip\ we find that the
integrand for the manifold $\widehat X$ is that of
the integrand for $X$ times:
\eqn\blwmatter{
{\alpha \over  \beta} 2^{9/4} t
\exp\biggl[-t^2 H(u) - \sum_{k=2}^\infty {t^{2k} \over  2k} {G_{2k}(\tau) \over
 (2 \pi h)^{2k} } \biggr]
}
where $h=da/du$.
The Eisenstein functions $G_{2k}(\tau)$ are related
to Eisenstein series $G_{2k}(L)$
 of the lattice $L=\omega_1 \IZ + \omega_2 \IZ$
by
\eqn\eisfuns{
{G_{2k}(\tau) \over  (2 \pi h)^{2k} } = 2^k G_{2k}(L)
}
The $G_{2k}(L)$  may be expressed as
polynomials in the coefficients $a_2,a_4, a_6 $ of \curvei.
Therefore,  there are universal polynomials
$\CB_k(u, a_2,a_4, a_6)$ defined by:
\eqn\blmtterii{
t \exp\biggl[-t^2 H(u) - \sum_{k=2}^\infty {(2 t^2)^k \over  2k}
G_{2k}(L) \biggr] =
\sum_{k=1}^\infty t^k \CB_k(u, a_2,a_4, a_6)
}
such that, for all $N_f$ the blowup formula is given by:
\eqn\blwvimtter{
\eqalign{
\biggl\langle \exp\bigl[ I(S) + t I(B) + p \CO \bigr] \biggr\rangle_{\hat{X}  }
& \qquad\qquad
\qquad\qquad\cr
 = {\alpha \over  \beta}
2^{9/4} \sum_{k\geq 1}
&
t^k
\biggl\langle \exp\bigl[ I(S) + p \CO \bigr]
\CB_k(u, a_2,a_4, a_6) \biggr\rangle_{
X } \cr}
}
(where $w_2(E)  = w_2(X)$ on both sides).
We conclude that the coefficients in the
generalized blowup formula are polynomial in the masses
with rational coefficients.

One can analyze
wall-crossing just as in section 4, with results just like
those of section 4 at zeroes of $\Delta$ (and some modifications, for reasons
explained in section 11.3, near $u=\infty$). At a
zero $u=u_*$ of $\Delta$, the wall crossing at
$W_\lambda$ is:
\eqn\wkcouplwc{
\eqalign{
Z_+ - Z_- =2 \sqrt{2}
e^{2 \pi i \lambda_0^2}
&
 (-1)^{(\lambda-\lambda_0)\cdot w_2(X)}
\alpha^\chi \beta^\sigma \qquad\qquad\cr
\Biggl[q^{-\lambda^2/2}
&
({d u \over  d \tau})({da \over  du})^{1-\half \chi}
\Delta^{\sigma/8} e^{2 p u + S^2 T - i {du \over  da} (S,\lambda)}
\Biggr]_{q^0}
\cr}
}
where we expand in the good local $q$-coordinate:
\eqn\wkcoupl{
\eqalign{
u & = u_* + \kappa_* q + \CO(q^2) \cr
{da \over  du} & = ({da \over  du} )_* + ({d^2 a \over  du^2} )_*
(u-u_*)+\cdots  \cr}
}
Similarly, the   wall-crossing at $\infty$ is given by
the contour integral:
\eqn\mwcinf{
\eqalign{
Z_+ - Z_- = 2 \sqrt{2}   (4-N_f)
&
e^{2\pi i \lambda \cdot \lambda_0}
\alpha^\chi \beta^\sigma \qquad\cr
\cdot \lim_{R \rightarrow \infty} \oint_{\vert u \vert = R}
du
 q^{-\lambda^2/2}
&
 ({da \over  du})^{1-\half \chi}
\Delta^{\sigma/8}
\exp\bigl[2 p u + S^2 T - i {du \over  da} (S,\lambda)\bigr]
\cr}
}

The above wall-crossing formulae are consistent
with RG flow. That is, if we take a quark mass to
infinity $m^2 \rightarrow \infty$ in a theory with
$N_f$ quarks then the wall-crossing at $u\cong m^2$
combines with the wall-crossing for $N_f$ quarks at
$u=\infty$ to produce the wall-crossing at $u=\infty$
for the theory with $N_f-1$ quarks. To prove this
one expresses \wkcouplwc\  as a contour integral
in the $u$-plane and shows that it combines
correctly with \mwcinf\ to produce the
corresponding expression at $N_f-1$ in the
limit $m^2 \rightarrow \infty$.

As in section 11.3,
the main qualitative difference from
what we have  seen in the case $N_f=0$
comes in the
analysis of wall-crossing   at
$u=\infty$  for the asymptotically conformally invariant theories
$N_f=4$ and ${\cal N}=4$.
In these cases, because the effective $\tau$ parameter
does not diverge at $u=\infty$, the behavior near $u=\infty$ is not at all like
what we encountered in section 4.
In these cases there  is never any wall-crossing at
$u=\infty$, but there is
continuous variation with the period point $\omega$,
except for those correlators satisfying \ghostbound.
For these correlators
the convergence is uniform at infinity and
independent of the value of the period point
$\omega$.
Hence, there is no variation at all.
Since all other $u$-plane wall-crossing
(localized at zeroes of $\Delta$)
will cancel SW contributions,
the result is that
for these correlators, one actually gets true
invariants for
four-manifolds of $b_2^+=1$,
in contrast to the usual
situation in Donaldson theory,
in which one gets invariants only
for $b_2^+>1$.

A similar discussion holds
for the other theories at $N_f<4$. Using
\mwcinf\ one finds that in this
case there is no wall-crossing, hence no
variation of the correlators of
ghost number $Q$ for:
\eqn\ghstboundii{
6 + {N_f \sigma \over  2} +Q < 0 .
}
Since $Q\geq 0$ this phenomenon does
not occur in Donaldson theory (i.e., for
$N_f=0$).

\subsec{SW Contributions}

Finally we turn to the generalization of the
results of section 7. The universal functions
$C,P,L$ are obtained from \wkcouplwc\
in exactly the same way as before and the
result is:
%%%
\eqn\univmatter{
\eqalign{
C & = (a-a_*)/q \cr
L & =- i \pi 2\sqrt{2} \alpha^4   \bigl({du \over  da}\bigr)^2\cr
P & = - 8 \pi^2 \beta^8   \Delta (a-a_*)^{-1} \cr}
}
Indeed, with the proper interpretation of
$da/du$ this is the general form for all cases, at all
zeroes $u_*$ of the discriminant.

As an application of these formulae we give the
detailed form of the invariants for four-manifolds of simple
type with $b_2^+>1$ and $b_1=0$
(thus generalizing  equation 2.17 of \monopole\ to
all $N_f$).
\foot{In \marinop, J. Labastida and M. Mari\~no
generalized the reasoning of
\monopole\ to give the result for
$N_f=1$ in the massless case for spin manifolds.
One can check that the expression given below
agrees with their result for this case.}
In this case the $u$-plane integral
vanishes and the entire path integral is a sum
over the SW basic classes $\lambda$ which obey:
$d_\lambda = \lambda^2 - {2 \chi + 3 \sigma \over  4} =0$.
The contribution of each class  $\lambda$
is a sum over the zeroes $u_*$ of the discriminant,
with a given zero contributing
\eqn\spltypei{
\eqalign{
SW(\lambda)
2^{1 + {3 \sigma + \chi \over  2}}
 e^{2\pi i \lambda \cdot \lambda_0} (-i)^\delta
\biggl({ \pi^2 \beta^8 \over 2^{5} }\biggr)^{\sigma/8}
\biggl(   \sqrt{2} \pi \alpha^4   \biggr)^{\chi/4}
&
\kappa_*^\delta  \biggl(  ({da \over  du})_*  \biggr)^{-(\delta+\sigma)  }\cr
\exp\biggl[ 2 p u_*  +
 S^2
&
T_*  - i ({du \over  da})_* (S,\lambda)
\biggr]
\cr}
}
where $\delta = {\chi + \sigma \over  4}$
and $\kappa_*$ was defined in \wkcoupl.
Note that we can simplify
\eqn\valuetee{
T_* = -{1 \over  24} \bigl( ({du \over  da})_*^2 - 8 u_*\bigr)
}
Thus,   the contribution is expressed
solely in terms of the positions of the zeroes and the
values of the periods there.
Because of this, we can be more explicit in terms of
the relation between \spltypei\ and the parameters in
the SW curve \curvei.
By standard
reduction techniques (see, e.g., section three of
\husemoller)
\curvei\  is equivalent to the curve:
\eqn\curveii{
\eqalign{
y^2 & = x^3  - {c_4 \over  48} x - {c_6 \over  864} \cr
c_4 & = 16(a_2^2 - 3 a_4) \cr
c_6 & =  -64 a_2^3 + 288\ a_2\ a_4 - 864\ a_6 \cr}
}
By comparing the values of Eisenstein functions we
can extract  $({da \over  du})_*$  up to sign from
\eqn\valueper{
({da \over  du})_*^2   = {c_4(u_*)  \over  2 c_6(u_*)}
}
Using
\eqn\discrm{
 \Delta   = {(2\pi)^{12} \over  \omega^{12}} \eta^{24}(\tau)   =  2^{-6}
{\eta^{24}(q) \over  (da/du)^{12} }
}
we then obtain:
\eqn\derivs{
\eqalign{
\kappa_* & = {c_4^3(u_*) \over  \Delta'(u_*) } \cr}
}

Note that \valueper\ only determines the value of
the period up to sign. In fact, we do not need to
resolve the squareroot. We must sum over
the contributions of the SW basic classs $\lambda$
and $-\lambda$ and hence
we may average \spltypei\ over $\lambda$ and
$-\lambda$. Since \monopole\
\eqn\chgnsign{
SW(-\lambda) = (-1)^\delta SW(\lambda)
}
the factor
$ \exp\left[ - i ({du \over  da})_* (S,\lambda)\right]$
averages to a cosine when $\delta + \sigma$ is
even and to a sine when $\delta+\sigma$ is odd.
This combines with the prefactor to produce a
series in even powers of $({du \over  da})_*$.

When we sum \spltypei\
over the zeroes $u_*$ we obtain an expression
totally symmetric in the roots of $\Delta$. Therefore,
at any order in $p,S$ the invariants are
rational expressions in $m_i, e_i(\tau_0) $ at $N_f=4$ and
in $m_i, \Lambda$ for $N_f < 4$.

Perhaps the simplest example of these new invariants
is the result for $X$ a K3 surface. In this case,
only $\lambda=0$ contributes. The sum over the
roots becomes:
\eqn\kthree{
\sum_{i=1,\dots, N_f + 2}
{c_4(u_i)^{4} \over  \Delta'(u_i)^2 c_6(u_i)}
\exp\bigl[ (2p + S^2/3) u_i - {S^2 \over  12} (c_6(u_i)/c_4(u_i))\bigr].
}
We have used
$c_4(u_i)^3 = c_6(u_i)^2$,
since $\Delta(u_i)=0$,  and have
omitted an overall function of the $m_i, \Lambda$.

\subsec{Other $u$-plane integrals}

Much of the discussion of this section,
and the $u$-plane integral  \hoggomatter\
in particular, makes
sense for more general families of elliptic curves.
Thus, for example, toroidally compactified
 tensionless string
theories provide a family of
 $d=4, \CN=2$ theories
which can be twisted to produce
topological field theories. The Coulomb branch
of these theories is described by
the   $E_8$ curve of \refs{\ganor,\gms}.
In this case there are $12$ singularities in the
$u$-plane and $\tau$ becomes a constant at
infinity.
This strongly suggests that
 as in the discussion of the
$N_f=4,\CN=4$ theories,
only  a finite set of correlators satisfying a
condition analogous to \ghostbound\ will give invariants.

Another extremely interesting generalization of
the $u$-plane integrals might be provided by
topological field theories associated with the
$D3$ probe \refs{\senpillow,\bds}
 in the context of $F$-theory
\refs{\vafa,\vfmr}. In
this case one would integrate over the
$u$-sphere, regarded as the base of an
elliptically fibered K3 surface. Various
quantities in \hoggomatter\ can be interpreted as
sections of line bundles over the $u$-sphere and
nonvanishing correlators can be identified from
combinations of operators leading to a globally
well-defined $(1,1)$ form. Nevertheless,
while certain correlators in \hoggomatter\
apparently make sense, several open problems
remain. It is not clear, for example, how to
define the topological field theory whose
Coulomb branch leads to \hoggomatter.
The discovery of this theory could be
particularly interesting because
the $u$-sphere
theory, if it really
exists, will obey all the axioms of topological field theory, with Hilbert
spaces associated to three-manifolds and complete cut and paste rules.
This is probably not the case for the
 other theories, even for the  $N_f=4$ and $\CN=4$ theories,
because of the noncompactness of the $u$-plane.

\newsec{Conclusions}

In this paper, we have obtained a more comprehensive understanding
of the relation between the Donaldson invariants and the physics
of $\CN=2$ supersymmetric Yang-Mills theory.  In particular, we have explained
the role of the $u$-plane in Donaldson theory
more thoroughly than
had been done before,
both for $b_2^+=1$ and for hypothetical four-manifolds of $b_2^+>1$ that
are not of simple type.
We hope that in the process the power of the
quantum field theory approach to Donaldson theory and the rationale
for the role of modular functions in Donaldson theory have become clearer.

Our results can be summarized by an admittedly rather complicated
formula for the
Donaldson invariants of {\it any} simply connected
compact four-manifold with $b_2^+>0$. It is:
\eqn\finalresult{
\langle e^{p\CO+I(S)}\rangle_{\xi }
= Z_{u,\xi} + \sum_{\lambda\in H^2(X;\IZ) + \half w_2(X)}
\biggl[
\langle e^{p\CO+I(S)}\rangle_{\xi,1,\lambda }
+
\langle e^{p\CO+I(S)}\rangle_{\xi,-1,\lambda }
\biggr]
}
where $Z_{u,\xi}$ is defined by equations \newtheta, \hoggo, and \measfact\
and the SW contribution at $u=1$ is defined by
\ollo\ and \lilpo, with a similar formula for $u=-1$.
The result can be extended to nonsimply connected
manifolds along the lines discussed in section 10.

The above considerations can be generalized and
extended in several interesting ways.

It is of some
interest to extend Donaldson invariants to
 invariants of a family of four-manifolds, valued in $H^*({\rm BDiff(X)} )$
\donrev, and some work on wall-crossing formulae
in this context has recently been done
\ref\liliu{T. Li and J. Liu, to appear.}.  It should be possible to study
family invariants, and their wall-crossing
formulae (which will occur for $b_2^+>1$)
by a relatively simple extension
of the above arguments. To do so, one would include in the analysis
a  BRST
partner of the metric $\{ \bar \CQ, g_{\mu\nu} \}
= \psi_{\mu\nu} $, giving rise to
differential forms on ${\rm BDiff(X)}$. Wall-crossing
formulae should be obtained from the corresponding
$u$-plane integral as above.

Another avenue for research is in the
generalization of the above results to other
$\CN=2$ systems. We have indicated in section 11
how the results generalize to $SU(2)$ theories with
hypermultiplets. It would also be interesting to
investigate  higher rank gauge groups,
and to study more thoroughly the
reductions of six-dimensional tensionless
string theories, and their hypothetical
$F$-theoretic generalizations.
Some of these generalizations are
currently under study.

It would also be of some interest to  connect these
results to nonperturbative string theory.
The above results will probably have some
use in working with wrapped D-branes.

\bigskip
\centerline{\bf Acknowledgements}\nobreak
\bigskip

We would like to thank
R. Borcherds, I. Frenkel, R. Friedman,
P. Kronheimer, T. Li, M. Mari\~no, J. Morgan, T. Mrowka, P. Sarnak,
N. Seiberg, R. Stern, and G. Zuckerman
for discussions and correspondence.   GM would like to thank
A. Losev, N. Nekrasov, and S. Shatashvili
for many discussions on Donaldson theory
over the years.
GM would like to thank
CERN  and the Aspen Center for Physics
for hospitality during the latter course of this
work. The work of GM is supported by
DOE grant DE-FG02-92ER40704.
The work of EW is supported by NSF grant PHY-9513835.

\appendix{A}{Elliptic curves, congruence subgroups,
and modular forms}

Here we collect some useful facts and notations
related to various modular forms in the paper.

The covariant Eisenstein function of weight two is
$\hat E_2$ where:
\eqn\cov{
\eqalign{
E_2 & = 1 - 24 q + \cdots\cr
\hat E_2 & \equiv E_2 - { 3 \over  \pi y} \cr}
}

Our conventions for theta functions are:
\eqn\frmpf{
\eqalign{
{\vt{\theta}{\phi}{0} \over  \eta}
& = {\sum q^{\half (n+ \theta)^2 } e^{ 2 \pi i (n+\theta)\phi} \over  \eta}\cr
 = e^{ 2 \pi i \theta \phi}
q^{({\theta^2 \over  2} - {1 \over  24}) }
&
\prod (1+ e^{2 \pi i \phi} q^{n-\half + \theta} )
(1+ e^{- 2 \pi i \phi} q^{n-\half -\theta} ) \cr
}
}
$0\leq \theta, \phi\leq 1$.

In particular the three Jacobian theta functions have
series and product representations:
\eqn\thtprd{
\eqalign{
\vartheta_2& = \vt{1/2}{0}{} = 2 q^{1/8}\prod (1-q^n)(1+q^n)^2\cr
& = \sum_{n\in\IZ} q^{\half(n+\half)^2} = 2 q^{1/8} + \cdots\cr
\vartheta_3& = \vt{0}{0}{} = \prod (1-q^n)(1+q^{n-\half} )^2\cr
& = \sum_{n\in\IZ} q^{\half n^2} = 1 + 2 q^\half + \cdots\cr
\vartheta_4& = \vt{0}{1/2}{} = \prod (1-q^n)(1-q^{n-\half} )^2\cr
& = \sum_{n\in\IZ} q^{\half n^2}(-1)^n = 1 - 2 q^\half + \cdots\cr}
}

The Seiberg-Witten curve is:
%%%
\eqn\swcurve{
y^2 = x^3 -   u x^2 + {\Lambda^4\over  4} x
}
If we set $\Lambda=1$ the  singularities will be at:
 $  u =1$ for the monopole cusp and $u=-1$ for the
dyon cusp. This is the modular curve of $\Gamma^0(4)$.

The group $\Gamma^0(4)$ is conjugate in $GL(2,{\bf Q})$ to the subgroup
$\Gamma(2)$ of $SL(2,\Z)$ which consists of matrices congruent to 1 modulo 2.
The $u$-plane could equally well be identified (as in \swi)
as the modular curve of $\Gamma(2)$,
which parametrizes a family of elliptic curves $C'_u$, defined by
$y^2=(x^2-1)(x-u)$, with a distinguished level two structure (the points with
$y=0$ and $x=1,-1,u$).  The two families
of elliptic curves differ by a two-isogeny.  We use here
(as in \swii) the $\Gamma^0(4)$ description, to make some formulas slightly
more natural and to facilitate comparison to the mathematical literature. The
translation between the two descriptions is
given by:
\eqn\relation{
\eqalign{
 u & = \tilde u \cr
 \tau & = 2 \tilde \tau \cr
 a & = \tilde a/2 \cr
 a_D & = \tilde a_D \cr}
}
where quantities in the $\Gamma(2)$ description
are denoted with a tilde.

In terms of theta functions we have:
\eqn\swquant{
\eqalign{
 u & = \half { \vartheta_2^4 + \vartheta_3^4 \over  (\vartheta_2 \vartheta_3)^2
} \cr
u^2- 1 & = {1 \over  4} {\vartheta_4^8 \over  (\vartheta_2 \vartheta_3)^4 } = {
\vartheta_4^8 \over  64 h^4(\tau)} \cr
{ i \over  \pi} {du \over  d \tau} & = {1 \over  4}
 {\vartheta_4^8 \over  (\vartheta_2 \vartheta_3)^2 } \cr
\Biggl( { ( {2i \over  \pi} {du \over  d \tau} )^2 \over  u^2-1 } \Biggr)^{1/8}
& =
\vartheta_4 \cr
h(\tau) & =
\dau = \half \vartheta_2 \vartheta_3\cr}
}

The following $q$-expansions are sometimes
useful:

\eqn\qyooexp{
\eqalign{
u = u_{(\infty,0)} & = { 1 \over  8 q^{1/4}} \biggl( 1 + 20 q^{1/2} - 62 q +
216 q^{3/2} + \cdots \biggr) \cr
& = {1\over {8\,{q^{{1\over 4}}}}} + {{5\,{q^{{1\over 4}}}}\over 2} -
  {{31\,{q^{{3\over 4}}}}\over 4} + 27\,{q^{{5\over 4}}} - {{641\,{q^{{7\over
4}}}}\over 8} +
  {{409\,{q^{{9\over 4}}}}\over 2} + \cdots \cr}
}

\eqn\qyooexpi{
\eqalign{
 u_{(\infty,1)} & = -{i \over  8 q^{1/4}}
   + {{5\,i}\over 2}\,{q^{{1\over 4}}} +
   {{31\,i}\over 4}\,{q^{{3\over 4}}} + 27\,i\,{q^{{5\over 4}}} +
   {{641\,i}\over 8}\,{q^{{7\over 4}}} + {{409\,i}\over 2}\,{q^{{9\over 4}}} +
\cdots  \cr}
}

\eqn\qyooexpii{
u_{M}(q_D) =
1 + 32\,q_D + 256\,{q_D^2} + 1408\,{q_D^3} + 6144\,{q_D^4} + 22976\,{q_D^5} +
76800\,{q_D^6}+\cdots
}

\eqn\qyooexpiii{
\eqalign{
 T(u) & = -{1 \over  24}  \biggl[ {  E_2\over  h(\tau)^2}  - 8 u\biggr]\cr
&
= q^{1/4} - 2\ q^{3/4} + 6\ q^{5/4} - 16\ q^{7/4} +
    37\ q^{9/4} - 78\ q^{11/4} \cr
& + 158\ q^{13/4} - 312\ q^{15/4} +
    594\ q^{17/4}  + \cdots\cr
T_M(q_D) & = \half + 8\ q_D + 48\ q_D^2 + 224\ q_D^3 + 864\ q_D^4 + 2928\ q_D^5
+ 9024\ q_D^6 + \cdots \cr }
}

\eqn\qyooexpiv{
\eqalign{
h & = \half \vartheta_2 \vartheta_3 = {1\over 4}\vartheta_2^2(\tau/2)\cr
&=  q^{1/8} + 2\ q^{5/8} + q^{9/8} + 2\ q^{13/8} +
    2\ q^{17/8} + 3\ q^{25/8} + 2\ q^{29/8} + \cdots\cr
h_M& = {1 \over  2i} \vartheta_3 \vartheta_4 = {1\over  2i}\vartheta_4^2(2
\tau_D) \cr
&={1\over  2i} (1- 4 q_D + 4 q_D^2 + 4 q_D^4 -8 q_D^5 + \cdots ) \cr}
 }

Finally, near the monopole cusp we have:
\eqn\adqyoo{
a_D(q_D) =
16i q_D( 1 + 6 q_D + 24 q_D^2 + 76 q_D^3 + \cdots)
}

\appendix{B}{Siegel-Narain  Theta functions}

Let $\Lambda$ be a lattice of signature $(b_+,b_-)$.  Let $P$
be a decomposition
of $\Lambda\otimes \IR$ as a sum of orthogonal subspaces
 of definite
signature:
\eqn\dfsign{ P:\Lambda \otimes \IR \cong
\IR^{b_+,0} \perp \IR^{0,b_-}
}
Let $P_\pm(\lambda)= \lambda_\pm $ denote the projections onto the two
factors.
We also write $\lambda = \lambda_+ + \lambda_-$.
 With our conventions $P_-(\lambda)^2 \leq 0 $.

Let  $\Lambda+ \gamma $ denote a translate of the lattice
$\Lambda$.
We define the Siegel-Narain theta function
\eqn\sglthet{
\eqalign{
\Theta_{\Lambda + \gamma} (\tau, \alpha,\beta; P, \xi)
\equiv
&
 \exp[{ \pi \over  2 y} ( \xi_+^2 - \xi_-^2) ] \cr
\sum_{\lambda\in \Lambda + \gamma}
\exp\biggl\{ i \pi \tau (\lambda+ \beta)_+^2 +
i \pi \bar \tau (\lambda+ \beta)_-^2
&
+ 2 \pi i (\lambda+\beta, \xi) - 2 \pi i
(\lambda+\half \beta, \alpha) \biggr\} \cr
= & e^{i \pi (\beta,\alpha)}
 \exp[{ \pi \over  2 y} ( \xi_+^2 - \xi_-^2) ] \cr
\sum_{\lambda\in \Lambda + \gamma}
\exp\biggl\{ i \pi \tau (\lambda+ \beta)_+^2 +
i \pi \bar \tau (\lambda+ \beta)_-^2
&
+ 2 \pi i (\lambda+\beta, \xi) - 2 \pi i
(\lambda+  \beta, \alpha) \biggr\} \cr}
}
We modify slightly  the definitions in
\borcherds\
for the present case: There is no essential simplification
in passing to an even sublattice. We also treat
insertions somewhat differently.

The main transformation law is:
\eqn\thetess{
\Theta_{\Lambda } (-1/\tau, \alpha,\beta; P, {\xi_+ \over  \tau} +
{\xi_- \over  \bar \tau} )
= \sqrt{\vert \Lambda \vert \over  \vert \Lambda' \vert}
(-i \tau)^{b_+/2} (i \bar \tau)^{b_-/2}
\Theta_{\Lambda' } ( \tau, \beta,-\alpha ; P, \xi )
}
where $\Lambda'$ is the dual lattice.
If there is a characteristic vector, call it $w_2$, such that
\eqn\characteristic{
(\lambda,\lambda) = (\lambda, w_2)~ \mod ~2
}
for all $\lambda$
then we have in addition:
\eqn\thettee{
\Theta_{\Lambda } (\tau+1, \alpha,\beta; P, \xi)
= e^{-i \pi(\beta,w_2)/2}
\Theta_{\Lambda } ( \tau, \alpha-\beta-\half w_2,\beta ; P, \xi )
}

\appendix{C}{Some details in the derivation of \compare }

The strategy to do the integral was explained below
\dfnmu. We must evaluate two terms separately, called
the degenerate and nondegenerate orbits.

\subsec{The nondegenerate orbit}

After transforming the integration to the
strip $\CS$ and setting $d=0$, we
 simplify the $y$-dependent terms in the
exponentials to get:
\eqn\nondegi{
\eqalign{
{1 \over   \sqrt{2  z_+^2}}
\int_{\CS}
{dx dy \over  y^{2}} \sum_I
&
f_{I} \sum_{j=-\infty}^\infty{}' \exp\Biggl\{- {1 \over  y} {\pi \over  2
z_+^2 h_I^2} \cdot \cr
\cdot \biggl(j h_I - {1\over  2 \pi} (\omega,z_+)[(S_+,\omega) -
{(S_-,z_-)\over  (\omega,z_+)} ]\biggr)
&
\biggl(j h_I + {1\over  2 \pi} (\omega,z_+)[(S_+,\omega) +
{(S_-,z_-)\over  (\omega,z_+)} ]\biggr)  \Biggr\}
\cr
\sum_{\lambda\in K}
\exp\biggl\{ -  i \pi \tau (\lambda+ \beta^I)^2
-  i (\lambda  +
&
 \beta^I , \tilde P(S)_-/h_I )
+ 2\pi i (\lambda  + \beta^I , \alpha^I + j \mu) \biggr\}  \cr
\Biggl[(S_+,\omega) -
&
 {2\pi \over  (\omega,z_+)}
[h_I j + { (S_-,z_-) \over  2 \pi} ] \Biggr]
\cr}
}
where we have renamed $c \rightarrow j$,
and the prime on the sum means we omit
the term $j=0$.

The integrand is now written as a function of
$q$ and of $y$. If we first integrate over the
$x$ variable then we isolate the power $q^0$.
We can then integrate over the $y$ variable.
There is a nice cancellation
and the integral becomes
\eqn\nondegii{
\eqalign{
- 2 \sqrt{2}
\Biggl[ \sum_I \sum_{\lambda\in K}
f_{I} h_I
\sum_{ j=-\infty}^{\infty}{}'
&
{\exp\biggl\{ 2\pi i (\lambda  + \beta^I , \mu) j \biggr\}
\over
j +  (S,z)/(2\pi h_I)   }
\cr
\exp\biggl\{ -  i \pi \tau (\lambda+ \beta^I)^2 -   i (\lambda  +
&
 \beta^I , \tilde P(S)_-) /  h_I
+ 2\pi i (\lambda  + \beta^I , \alpha^I  ) \biggr\}  \Biggr]_{q^0} \cr
}
}

The sum over $j$ can be done using
the identity:
\eqn\triide{
\eqalign{
\sum_{ j=-\infty}^\infty{}' {e^{i \theta j} \over  j+A} &=
-{1 \over  A} + {2 \pi i } {e^{-i A \theta} \over  1-e^{-2 \pi i A} }\cr}
}
which is valid for
\eqn\validfor{
0 < \theta < 2 \pi \qquad A \notin \IZ\qquad .
}
In our case we can
apply this formula with the  angle:
\eqn\defangl{
\theta = 2 \pi\bigl( (\lambda + \beta, \mu) - [(\lambda + \beta, \mu)
]\bigr) \quad \qquad 0 \leq \theta < 2 \pi
}
where $[ \cdot ]$ is the greatest integer. Moreover, define
\eqn\defaeye{
A_I \equiv {   (S, z) \over  2 \pi h_I} \quad .
}

Now we can begin to see some topological
invariance. We can combine exponentials
using the identity:
\eqn\combexp{
\eqalign{
\tilde P(S)_- +(S,z) \mu & = S^K  +
{(S, z_+ - z_-) \over  2 z_+^2} z\cr
S^K & \equiv S - (S,z)z' \cr}
}
On the right hand side of
\combexp\  $S^K$  is topological, and
projects to $K$. The second term is metric dependent
and changes continuously within chambers, but has
zero inner product with all vectors in $K$. Define an
angle:
\eqn\defpsieye{
\psi_I \equiv  { 1  \over  h_I}
\biggl[ (\lambda + \beta_I, S^K ) - [(\lambda + \beta_I, \mu)](S, z)   \biggr]
}

Using this \nondegii\ becomes:
\eqn\nondegiii{
\eqalign{
- 2 \sqrt{2}
\Biggl[ \sum_I \sum_{\lambda\in K}
f_{I} h_I
&
\exp\bigl[ -  i \pi \tau (\lambda+ \beta^I)^2
+ 2\pi i (\lambda  + \beta^I , \alpha^I  ) \bigr] \cr
\Biggl\{ - {2\pi h_I \over   (S,z)}
\exp\biggl(-  i (\lambda +
 \beta^I , \tilde P(S)_- )/h_I \biggr)  &
+ { 2 \pi i \over  1- e^{ - 2 \pi i A_I} } e^{- i \psi_I}
\Biggr\}   \Biggr]_{q^0} \cr
}
}
The expression  in curly brackets in
\nondegiii\ is a sum of two terms. The
first is {\it not} topological and varies continuously
with metric within chambers, while the second term is a nice
topological expression, within each chamber.
Equation \nondegiii\
should be regarded as a formal series in
$S$. The pole terms cancel between topological
and nontopological pieces.

\subsec{The degenerate orbit}

Returning to \cdint\ we consider the term with
$c=d=0$:
\eqn\dgcdint{
 {1 \over
\sqrt{2  z_+^2}} \int_{\CF }
{dx dy \over  y^{2}} \sum_I
\hat f_{I}
\bar \Theta_{(\Gamma+\beta^I)\cap z^\perp/z}
(\tau,   \alpha^I, 0; \tilde P(\xi^I))
\Biggl[(S_+,\omega) -
{ (S_-,z_-)  \over  (\omega, z_+) } \Biggr]
 }

Using the identities
\eqn\sigsqr{
S^2 - (\tilde P(S)_-)^2 = S_+^2 - {(S_-,z_-)^2\over  z_+^2}
=(S,z) (S, z_+ - z_-){1 \over  z_+^2}
}
and
\eqn\sigiii{
\Biggl[(S_+,\omega) -
{ (S_-,z_-)  \over  (\omega, z_+) } \Biggr]
= {1 \over  \sqrt{z_+^2}} (S, z_+ - z_-)
}
and isolating the $1/y$ dependence in the
exponential we find that
the integrand can be written as a
total derivative ${d \over  d \bar \tau}$
of a modular invariant expression. In the
standard way only the constant term at
$\tau \rightarrow i \infty$ contributes.
Note that we must work with formal series
expressions in $S$.

Doing the integral by parts we find a
metric-dependent expression:
\eqn\degiii{
\eqalign{
-  4\pi \sqrt{2}   {1 \over  (S,z)}
\Biggl[ \sum_I &
f_{I} h_I^2 \sum_{\lambda\in K}
 \cr
\exp\biggl\{ -  i \pi \tau (\lambda+ \beta^I)^2
+2\pi i (\lambda  + \beta^I , \alpha^I  )
&
-  i (\lambda  +
 \beta^I , \tilde P(S)_- )/h_I  \biggr\}
\Biggr]_{q^0}
\cr}
}
This metric-dependent expression exactly
cancels
cancels the metric-dependent term in the
nondegenerate orbit!

\listrefs
\end